\documentclass[fleqn,usenatbib]{mnras}

\usepackage{newtxtext,newtxmath}

\usepackage[T1]{fontenc}

\DeclareRobustCommand{\VAN}[3]{#2}
\let\VANthebibliography\thebibliography
\def\thebibliography{\DeclareRobustCommand{\VAN}[3]{##3}\VANthebibliography}

\usepackage{graphicx}	
\usepackage{amsmath}	
\usepackage{amssymb}	

\title[Characteristic Time Variability of GW and $\nu$]{Characteristic Time Variability of Gravitational-Wave and Neutrino Signals from Three-dimensional Simulations of Non-Rotating and Rapidly Rotating Stellar Core-Collapse}

\author[Shibagaki, Kuroda, Kotake, \& Takiwaki]{
Shota Shibagaki,$^{1}$
Takami Kuroda$,^{2,3}$
Kei Kotake,$^{1,4}$ and 
Tomoya Takiwaki$^{5}$
\\
$^1$Department of Applied Physics, Fukuoka University, 8-19-1, Nanakuma, Fukuoka, 814-0180, Japan\\
$^2$Institut f{\"u}r Kernphysik, Technische Universit{\"a}t Darmstadt, Schlossgartenstrasse 9, D-64289 Darmstadt, Germany\\
$^3$Max-Planck-Institut f{\"u}r Gravitationsphysik, Am M{\"u}hlenberg 1, D-14476 Potsdam-Golm, Germany\\
$^4$Research Insitute of Stellar Explosive Phenomena (REISEP), Fukuoka University, Nanakuma 8-19-1, Johnan, Fukuoka 814-0180, Japan\\
$^5$Division of Science, National Astronomical Observatory of Japan (NAOJ), 2-21-1, Osawa, Mitaka, Tokyo, 181-8588, Japan
}

\date{Accepted XXX. Received YYY; in original form ZZZ}

\pubyear{2020}

\begin{document}
\label{firstpage}
\pagerange{\pageref{firstpage}--\pageref{lastpage}}
\maketitle

\begin{abstract}
We present results from full general relativistic three-dimensional hydrodynamics simulations of stellar core collapse of a 70\,M$_\odot$ star with spectral neutrino transport. To investigate the impact of rotation on non-axisymmetric instabilities, we compute three models by parametrically changing the initial strength of rotation. The most rapidly rotating model exhibits a transient development of the low-$T/|W|$ instability with one-armed spiral flow at the early postbounce phase. Subsequently, the two-armed spiral flow appears, which persists during the simulation time. The moderately rotating model also shows the growth of the low-$T/|W|$ instability, but only with the two-armed spiral flow. In the nonrotating model, a vigorous activity of the standing accretion-shock instability (SASI) is only observed. The SASI is first dominated by the sloshing mode, which is followed by the spiral SASI until the black hole formation. We present a spectrogram analysis of the gravitational waves (GWs) and neutrinos, focusing on the time correlation. Our results show that characteristic time modulations in the GW and neutrino signals can be linked to the growth of the non-axisymmetric instabilities. We find that the degree of the protoneutron star (PNS) deformation, depending upon which modes of the non-axisymmetric instabilities develop, predominantly affects the characteristic frequencies of the correlated GW and neutrino signals. We point out that these signals would be simultaneously detectable by the current-generation detectors up to $\sim10$\,kpc. Our findings suggest that the joint observation of GWs and neutrinos is indispensable for extracting information on the PNS evolution preceding the black hole formation.
\end{abstract}




\section{INTRODUCTION}
Core-collapse supernovae (CCSNe) are triggered by the gravitational collapse of the central iron core of the massive stars.
Zero-age main sequence (ZAMS) masses of the progenitors are in the range roughly between 10--150 solar masses \citep[e.g.][]{Heger05}.
In the collapsing massive stars, a shock wave is launched at bounce, however stalls in the core. In the neutrino mechanism, the stalled shock is considered to be revived by neutrino heating of material behind the shock
(e.g. \citet{janka16} for a review).
Although what progenitor finally explodes as a CCSN is an open question, recent observations have been successful in the direct identification of supernova progenitors, which indicates that their progenitor masses would be less than about 18 solar masses \citep{smartt15}.
By contrast, if the shock revival is failed, or if the revived shock is too weak to unbind the stellar mantle and a substantial amount of material falls back on to a central protoneutron star (PNS), the PNS collapses to a black hole (BH).
Recent monitoring of a million supergiants has confirmed a disappearing star without an SN \citep{Kochanek08,Gerke15,Adams17a,Adams17b,Basinger20}.
These observational findings encourage investigation of the underlying physics of the gravitational collapse for failed supernovae as well as successful supernovae.

Many recent supernova simulations have been performed in three dimensions (3D) with sophisticated neutrino transport schemes, and a growing number of the models report the onset of neutrino-driven explosions in 3D \citep{takiwaki12,takiwaki14,takiwaki16,Melson15a,BMuller15a,lentz15,roberts16,Ott18,Chan&Muller18,BMuller18,Nakamura19,radice19,Vartanyan19a,Burrows19,Glas19a,Vartanyan19b,Nagakura19,Burrows20}.
In these models, multi-dimensional (multi-D) fluid motions such as neutrino-driven convection play a key role to facilitate the multi-D neutrino heating  mechanism. Moreover, some models with high mass accretion rates showed the emergence of standing accretion shock instability (SASI) \citep{blon03,Foglizzo06} which globally deforms the shock morphology.
Several 3D simulations also explored the impact of rotation of progenitor star on the SASI and the explosion.
In rapidly rotating progenitor models, \citet{Summa18} found a vigorous spiral SASI activity that expands the gain region and facilitates the neutrino heating.
\citet{takiwaki16} observed a spiral flow due to low-$T/|W|$ instability \citep[e.g.][]{Ott05}, which directly transports energy from the PNS core to the post-shock region.
Both of them showed the successful explosions for the rapidly rotating progenitors that otherwise did not explode in the absence of rotation.

Gravitational waves (GWs) and neutrinos emitted from the SN core carry the imprints of the dominant multi-dimensional dynamics inside the PNS.
Some of the previous studies reported that the SASI-dominated models showed a time variability with relatively low frequency ($\lesssim 200$\,Hz) in both GWs \citep{andresen19,Mezzacappa20} and neutrinos \citep{Walk18,Walk20,Nagakura21}, meanwhile there were no clear low-frequency feature for the convection-dominated models \citep{KurodaT18,Vartanyan19b,powell19,Powell20}.
The SASI modulation in neutrino emission is also known to be highly directionally dependent \citep[e.g.][]{Tamborra14}.
The rapidly rotating model that explodes due to the low-$T/|W|$ instability in \citet{takiwaki18} exhibited a quasi-periodic modulation of the neutrinos for an observer on the equatorial plane perpendicular to the rotation axis, while the GW frequency  is about two times higher that the neutrino modulation frequency.
The lepton emission self-sustained asymmetry (LESA) first found in \citet{Tamborra14ApJ} also potentially leads to anisotropic emission of (anti-)electron-type neutrinos \citep{O'Connor18,Glas19b,Walk19,Vartanyan19b,Walk20,Stockinger20}.
Measuring these signals would provide us observational constraints on the successful/failed explosion mechanisms of CCSNe.

Many of BH formation simulations, in which the shock revival failed have been carried out in one dimension (1D) so far \citep[e.g.][]{Liebendorfer04,sumi06BH,sumi07BH,sumi08BH,Fischer09,Nakazato13}.
In the successful explosion models, the neutrino luminosities generally decrease when the mass accretion rate drops due to the shock revival.
Meanwhile, the neutrino luminosities in the failed supernova models remain high due to the continuous mass accretion.
In the failed explosion models, it is expected that, when the central PNS gravitationally collapses into a BH, two competing effects, namely the gravitational redshift and rapid increase of the PNS core temperature, determine the drastically changing emergent neutrino energy.
The drastic change, however, suddenly ceases by the formation of the BH horizon and disappearance of neutrino spheres swallowed by it \citep{sumi06BH}.
Recent systematic studies in spherical symmetry have explored the dependencies of various progenitor models and nuclear equations of state (EOSs) on the explodability and the remnant properties such as newly born BH masses \citep[e.g.][]{O'Connor11,Ugliano12,Sukhbold16,Ebinger19,Schneider20,Warren20}.
\citet{Pan18} performed two-dimensional (2D) axisymmetric core-collapse simulations for a BH-forming progenitor with various EOSs to investigate the impacts of EOSs on the BH formation as well as on the GW and neutrino emission.
These previous studies in lower dimensions provide a basis to understand the results of full 3D simulations of the failed CCSNe.

Only a few studies have followed the BH formation in full 3D hydrodynamic simulations with sophisticated neutrino transports \citep{KurodaT18,Chan&Muller18,Walk20}.
Aside from these BH-forming models, some of previous studies showed neither a clear BH formation nor a successful shock revival during the simulation time.
For these models, we can naively expect that they may eventually form a BH after the simulation time.
Similarly to the successful explosion models, the characteristic modulation due to the SASI was seen in the GWs \citep{KurodaT17,Andresen17,O'Connor18,andresen19,Vartanyan19b,Powell20} and the neutrinos \citep{Tamborra13,Tamborra14,KurodaT17,O'Connor18,Walk18,Vartanyan19b,Walk20}.
\citet{KurodaT18} performed 3D general relativistic (GR) simulations for a 70\,M$_\odot$ zero-metallicity progenitor until the BH formation.
In this model, the dynamics just prior to the BH formation was dominated by both the SASI and convection, which produce broad-band GW signals ($\sim100$--2000\,kHz).
\cite{Walk20} carried out 3D BH-forming simulations for 40 and 75\,M$_\odot$ progenitors.
The 40\,M$_\odot$ model showed dipolar SASI activity while both dipolar and quadrupolar SASI was seen in the 75\,M$_\odot$ model.
They pointed out that the modulations characterized by each of the SASI modes may be observed in the neutrino signals.

In this work, we explore the impacts of fluid instabilities on the GW and neutrino signals from a BH-forming progenitor with various rotations, and reveal characteristic features of GWs and neutrinos and their correlation.
We perform full 3D-GR three-flavour spectral neutrino transport simulations of stellar core collapse of a 70\,M$_\odot$ star with three different initial rotation.
We carry out the time-frequency analysis of the GW and neutrino signals obtained from our simulations and make a comparison between them to find their correlation.
We also discuss the detectability of the characteristic features of the GWs, including their circular polarization, and those of the neutrinos.

Our paper is structured as follows. 
We briefly introduce our radiation-hydrodynamics code, input physics, and initial conditions in Section \ref{sec2}.
In Section \ref{sec3}, we present the results of our simulations.
We compare the dynamics, GWs, and neutrinos between our models.
We also explore the detectability of their signals.
We summarize our results in Section \ref{sec4}.

\section{NUMERICAL METHOD AND Initial Models}\label{sec2}
 Our 3D-GR code is essentially the same as the one developed by \citet{KurodaT16}, 
 except several updates in the directionally unsplit predictor-corrector scheme instead of the Strang splitting scheme. 
The metric evolution is solved in fourth-order accuracy 
 by a finite-difference scheme in space and with a Runge-Kutta method in time.
 Using an M1 analytical closure scheme \citep{Shibata11},
 we solve spectral neutrino transport of the radiation energy and momentum including all the gravitational redshift and Doppler terms. In this work, 12 energy bins that logarithmically
spread from 1 to 300 MeV are employed. We consider three-flavour of neutrinos ($\nu \in \nu_e,\bar\nu_e,\nu_x$) with 
$\nu_x$ being the heavy-lepton neutrinos.
  Concerning neutrino opacities, the standard weak interaction set in \citet{Bruenn85} plus nucleon-nucleon Bremsstrahlung is included
 (see \citet{KurodaT16} for more detail).

We use a 70\,M$_{\odot}$ zero-metallicity star of \cite{Takahashi14}.
At the precollapse phase, the mass of the central iron core is $\sim4.6$\,M$_{\odot}$ and the enclosed mass up to the helium layer is $\sim 34$\,M$_{\odot}$.
The central density profile of this progenitor is similar to that of 75\,M$_\odot$ ultra metal-poor progenitor model of \citet{WHW02} that was used as a collapsar model in \citet{Ott11} (see also \citet{Walk20}).
The central angular velocity of the original progenitor model is $\sim 0.03$\,rad\,s$^{-1}$. The ratio of the 
rotational to gravitational energy, $T/|W|$,  at bounce is $\sim10^{-5}$.
For such slow rotation, non-axisymmetric rotational instabilities
 are unlikely to develop, and the postbounce dynamics would not be significantly deviated from that of the non-rotating model.
To clearly see the impact of rotation, we parametrically change the initial strength of rotation. Assuming a cylindrical rotation profile, we impose the initial angular momentum of the core as
\begin{equation}
    u^t u_{\phi}=\varpi^2_0 (\Omega_0-\Omega),
\end{equation}
where $u^t$ is the time component of the contravariant four-velocity, $u_{\phi}\equiv \varpi^2\Omega$ with $\varpi=\sqrt{x^2+y^2}$, and $\varpi_0$ is set as $10^8$\,cm.
We simulate three models by changing the initial central angular velocity $\Omega_0=0, 1 ,2$\,rad\,s$^{-1}$.
For reference, the initial rotational energy and $T/|W|$ are $\sim1.3\times10^{45}$\,erg and $\sim2\times10^{-7}$ for the original progenitor model and are $\sim2.6\times10^{49}$ $(\sim6.5\times10^{48})$\,erg and $\sim3\times10^{-3}$ $(8\times10^{-4})$ for the $\Omega_0 =2$ $(1)$\,rad\,s$^{-1}$ models, respectively.
We use the EOS by \citet{LSEOS} with nuclear incompressibility of $K = 220$\,MeV (LS220). 
The 3D computational domain is a cubic box
with 15,000\,km width, and nested boxes with nine refinement
levels are embedded in the Cartesian coordinates. Each box contains $64^3$ cells and the
minimum grid size near the origin is $\Delta x=458$\,m.
The PNS core surface ($\sim10$\,km) and stalled shock ($\sim110$--300\,km) are resolved by $\Delta x=458$\,m and $7.3$\,km, respectively. 
Our simulations proceed about 100~ms per a month with 512 cores of Cray XC50 at the Center for Computational Astrophysics, National Astronomical Observatory of Japan, and of Cray XC40 at YITP in Kyoto University.
$t_{\mathrm{pb}}$ represents the time measured after core bounce.

\section{RESULT}\label{sec3}

In this section, we start to overview the shock evolution of the computed models in Section \ref{sec:shock}.
Then in Section \ref{sec:TW}, we focus on the development of the low-$T/|W|$ instability in our rotating models.
In Sections \ref{sec:GW} and \ref{sec:neu}, we  investigate the characteristic time 
modulation in the GW and neutrino emission and discuss their detectability.

\subsection{Shock Evolution} \label{sec:shock}
\begin{figure}
\centering
\includegraphics[width=0.475\textwidth]{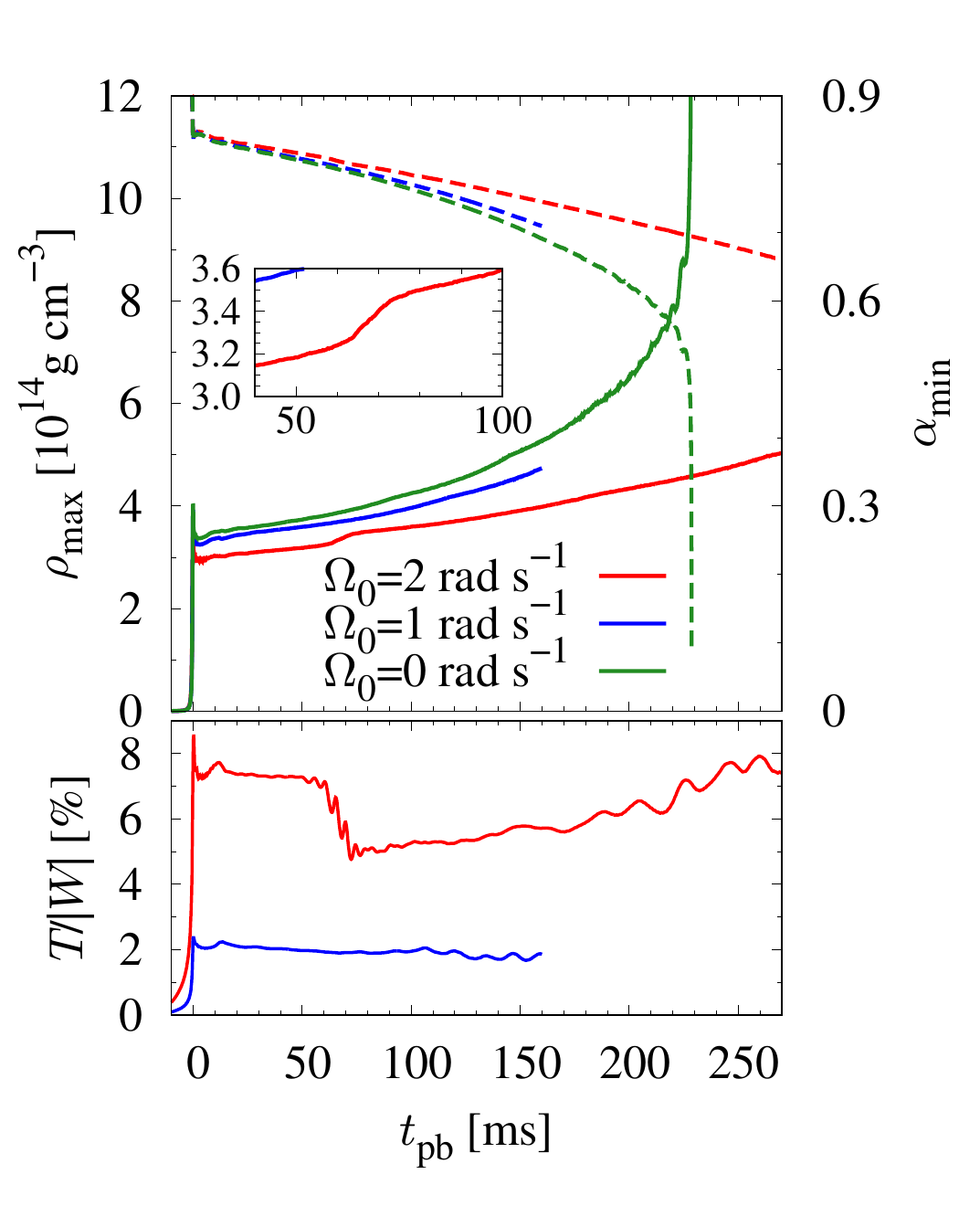}
\caption{Evolution of the maximum density (top panel; solid lines), minimum lapse function (top panel; dashed lines), and the ratio of the rotational energy to the gravitational energy (bottom panel) for the $\Omega_{0}=2$\,rad\,s$^{-1}$ (red lines), $\Omega_{0}=1$\,rad\,s$^{-1}$ (blue lines), and $\Omega_{0}=0$\,rad\,s$^{-1}$ (green lines) models. The inset minipanel highlights  a steep rise of the maximum density for the $\Omega_{0}=2$\,rad\,s$^{-1}$ model due to energy transport by the low-$T/|W|$ instability (see text).} \label{fig:central}
\end{figure}

The postbounce shock evolution is significantly affected by the PNS contraction.
To overview the process of the central PNS core evolution, we show the evolution of the maximum density, minimum lapse function, and  ratio of the rotational energy to the gravitational energy in Fig.~\ref{fig:central}. 
In this figure, the solid and dashed lines indicate the maximum density and minimum lapse function shown by the left and right vertical axes, respectively.
The colour of each line denotes the model difference.

The $\Omega_{0} = 0$\,rad\,s$^{-1}$ model (green line) always shows the largest maximum density among the three models during the simulation time.
At the final simulation time $t_{\mathrm{pb}} \sim 230$\,ms, the maximum density and minimum lapse function exhibit a steep increase and decrease, respectively, indicating the BH formation\footnote{Since \citet{KurodaT18} underestimated the cooling/heating rates of $\nu_x$ due to a loose convergence criterion in the implicit method, the simulation using the same progenitor showed the later BH formation time $t_{\mathrm{pb}}\sim300$\,ms.}.
Compared with the non-rotating model, the maximum density of our rapidly rotating model (with $\Omega_0 = 2$\,rad\,s$^{-1}$, red line) shows a slower increase after bounce due to the centrifugal forces. The evolution of the moderately rotating model ($\Omega_0 = 1$\,rad\,s$^{-1}$, blue line) lies in-between.

At the end of the simulations, the $\Omega_{0} = 0$\,rad\,s$^{-1}$ model finally reaches the PNS mass of $\sim2.5$\,M$_{\odot}$, here the PNS is defined where the density is larger than $10^{11}$\,g\,cm$^{-3}$, while the PNS mass of the $\Omega_{0} = 2\ (1)$\,rad\,s$^{-1}$ model is $\sim2.6\ (2.2)$\,M$_{\odot}$. At this time the rotation periods of the PNS for the $\Omega_{0} = 2$ and 1\,rad\,s$^{-1}$ models, estimated from the moment-of-inertia-weighted mean angular velocity \citep[see e.g.][]{ott_birth}, are $\sim$2\,ms and $\sim$6\,ms, respectively.

\begin{figure}
\centering
\includegraphics[width=0.475\textwidth]{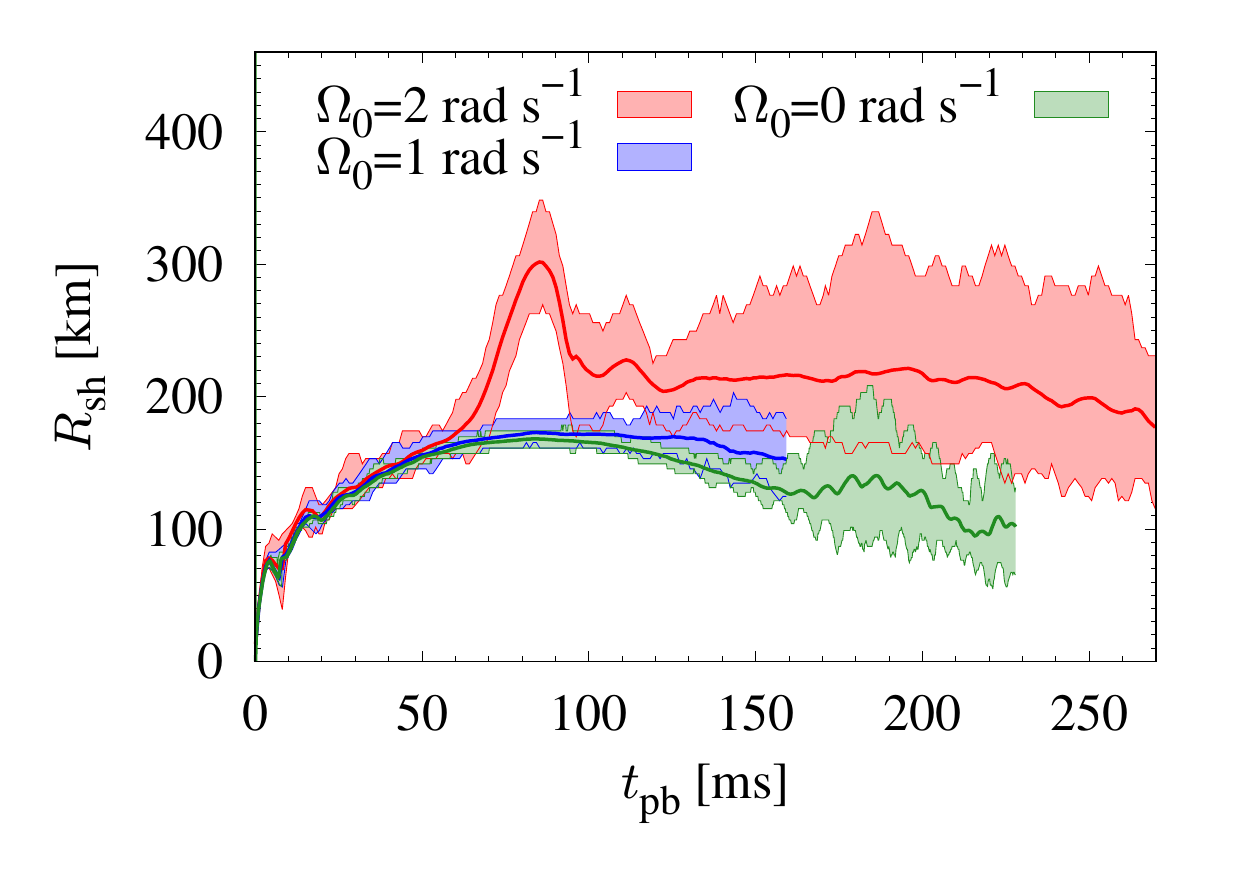}
\caption{Evolution of shock radius for the $\Omega_{0}=2$\,rad\,s$^{-1}$ (red line), $\Omega_{0}=1$\,rad\,s$^{-1}$ (blue line) and $\Omega_{0}=0$\,rad\,s$^{-1}$ (green line) models. The coloured bands indicate the range of the shock radius from minimum to maximum.
The central thick lines in the colour bands indicate the mean shock radius. \label{fig:shock_r}}
\end{figure}

\begin{figure}
\centering
\includegraphics[ width=0.475\textwidth]{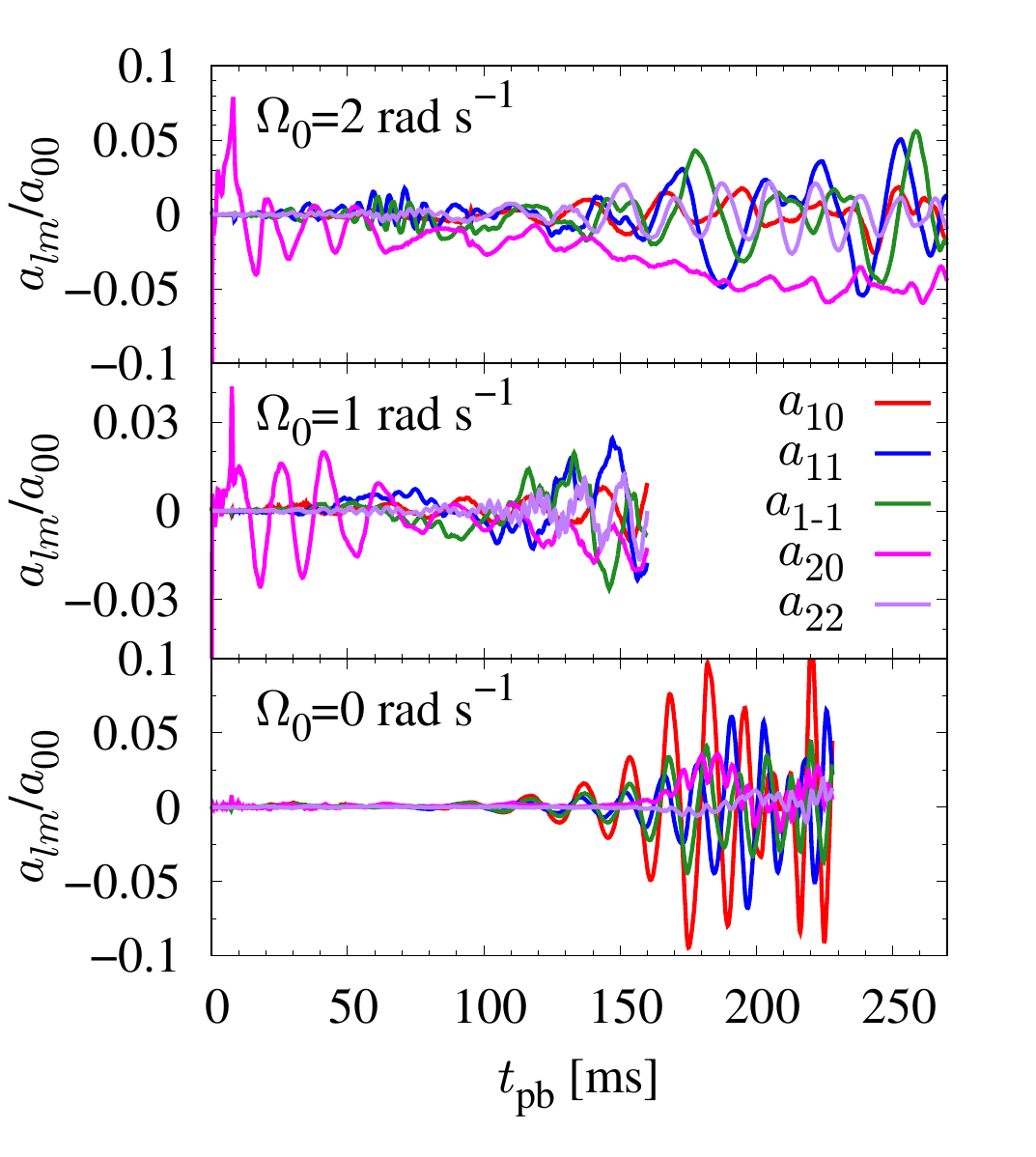}
\caption{Evolution of the selected (normalized) spherical harmonic coefficients of the shock surface, $a_{lm}/a_{00}$, for the $\Omega_{0}=2$\,rad\,s$^{-1}$ (top panel), $\Omega_{0}=1$\,rad\,s$^{-1}$ (middle panel) and $\Omega_{0}=0$\,rad\,s$^{-1}$ (bottom panel) models. \label{fig:shock_Ylm}}
\end{figure}

To investigate more quantitatively the shock evolution of each model, we next present the evolution of the minimum, maximum, and averaged shock radii ($R_{\rm{sh}}$) in Fig.~\ref{fig:shock_r}.
The maximum and minimum shock radii correspond to the upper and lower edge of the coloured bands, while the average shock radii are indicated by the thick solid lines. In Fig.~\ref{fig:shock_Ylm}, we also present the normalized spherical harmonics coefficients of the deformed shock surfaces ($a_{lm}/a_{00}$). Following \citet{burrows12}, $a_{lm}$ is defined as 
\begin{eqnarray}
  a_{lm}(t) = \frac{(-1)^m}{\sqrt{4\pi(2l+1)}}\oint_\Omega R_{\rm{sh}}(t,\theta,\phi) Y^m_l(\theta,\phi) d\Omega \label{eq:alm},
\end{eqnarray}
where $Y^m_l$ are the orthnormalized spherical harmonics described by the constants $N^l_m$ and the associated Legendre polynomials $P^l_m(\cos \theta)$ as follows:
\begin{eqnarray}
  Y^m_l(\theta,\phi) &=& \left\{ 
    \begin{array}{ll}
        \sqrt{2}N^m_lP^m_l\cos (m\phi) & m>0,\\
        N^0_lP^0_l & m=0,\\
        \sqrt{2}N^{|m|}_lP^{|m|}_l\sin (|m|\phi) & m<0,
    \end{array}
  \right.
  \label{eq:ylm}\\
  N_l^m&=&\sqrt{\frac{2l+1}{4\pi}\frac{(l-m)!}{(l+m)!}}.
\end{eqnarray}
In this paper, we focus on the dipole ($l=1$) and some representatives of the quadrupole ($l=2$) modes.

We begin with the shock evolution of the $\Omega_{0} = 2$\,rad\,s$^{-1}$ model.
In this model, the shock radius starts to deviate from the other models at $t_{\mathrm{pb}} \sim50$\,ms and shows a rapid expansion from $t_{\mathrm{pb}} \sim 60$\,ms (see the red band in Fig.~\ref{fig:shock_r}).
At the same time, the modes of $a_{1\pm1}$ show the out-of-phase oscillations with moderate amplitudes as indicated by the blue and green lines in the top panel of Fig.~\ref{fig:shock_Ylm}.
As we will see later in Section \ref{sec:TW}, this transient shock expansion results from the so-called low-$T/|W|$ instability, which produces a one-armed spiral flow and pushes the shock surface outward due to energy transport.
This energy transport decreases $T/|W|$ as shown in the bottom panel of Fig.~\ref{fig:central} (see also, \cite{Ott05}).
As a consequence of the weakening centrifugal force, the PNS contracts noticeably, leading to a steep increase of the maximum density as shown in the inset minipanel of Fig.~\ref{fig:central} ($t_{\rm pb} \sim 70$\,ms, red line).
After the low-$T/|W|$ instability ceases at $t_{\mathrm{pb}} \sim 85$\,ms, the shock expansion slows down and eventually turns into the recession.

As we will see in the next section \ref{sec:TW}, the low-$T/|W|$ instability sets in again at $t_{\mathrm{pb}} \sim 110$\,ms.
After the onset of the second low-$T/|W|$ instability, the maximum shock surface again starts expansion, whereas the position of the averaged shock radius hardly changes up to $t_{\mathrm{pb}} \sim 220$\,ms, when the shock starts to shrink.
In addition, the shock surface significantly deforms with more pronounced oblate deformation than the first period.
This feature is shown in the negatively large spherical harmonic coefficient of the shock surface with the $l=2$ and $m=0$ mode (namely, $a_{20}$) in the top panel of Fig.~\ref{fig:shock_Ylm}.
As we have already shown in \cite{Shibagaki20}, the onset of the second low-$T/|W|$ instability is associated with the generation of the two-armed spiral flow in the vicinity of the PNS.

\begin{figure}
\center
\includegraphics[ width=0.45\textwidth]{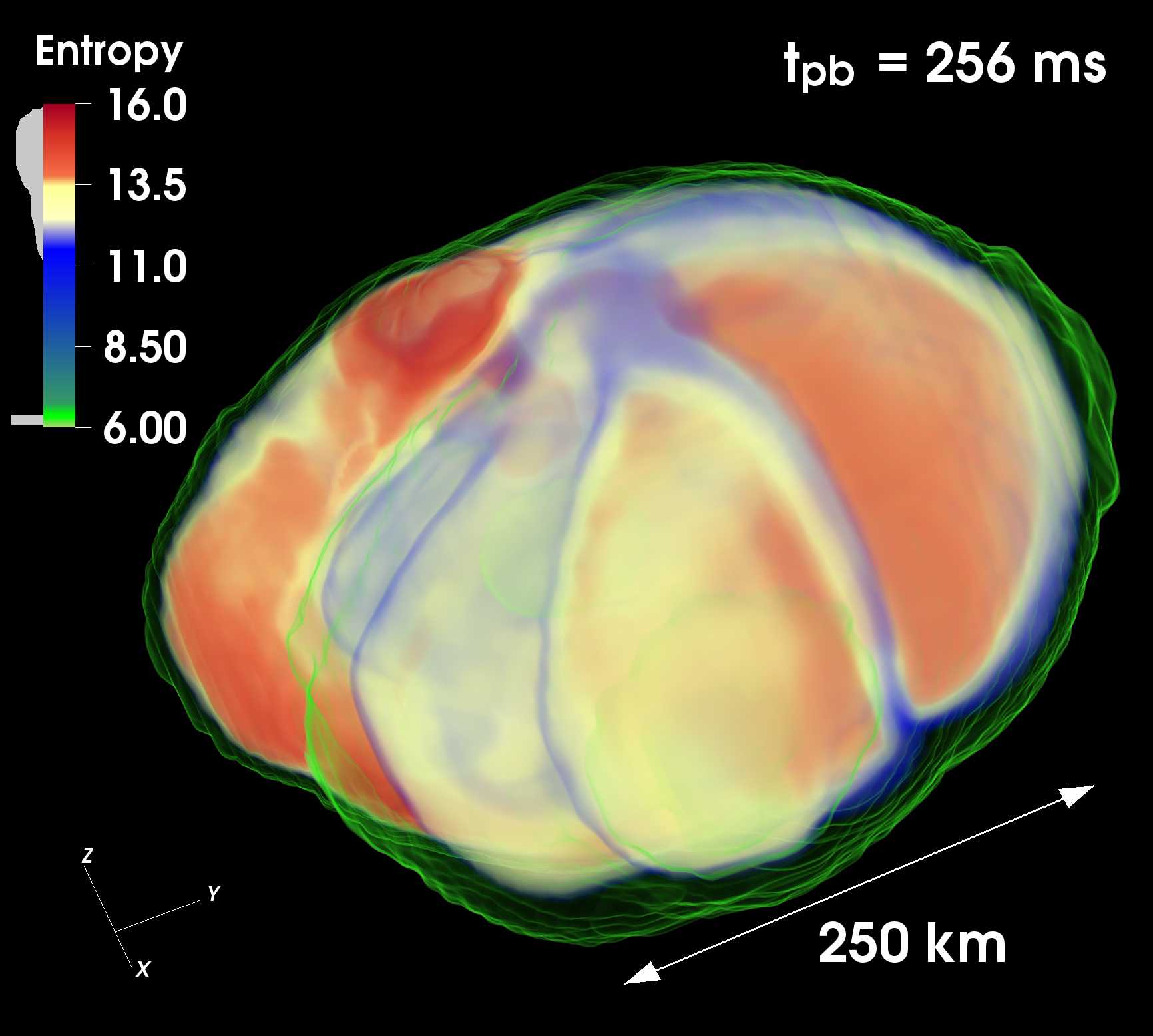}
\includegraphics[ width=0.45\textwidth]{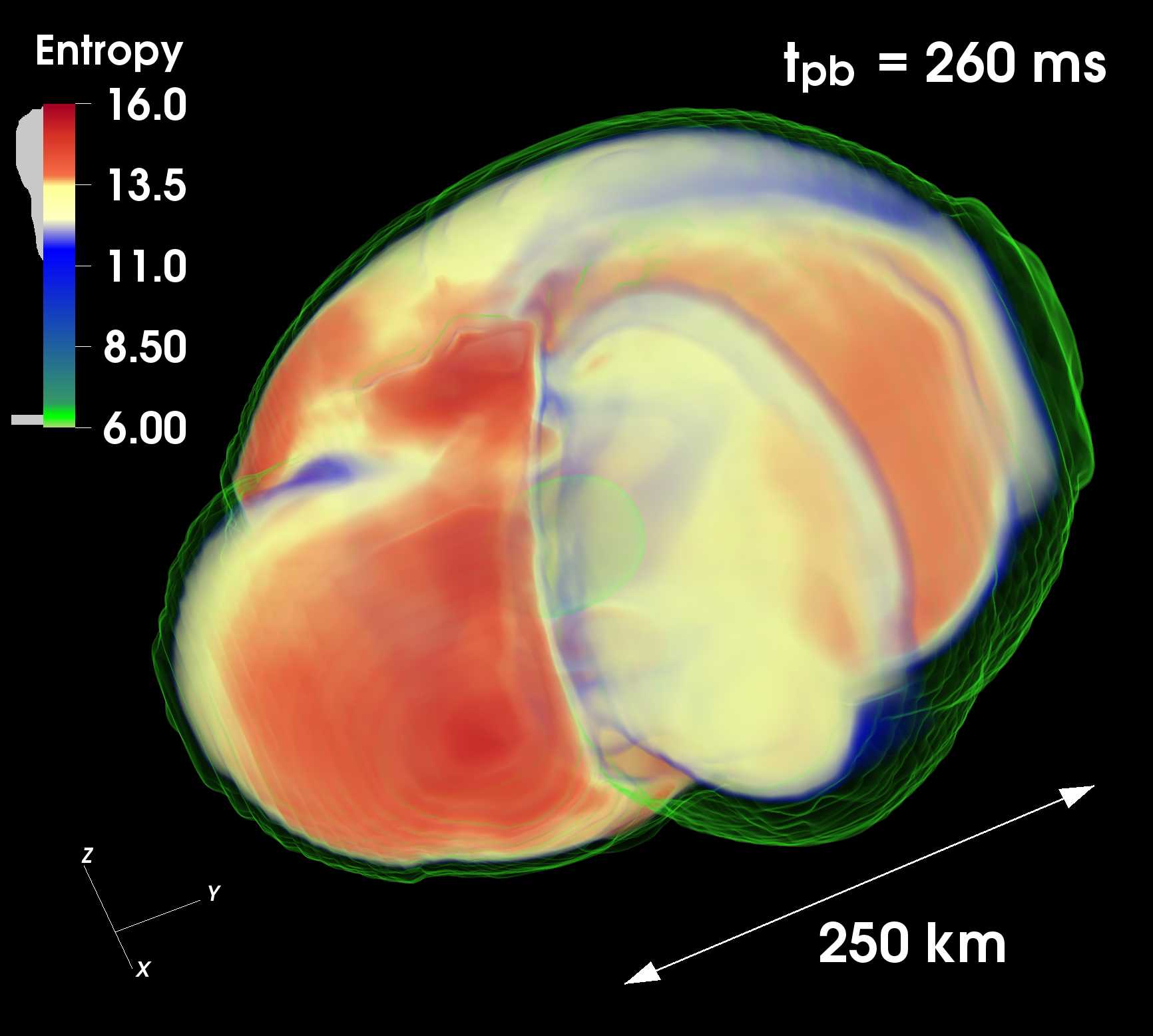}
\includegraphics[ width=0.45\textwidth]{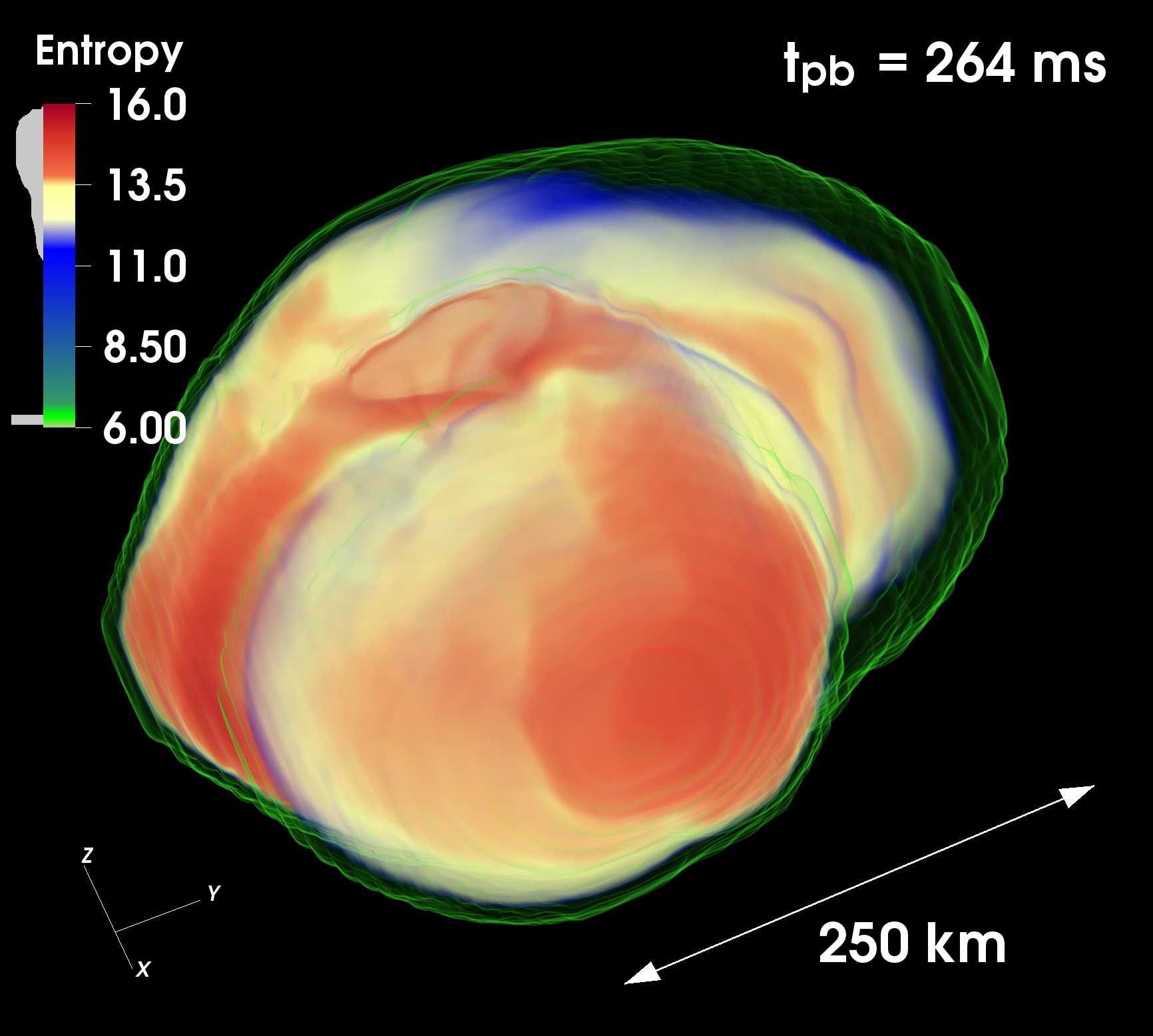}
\caption{3D entropy plots for the $\Omega_{0}=2$\,rad\,s$^{-1}$ model. The three sequential snapshots roughly corresponds to a half of the rotation period of the shock surface. Note that the $z$ axis corresponds to the spin axis. \label{fig:3dent_rot}}
\end{figure}

To clearly show the shock morphology of the model $\Omega_0 = 2$\,rad\,s$^{-1}$, we present volume renderings of the entropy at $t_{\mathrm{pb}} = 256$, $260$, and $264$\,ms (i.e. roughly at intervals of 4\,ms, from top to bottom) in Fig.~\ref{fig:3dent_rot}.
The snapshots present an oblate spheroid with an evident $|m|=1$ deformation, which can be seen particularly at near the end of simulation time $t_{\mathrm{pb}}\gtrsim256$ ms.
It is consistent with the large values of $a_{1\pm1}$ plotted in the top panel of Fig.~\ref{fig:shock_Ylm}. Seen 
from the positive $z$-axis, the $m=1$ mode rotates counterclockwise. Note that the $z$-axis corresponds to the spin axis.
The rotation period of the distorted shock surface is $\sim 20$\,ms.

The $\Omega_{0} = 1$\,rad\,s$^{-1}$ and $\Omega_{0} = 0$\,rad\,s$^{-1}$ models show different shock evolution from that of the $\Omega_{0} = 2$\,rad\,s$^{-1}$ model.
 The shock evolution between the $\Omega_{0} = 1$\,rad\,s$^{-1}$ and $\Omega_{0} = 0$\,rad\,s$^{-1}$ models in Fig.~\ref{fig:shock_r}
 is not significantly different up to $t_{\mathrm{pb}} = 160$\,ms at when the simulation of the $\Omega_{0} = 1$\,rad\,s$^{-1}$ model is stopped (because of our limited, available, computational resources).
The shock of this model stalls at a radius of $\sim170$\,km around $t_{\mathrm{pb}} \sim 80$\,ms.

 As shown in the middle panel of Fig.~\ref{fig:shock_Ylm},  the shock deformation of the $\Omega_{0} = 1$\,rad\,s$^{-1}$ model becomes larger around $t_{\mathrm{pb}} \sim 100$\,ms,
The secular drift of the $a_{20}$ mode toward the negative values indicates that the shock is deformed to be more oblate.
Furthermore, the clear modulation of the $a_{22}$ mode shows a quadrupole shock deformation along the $z=0$ plane and the deformed shock rotates along the $z$-axis with a rotation period of $\sim15$\,ms. In the meantime, the $a_{11}$ and $a_{1-1}$ modes, which are the largest amplitudes among the modes, changes stochastically with time.
The maximum amplitudes of the spherical harmonic coefficients in the $\Omega_{0} = 1$\,rad\,s$^{-1}$ model are roughly a factor of two smaller than in the other models.
This is likely due to the short simulation time.

\begin{figure}
\center
\includegraphics[ width=0.45\textwidth]{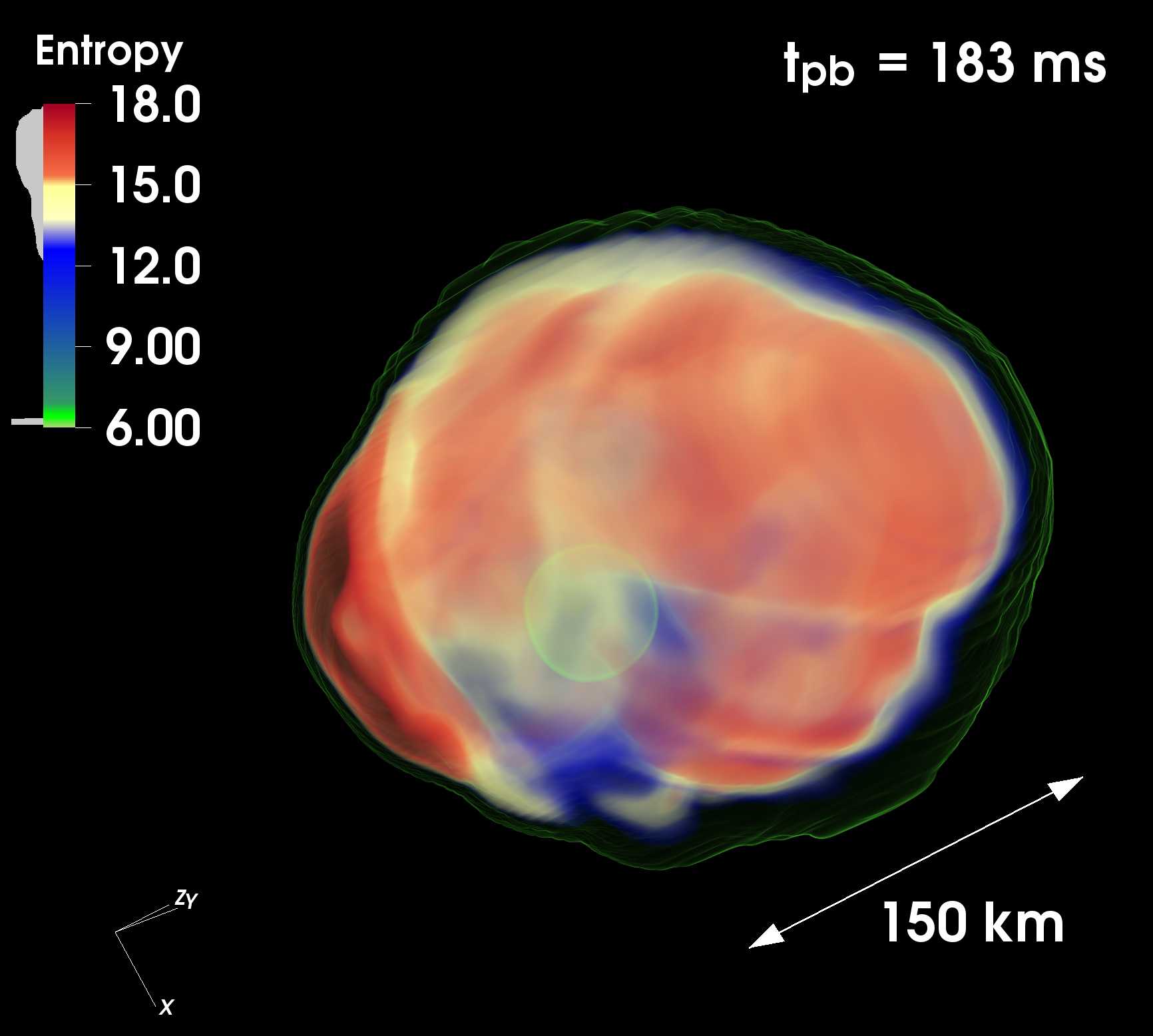}
\includegraphics[ width=0.45\textwidth]{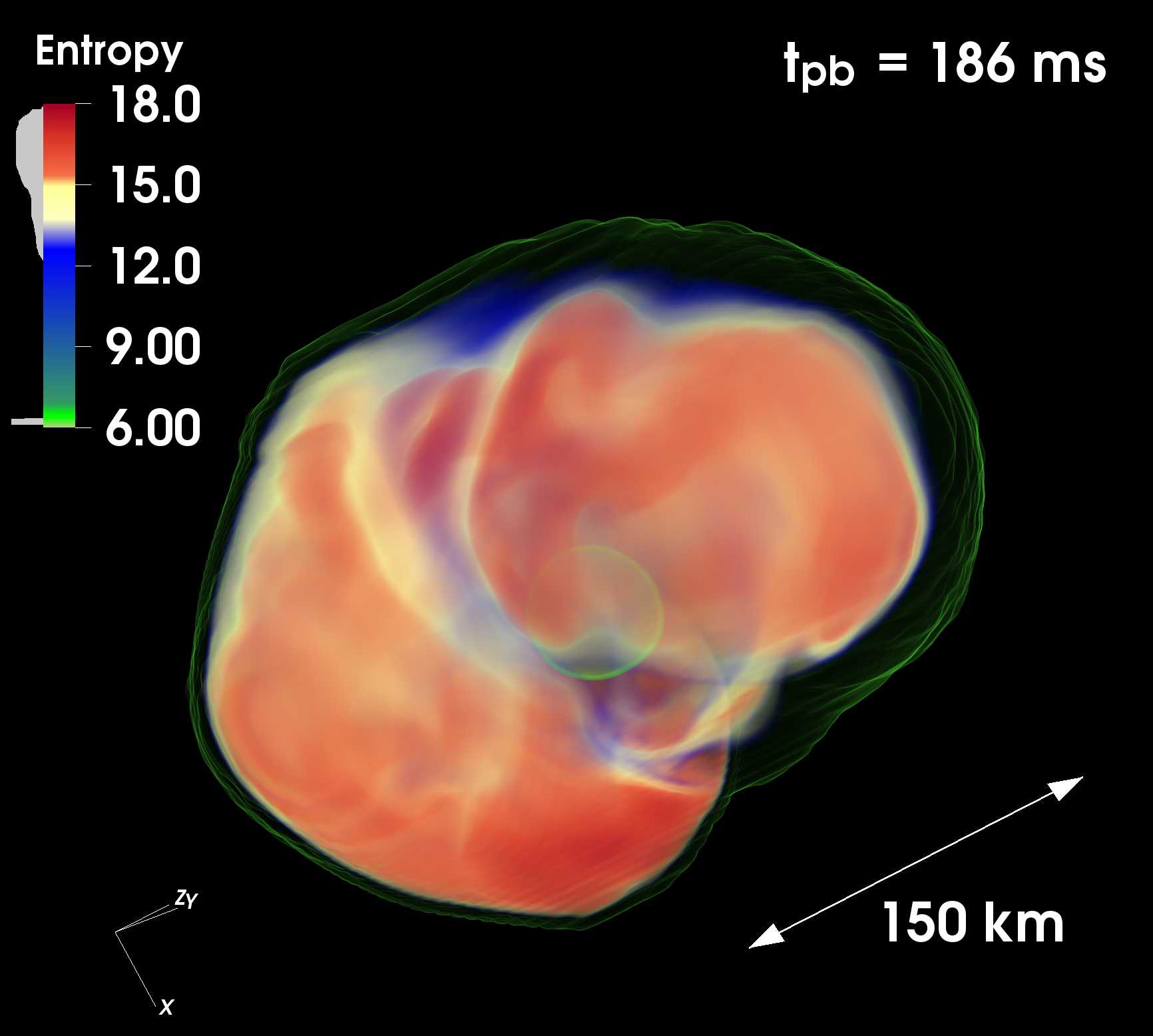}
\includegraphics[ width=0.45\textwidth]{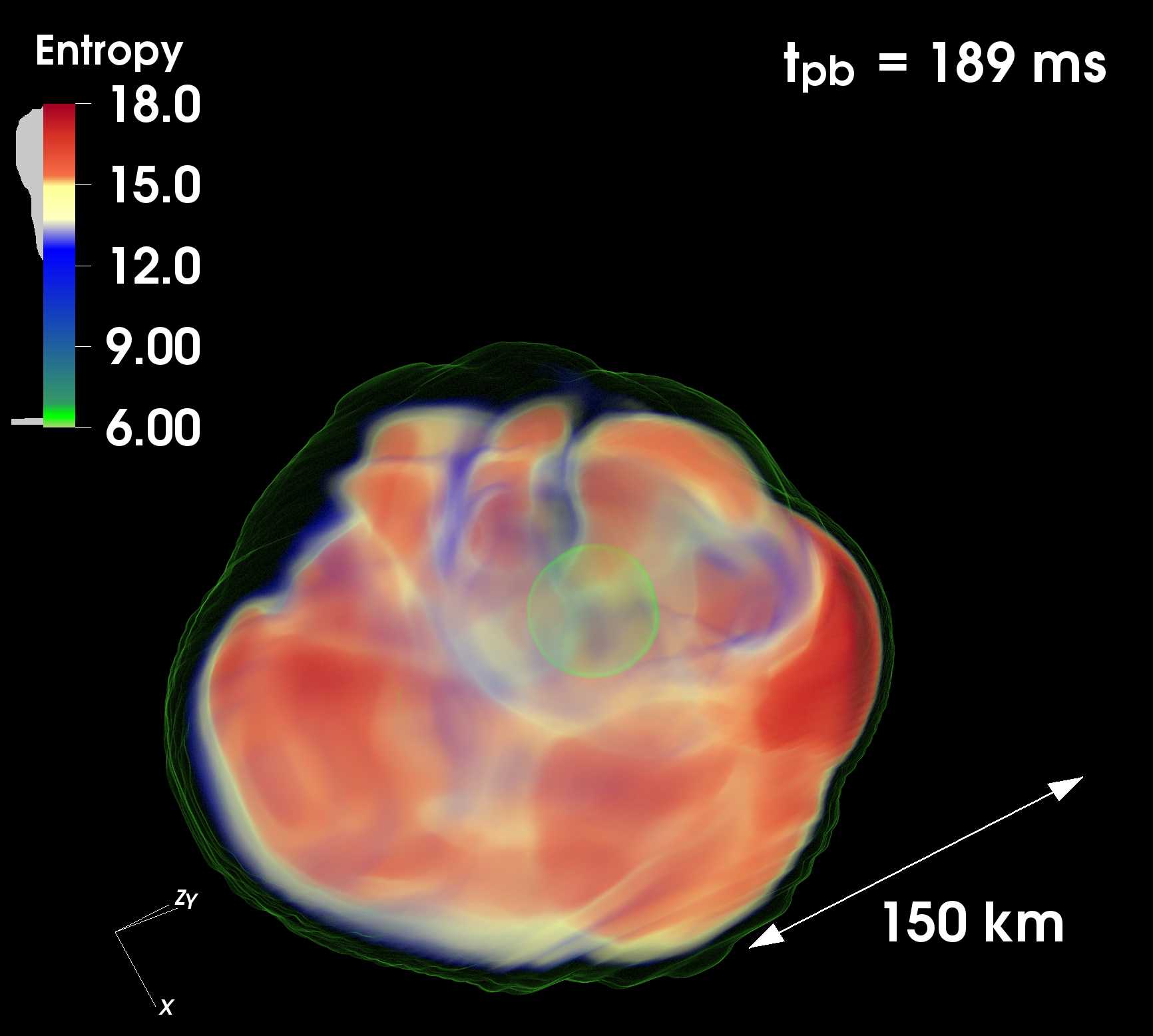}
\caption{Same as Fig.~\ref{fig:3dent_rot} but for the $\Omega_{0}=0$\,rad\,s$^{-1}$ model. The three sequential snapshots roughly corresponds to a half of the rotation period of the shock surface.\label{fig:3dent_norot}}
\end{figure}

In the $\Omega_{0} = 0$\,rad\,s$^{-1}$ model, all the coefficients with $l=1$ mode shown in the bottom panel of Fig.~\ref{fig:shock_Ylm} start to oscillate approximately in phase from $t_{\mathrm{pb}} \sim100$\,ms.
This feature indicates that the sloshing SASI mode is dominant at this time.
At later times, all the coefficients show phase-shifted oscillations from $t_{\mathrm{pb}}\sim150$\,ms, indicating that the spiral SASI takes over the
leading mode.
To explicitly exhibit the shock evolution with the spiral SASI activity, we show 3D volume renderings of entropy in Fig.~\ref{fig:3dent_norot}.
Similarly to Fig.~\ref{fig:3dent_rot}, Fig.~\ref{fig:3dent_norot} shows the time evolution from top to bottom.
We can see that the rotation axis of the spiral SASI is directed to $\theta\sim60^{\circ}$, $\phi\sim-60^{\circ}$, and the spiral motion rotates counterclockwise around the axis.
The rotational period of the spiral SASI is about $\sim10$\,ms.

\subsection{The low \texorpdfstring{$T/|W|$}{T/W} instability} \label{sec:TW}
In this section, we discuss the low-$T/|W|$ instability of our rotating models in detail.
The low-$T/|W|$ instability is a dynamical instability that drives non-axisymmetric deformation of a differentially rotating star when the ratio $T/|W|$ exceeds a certain value as low as the order of one per cent \citep{shibata02,Shibata03a}.
\citet{Watts05} firstly pointed out that a potential requirement for the occurrence of the low-T/|W| instability is the existence of the corotation radius where the angular velocity is equal to the pattern speed of the unstable mode.
In the context of rotating PNS, \citet{Ott05} was the first to report the development of the low-$T/|W|$ instability that leads to a formation of spiral flow.
Subsequent works have confirmed the similar results with more sophisticated simulations \citep{Scheidegger08,Scheidegger10,Kuroda14,takiwaki16,takiwaki18,Shibagaki20}.

Little is known about the ways to clearly distinguish the two spiral instabilities, i.e. the spiral SASI and low-T/|W| instability, that are driven by physically different mechanisms \citep[see e.g.][]{KurodaT14,Kazeroni17}. However, a comparison between the oscillation timescales of the SASI and of the low-T/|W| instability may give us a hint to recognize which instability has been likely excited.
While the timescale of SASI is $\lesssim$100\,Hz \citep[see e.g.][]{BMuller14,BMueller19}, the timescale of the low-T/|W| instability is determined by the rotation period at the corotation radius. For example, our rotating models show the rotation frequency of $\sim150$\,Hz at the corotation radius, which could take even a higher value as the PNS contracts \citep{Shibagaki20}.
Such a value is significantly higher than the expected typical SASI frequency.
Furthermore, the characteristic frequency of neutrinos as well should be characterized by the timescale of the spiral instability.
Like these, monitoring their characteristic frequencies would allow us to infer which spiral instability is developed (see Section \ref{sec:GW} and \ref{sec:neu}).

\begin{figure}
\centering
\includegraphics[width=0.46\textwidth]{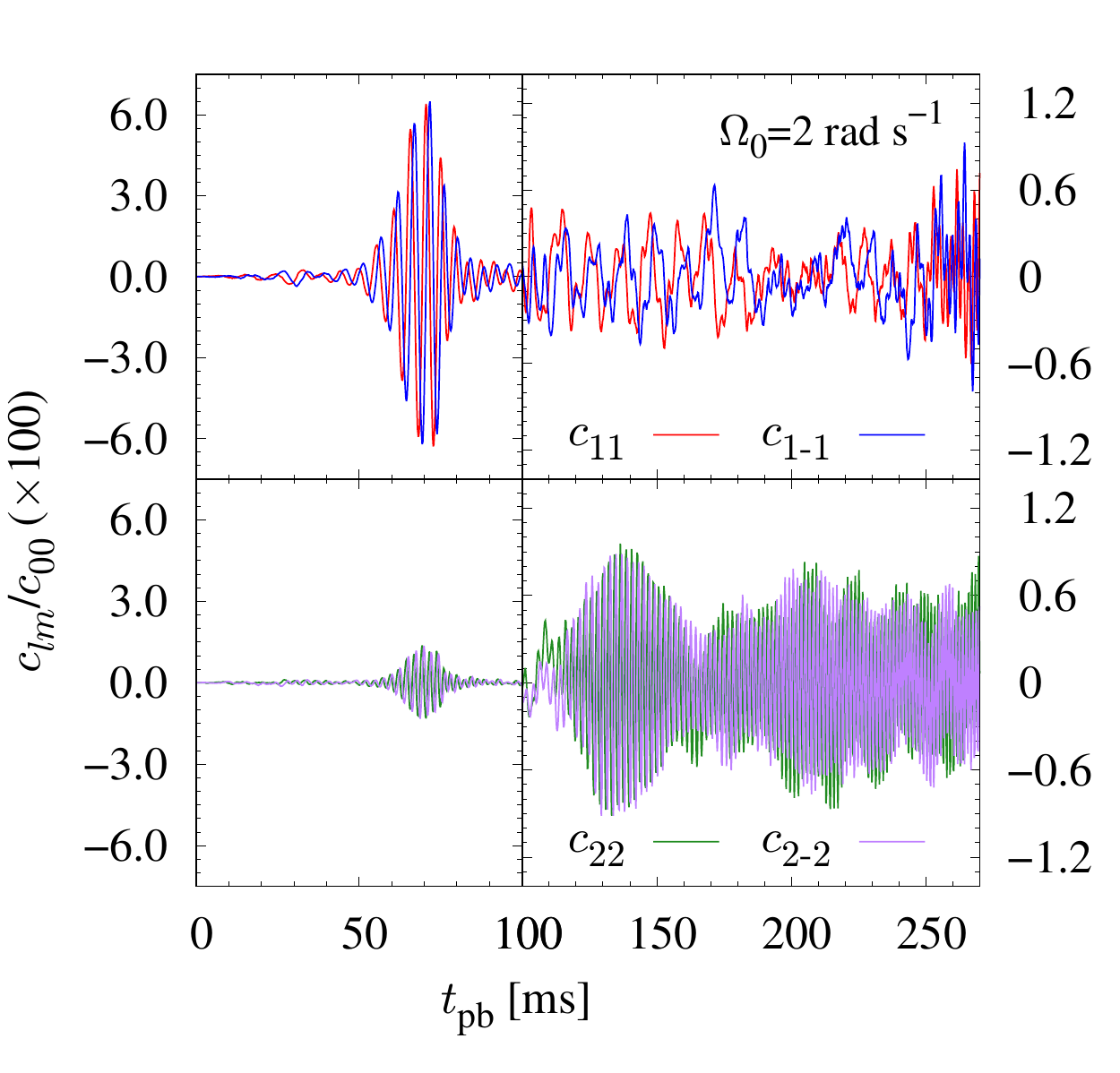}
\includegraphics[width=0.44\textwidth]{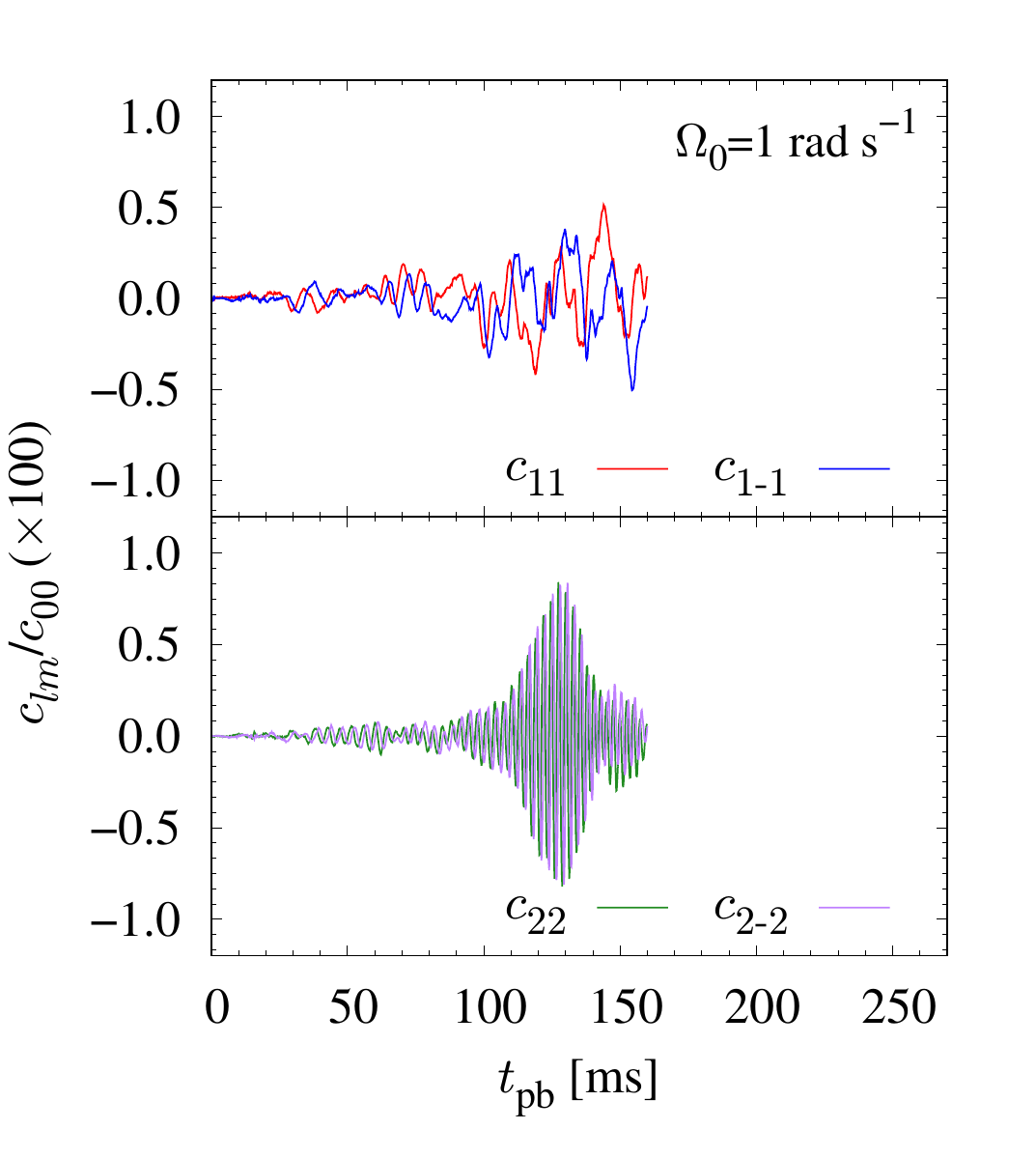}
\caption{
Evolution of the selected (normalized) spherical harmonic coefficients of the PNS surface $c_{lm}/c_{00}$, defined at the fiducial density of $\rho=10^{  11}$\,g\,cm$^{-3}$, for the $\Omega_{0} =2$\,rad\,s$^{-1}$ (top panel) and $\Omega_{0} =1$\,rad\,s$^{-1}$ (bottom panel) models, respectively. Note that the vertical scales of the top left and right panels are different. \label{fig:Rnu_Ylm}}
\end{figure}

We begin by looking at the spiral deformation of the rotating PNS as a clue to identify the onset of the low-$T/|W|$ instability.
Fig.~\ref{fig:Rnu_Ylm} displays the temporal evolution of normalized spherical harmonic coefficients of PNS radii $c_{lm}/c_{00}$.
Here $c_{lm}$ is expressed similarly to \citet{KurodaT16ApJL} as,
\begin{eqnarray}
  c_{lm}(t) = \frac{(-1)^m}{\sqrt{4\pi(2l+1)}}\oint_\Omega R_{\mathrm{PNS}}(t,\theta,\phi) Y^m_l(\theta,\phi) d\Omega, \label{eq:clm}
\end{eqnarray}
with $R_{\mathrm{PNS}}$ being the PNS radius defined at the fiducial density of $\rho = 10^{11}$\,g\,cm$^{-3}$.
Note that the fluctuations of the angle between the spin axis of the PNS and the $z$-axis is on the order of 0.1\% at most in our rotating models. Therefore we can safely see the $z$-axis as the spin axis of the rotating PNS.

In the $\Omega_{0} = 2$\,rad\,s$^{-1}$ model (top panel), the $c_{11}$ and $c_{1-1}$ modes show large amplitudes at $t_{\mathrm{pb}} = 50$--90\,ms with a phase difference of $\pi/2$.
This indicates that the PNS rotates with $m=1$ deformation around the $z$-axis (i.e. the spin axis).
After $t_{\mathrm{pb}} \sim 110$\,ms, the $m=1$ mode diminishes and is taken over by the $m=2$ mode as shown by the larger values in the $c_{22}$ and $c_{2-2}$ modes.
We note that these two modes have again a phase difference of $\pi/2$.
This means that the PNS rotates with the $m=2$ deformation around the $z$-axis.
For the $\Omega_{0} = 1$\,rad\,s$^{-1}$ model (bottom panel), on the other hand, the $c_{22}$ and $c_{2-2}$ modes show large amplitude oscillations with a phase difference of $\pi/2$ from $t_{\mathrm{pb}} \sim 90$\,ms.

Let us compare the degree of the PNS deformation in the rotating models.
The highest deformation can be seen for $c_{11}$ and $c_{1-1}$, $\sim 6 \times 10^{-2}$, of the $\Omega_{0} = 2$\,rad\,s$^{-1}$ model at $t_{\mathrm{pb}} \sim 70$\,ms.
In contrast, when $m=2$ coefficients, $c_{22}$ and $c_{2-2}$, are the dominant terms, the largest amplitudes of these modes in the $\Omega_{0} = 2$\,rad\,s$^{-1}$ model is $\sim 9 \times 10^{-3}$ at $t_{\mathrm{pb}} \sim 135$\,ms, comparable to the one ($\sim 8 \times 10^{-3}$) in the $\Omega_{0} = 1$\,rad\,s$^{-1}$ model at $t_{\mathrm{pb}} \sim 130$\,ms.

\begin{figure}
\center
\includegraphics[ width=0.40\textwidth]{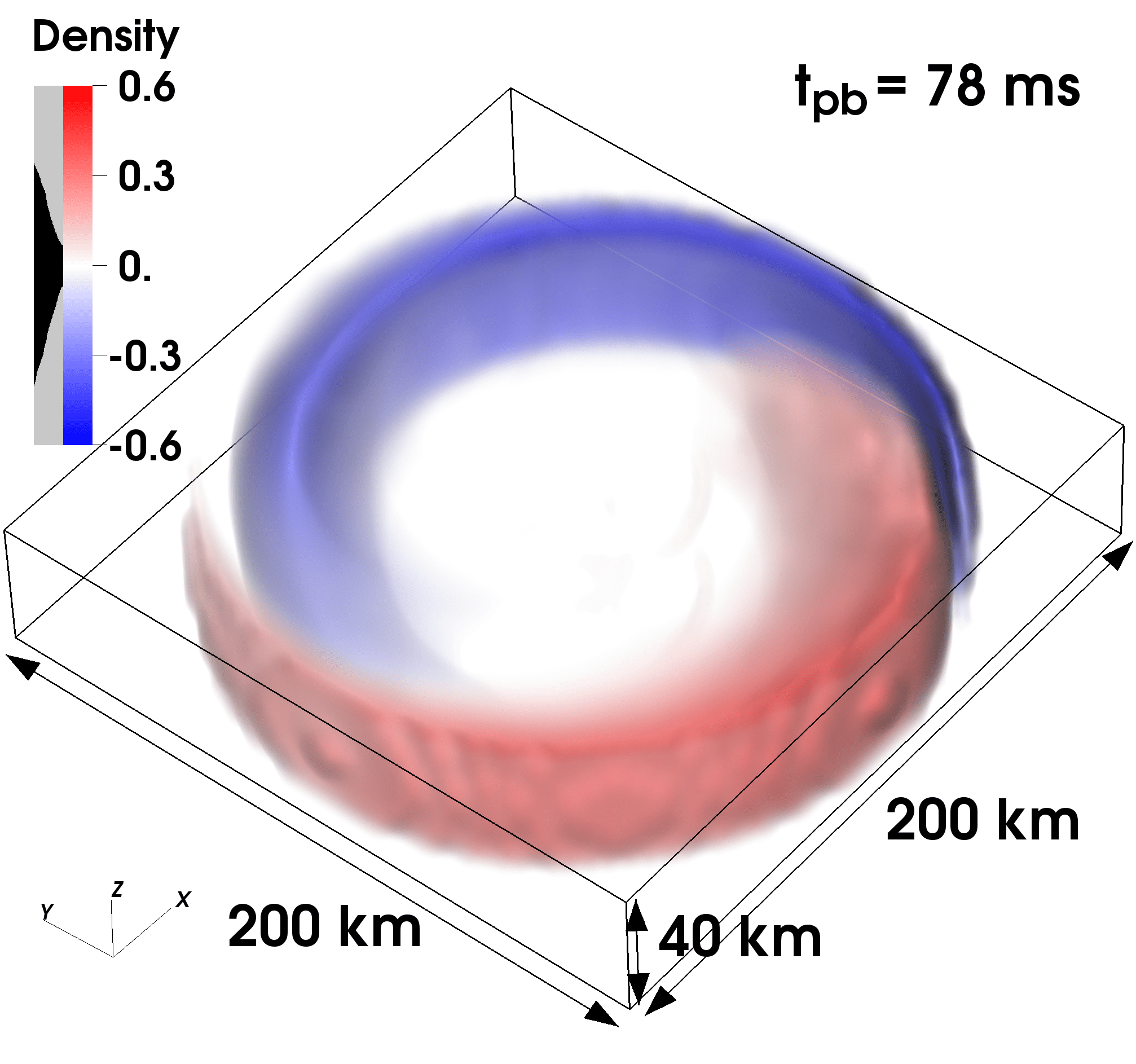}
\includegraphics[ width=0.40\textwidth]{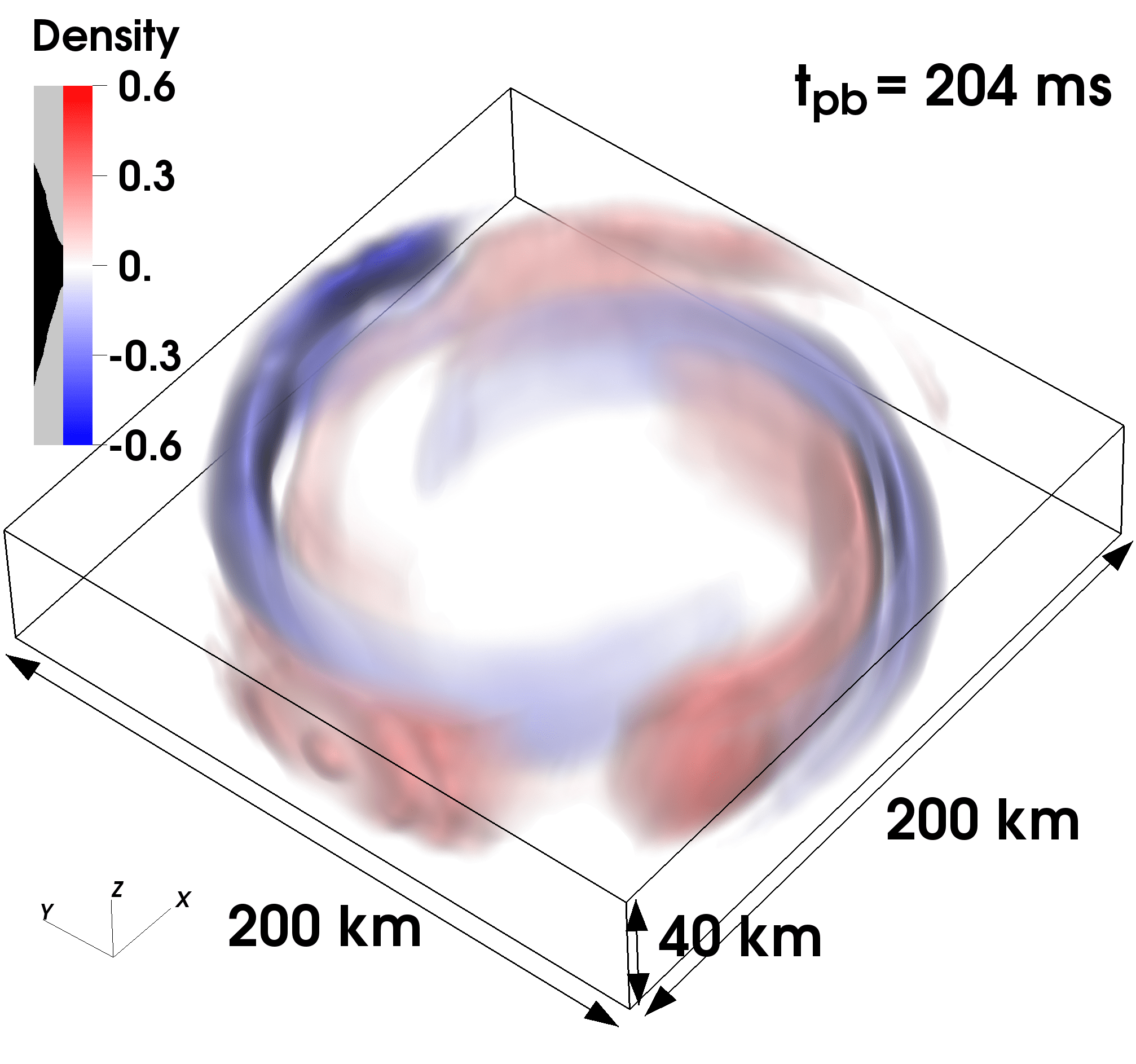}
\includegraphics[ width=0.40\textwidth]{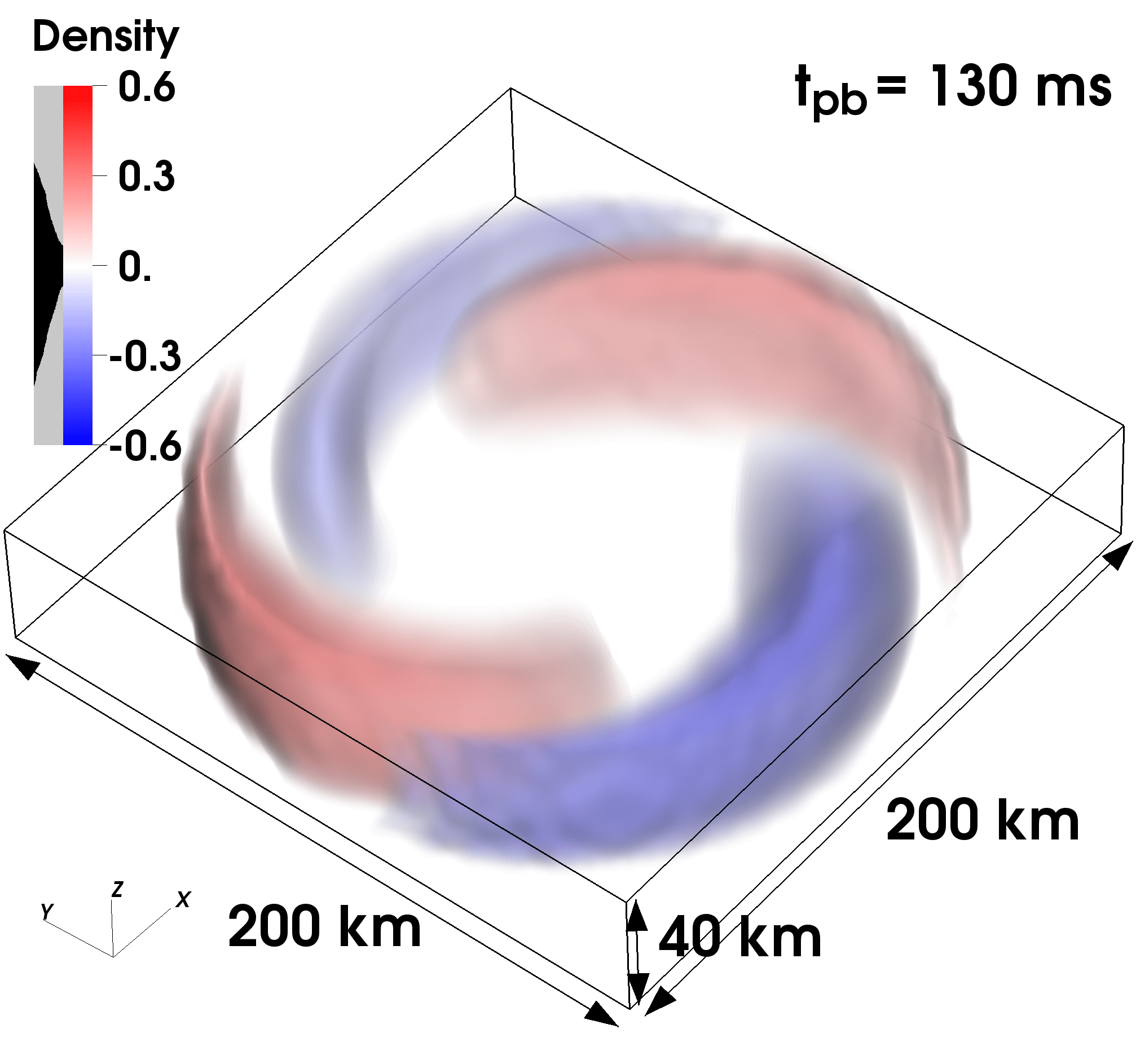}
\caption{Snapshots of normalized density deviation from the angle-averaged density at $t_{\mathrm{pb}}=78$\,ms (top panel) and at $t_{\mathrm{pb}}=204$\,ms (middle panel) for the $\Omega_{0} = 2$\,rad\,s$^{-1}$ model, and at $t_{\mathrm{pb}}=130$\,ms (bottom panel) for the $\Omega_{0} =1$\,rad\,s$^{-1}$ model. The one- or two-armed spiral pattern, coloured by red and blue, can be clearly seen in each snapshot. \label{fig:3d_arm}}
\end{figure}

Once the low-$T/|W|$ instability sets in, it is known that the PNS deformation becomes significant, leading to the formation of the one- or two- armed flow extending outwards from the PNS surface \citep{Ott05,Scheidegger08,Scheidegger10,takiwaki16}.
To see this phenomenon clearly, we present 3D volume renderings of normalized density deviation from the angle-averaged value, $(\rho-\left<\rho\right>)/\left<\rho\right>$, in Fig.~\ref{fig:3d_arm}.
The top panel of Fig.~\ref{fig:3d_arm} clearly shows the development of a one-armed spiral flow in the $\Omega_{0} = 2$\,rad\,s$^{-1}$ model at $t_{\mathrm{pb}}=78$\,ms when the $m=1$ PNS deformation occurs.
At a later time $t_{\mathrm{pb}}=204$\,ms, when the $m=2$ PNS deformation becomes dominant, the two-armed spiral flow can be clearly seen as indicated by the middle panel.
While in the $\Omega_{0} = 1$\,rad\,s$^{-1}$ model, there is no phase during which the $m=1$ mode is dominant and the two-armed spiral flow emerges as shown in the bottom panel.

\begin{figure}
\centering
\includegraphics[width=0.47\textwidth]{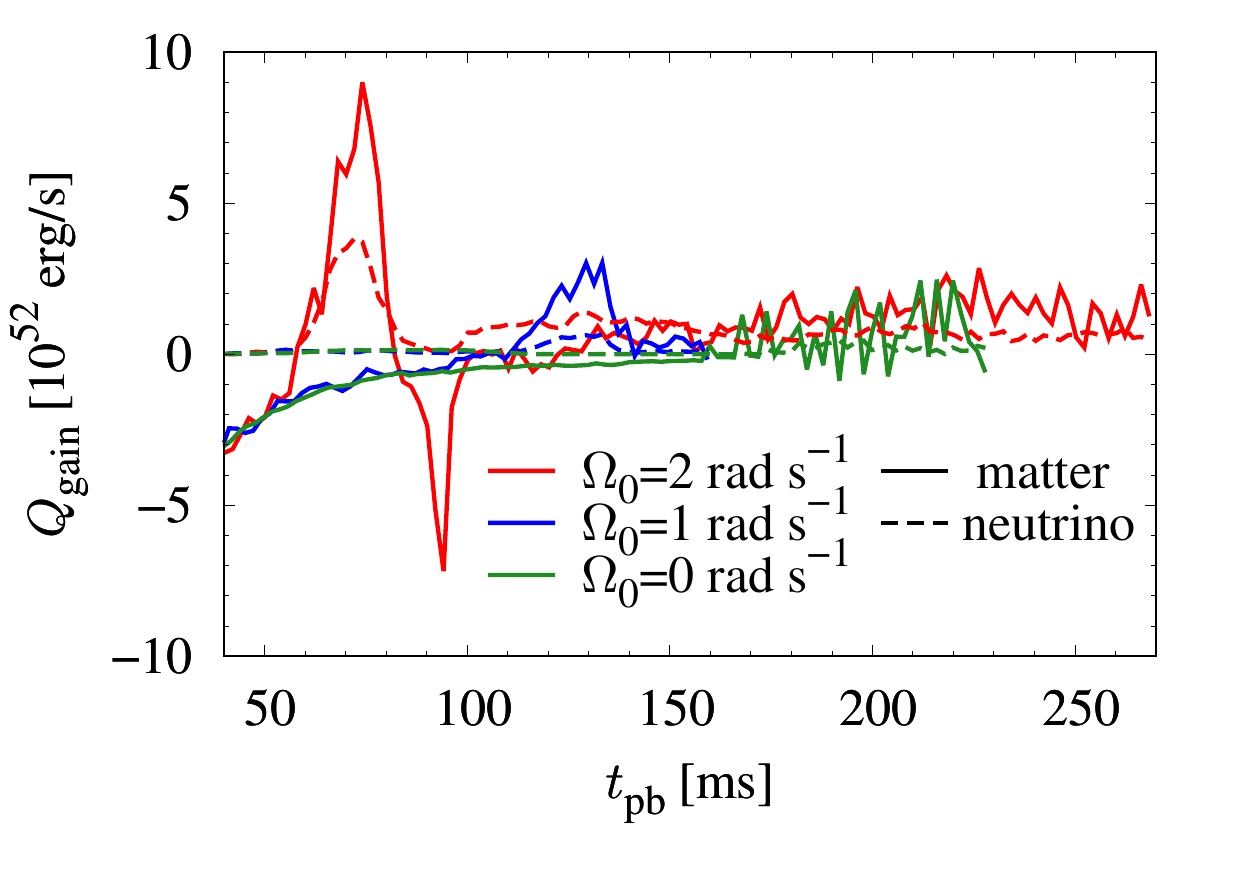}
\includegraphics[width=0.47\textwidth]{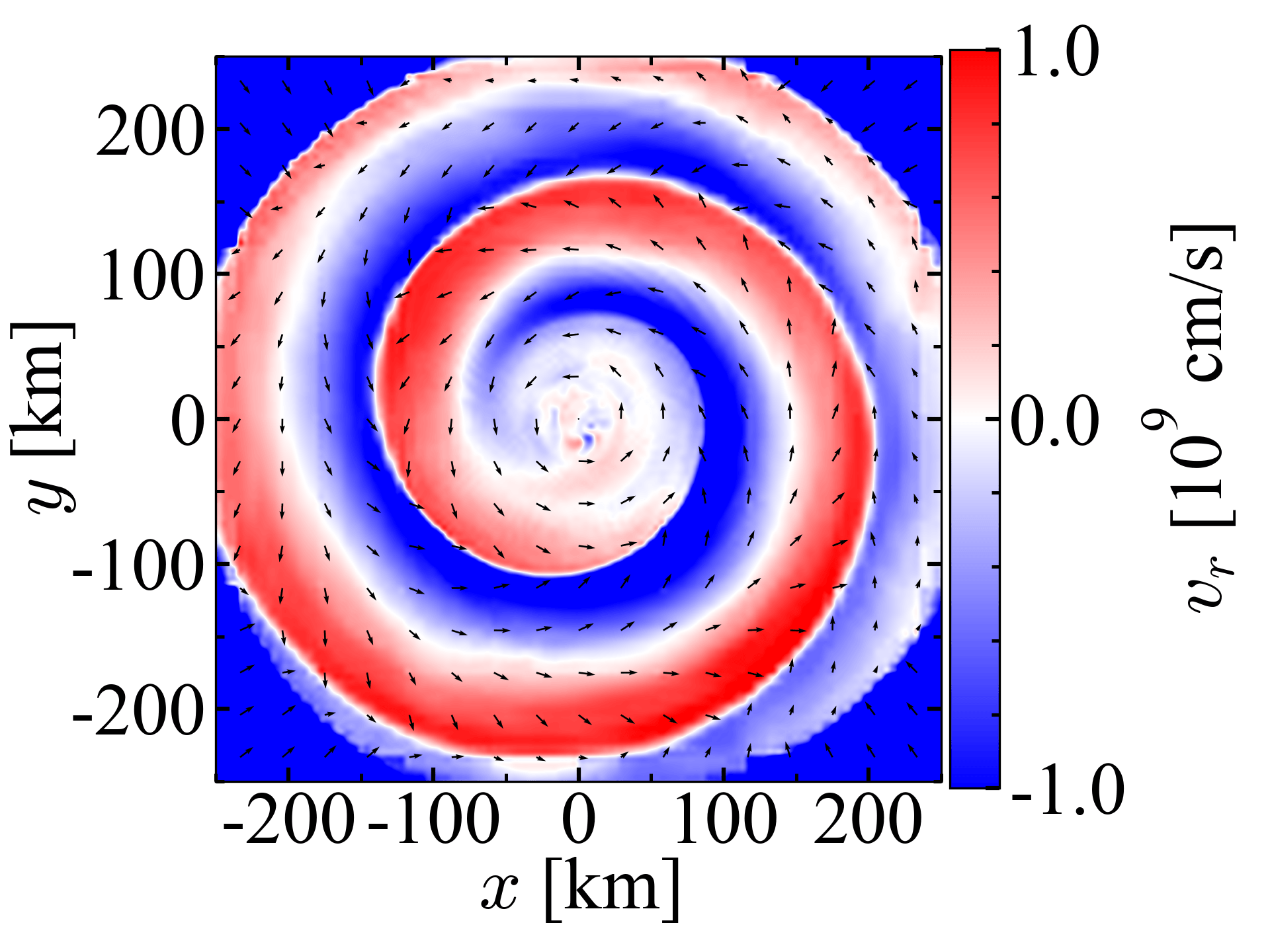}
\caption{Evolution of the energy gain in the gain region (top panel) evaluated by the net energy flux of matter at the surface of the gain region (solid lines) and by the net neutrino heating/cooling rate in the gain region (dashed lines) for the $\Omega_{0} =2$\,rad\,s$^{-1}$ (red lines), $\Omega_{0} = 1$\,rad\,s$^{-1}$ (blue lines) and $\Omega_{0} = 0$\,rad\,s$^{-1}$ (green lines) models, and radial velocity distribution in the equatorial plane at $t_{\mathrm{pb}}=78.2$\,ms for the $\Omega_{0}=2$\,rad\,s$^{-1}$ model (bottom panel). The black arrow indicates a vector of 3-velocity. Note that the 3-velocity vector is limited so that its norm does not exceed 10$^9$\,cm\,s$^{-1}$. \label{fig:qnet}}
\end{figure}

Next, we move on to investigate the dynamical impact of the spiral arm on the shock evolution.
To explore the transferred energy to the post-shock region by the spiral arm, we evaluate the net inflow rate of the matter (total) energy to the gain region \citep{takiwaki16} as follows:
\begin{eqnarray}
   Q_{\mathrm{gain,matter}}=\int d\Omega \left[\left(\alpha \left(\rho h W^2 - P\right) -\rho W c^2\right)r^2v_r\right]^{r=R_{\mathrm{gain}}}_{r=R_{\mathrm{sh}}},
\end{eqnarray}
where the surface integral is performed at the gain radius $R_{\mathrm{gain}}$ and the shock radius $R_{\mathrm{sh}}$, and $\alpha$, $\rho$, $h$, $W$, $P$, $c$, and $v_r$ are the lapse function, the density, the specific enthalpy, the Lorentz factor, the pressure, the speed of light, and the radial velocity, respectively.
In the top panel of Fig.~\ref{fig:qnet}, we show the temporal evolution of the $Q_{\mathrm{gain,matter}}$ and the net heating rate in the gain region, $Q_{\mathrm{gain,neutrino}}$, for each model.
The $\Omega_{0} = 2$\,rad\,s$^{-1}$ model shows that a large amount of the matter energy is transferred to the gain region at $t_{\rm pb} \sim60$--80\,ms ($\sim 9\times10^{52}$\,erg\,s$^{-1}$ at most).
The injection of the matter energy overwhelms the net neutrino heating ($\sim 4\times10^{52}$\,erg\,s$^{-1}$ at most).
As can be seen in Figs.~\ref{fig:shock_r} and \ref{fig:Rnu_Ylm}, this period corresponds to the significant increase of the shock radius and the strong $m=1$ PNS deformation.
As previously identified \citep{takiwaki16}, the one-armed flow extending from the PNS injects the matter energy to the gain region that results in the rapid shock expansion.

To support the idea that the matter energy
 is transferred to the gain region via the spiral arm, we also present a snapshot of the radial velocity distribution on the equatorial plane at $t_{\mathrm{pb}}=78.2$\,ms in the bottom panel of Fig.~\ref{fig:qnet}.
One can see that the one-armed spiral arm has a flow channel with the positive radial velocity (see the region coloured in red), which carries the energy from the PNS surface outwards.
These facts indicate that the first low-$T/|W|$ instability efficiently transfers angular momentum from the corotation point 
(see also, \citet{Shibagaki20})
to the outer region, leading to the energy transfer to the gain region.

On the other hand, after $t_{\mathrm{pb}} \sim 110$\,ms when the PNS exhibits the $m=2$ deformation, $Q_{\mathrm{gain,matter}}$ is not so large and comparable to $Q_{\mathrm{gain,neutrino}}$ $\sim 1$--$2\times10^{52}$\,erg\,s$^{-1}$.
This moderate $Q_{\mathrm{gain,matter}}$ helps to sustain the shock radius of 210\,km seen at $t_{\rm pb} \sim110$--220\,ms in Fig.~\ref{fig:shock_r}, but the shock finally starts to shrink at $t_{\rm pb} \sim 220$\,ms.
Similarly, the $\Omega_{0} = 1$\,rad\,s$^{-1}$ model shows the enhancement of $Q_{\mathrm{gain,matter}}$ at $t_{\rm pb} \sim110$--140\,ms, $\sim 3\times10^{52}$\,erg\,s$^{-1}$.
This may account for the deviation of the shock radius in the $\Omega_{0} = 1$\,rad\,s$^{-1}$ model from the one in the
 non-rotating model (see Fig.~\ref{fig:shock_r}).

\subsection{Characteristic time variability of GW signals} \label{sec:GW}
In this section, we focus on the GW signatures of our models.
We extract the GWs with a standard quadrupole formula \citep{Shibata&Sekiguchi03,KurodaT14}.
To investigate the spectral evolution of the GWs, we evaluate the viewing-angle dependent characteristic strain for an optimally oriented source, $h_{\rm char}$, as 
\begin{eqnarray}
      h_{\mathrm{char}}^2(\theta,\phi,t,f) &\equiv& \frac{8}{\pi}\frac{G}{c^3}\frac{1}{D^2}\frac{dE}{d\Omega df}, \label{eq:hchar}\\
    \frac{1}{D^2}\frac{dE}{d\Omega df} &\equiv& \frac{\pi f^2}{2}\frac{c^3}{G} 
    \left[|\Tilde{h}_{+}|^2+ |\Tilde{h}_{\times}|^2\right]. \label{eq:dedomegadf}
\end{eqnarray}
Here $D$ is the distance to the source, $f$ is the frequency of the GW, $\frac{dE}{d\Omega df}$ is the GW spectral energy density per unit solid angle, and $\Tilde{h}$ is the short-term Fourier transform of the GW strain $h$ with a Hann window, whose width is set as $20$\,ms. 
The characteristic GW strain has been often estimated by averaging the GW spectral energy density over the emission angles \citep{Flanagan:1998a}
\footnote{Note that the evaluation of $h_{\mathrm{char}}$ in this paper is different from that in \citet{KurodaT14} and \citet{Shibagaki20}. The difference is, for example, that the angle-average was partly taken in the previous estimate.}.
In order to discuss the viewing-angle dependence, we  estimate
 the characteristic strain (Eq. (\ref{eq:hchar})) without the angle average in the following.

\begin{figure*}
\center
\begin{center}
\large{$\Omega_0 = 2$\,rad\,s$^{-1}$}
\vspace{0.2cm}
\end{center}
\includegraphics[ width=0.47\textwidth]{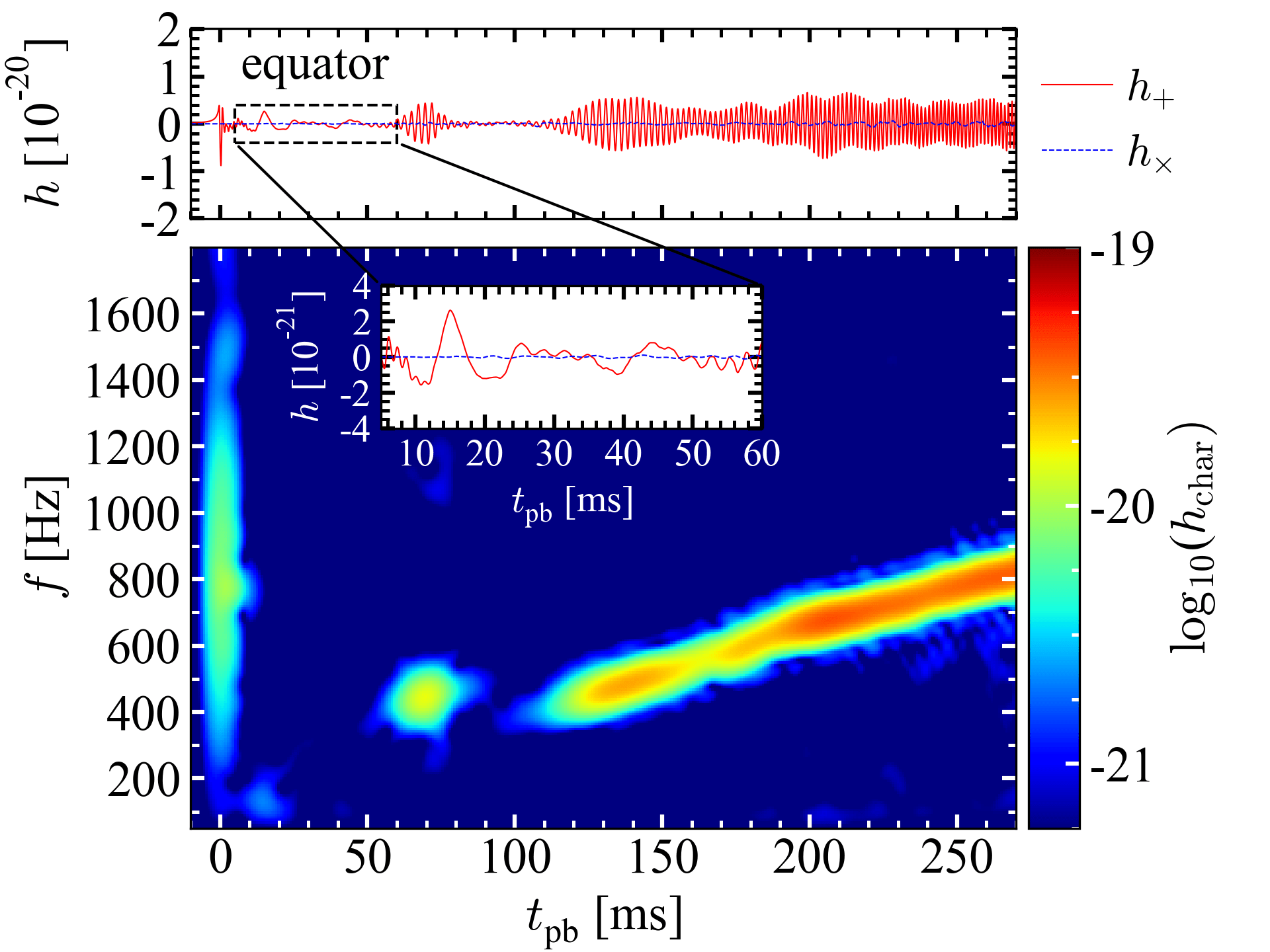}
\includegraphics[ width=0.47\textwidth]{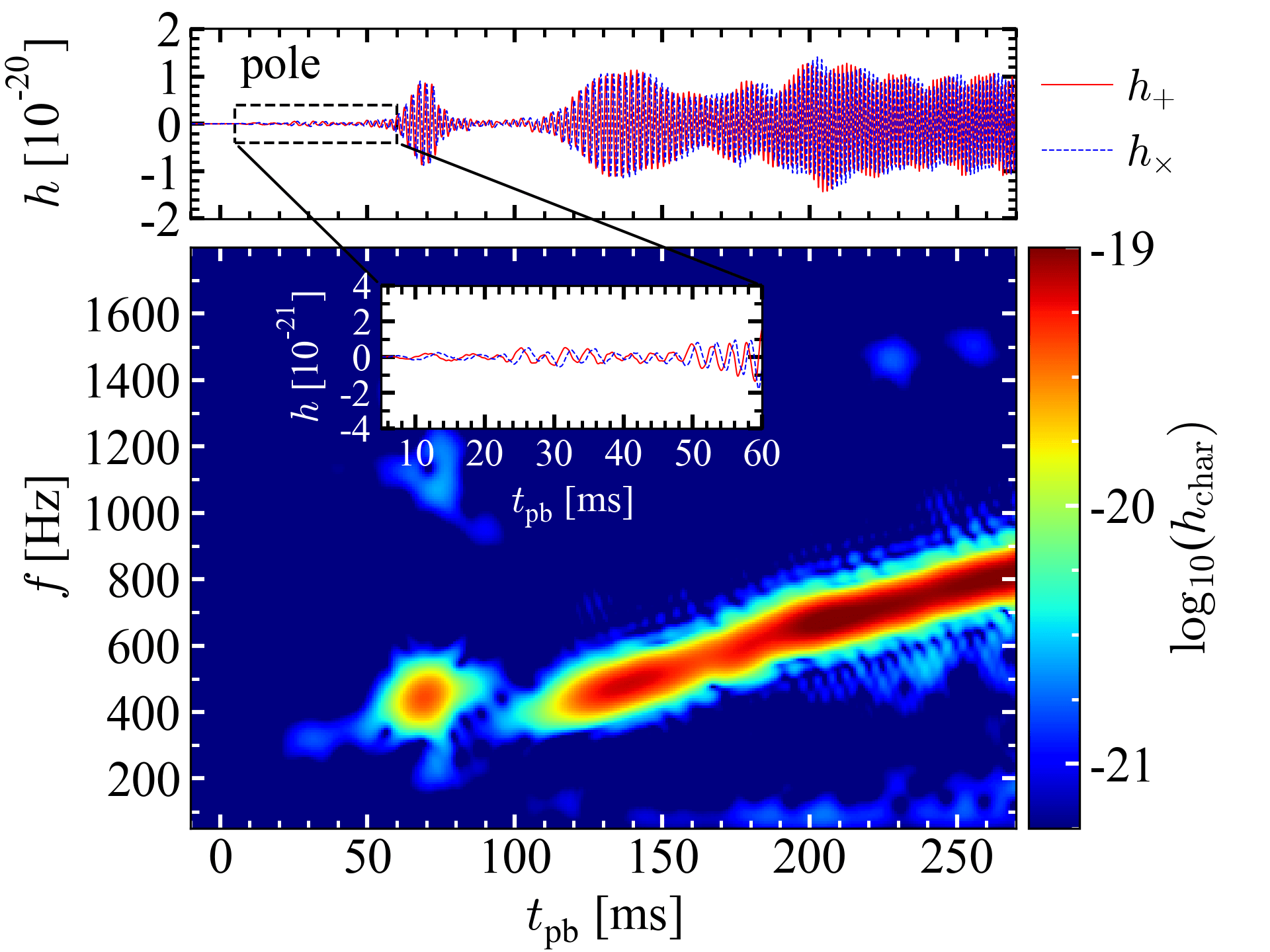}
\begin{center}
\vspace{-0.2cm}
\large{$\Omega_0 = 1$\,rad\,s$^{-1}$}
\end{center}
\includegraphics[ width=0.47\textwidth]{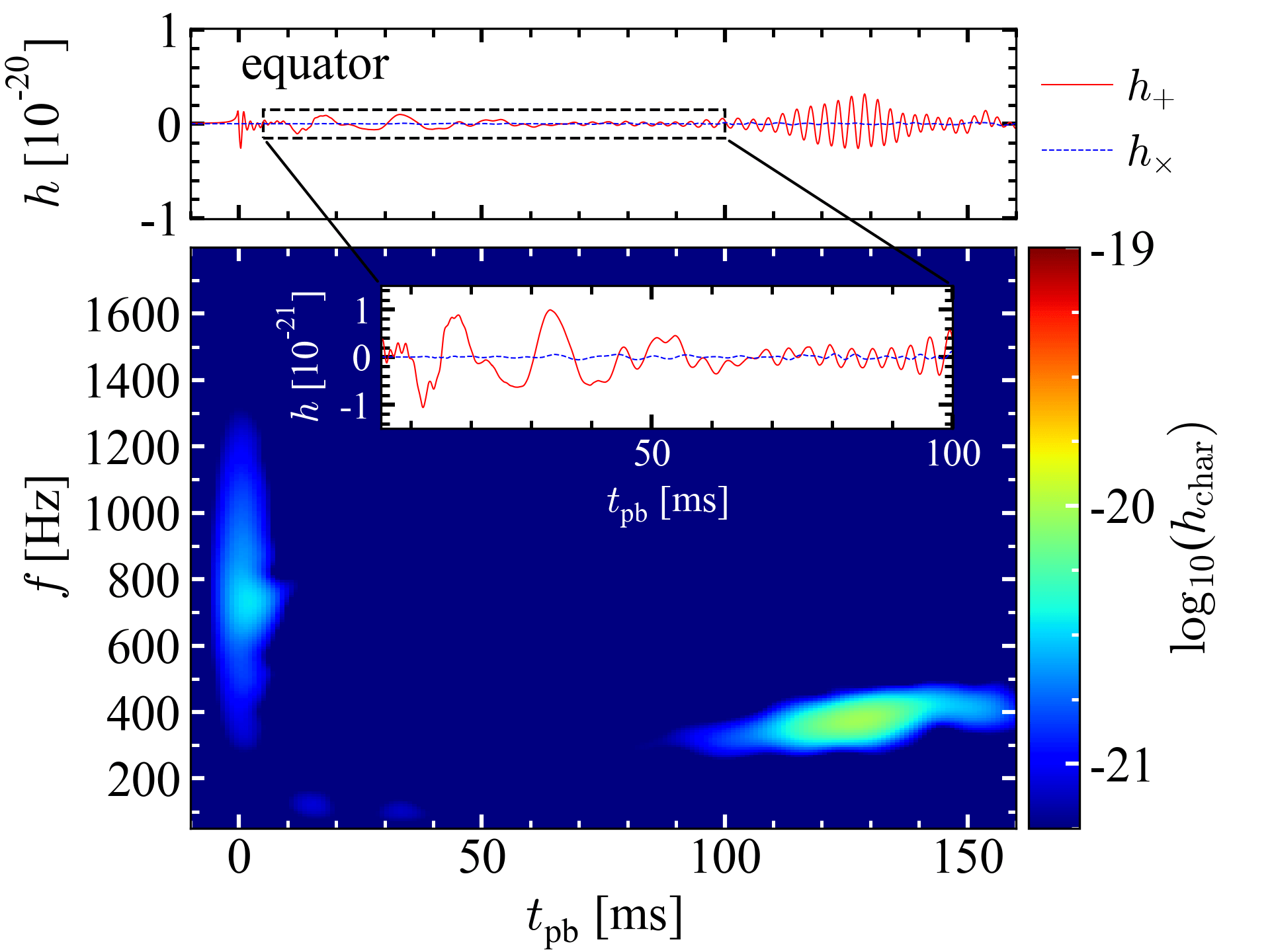}
\includegraphics[ width=0.47\textwidth]{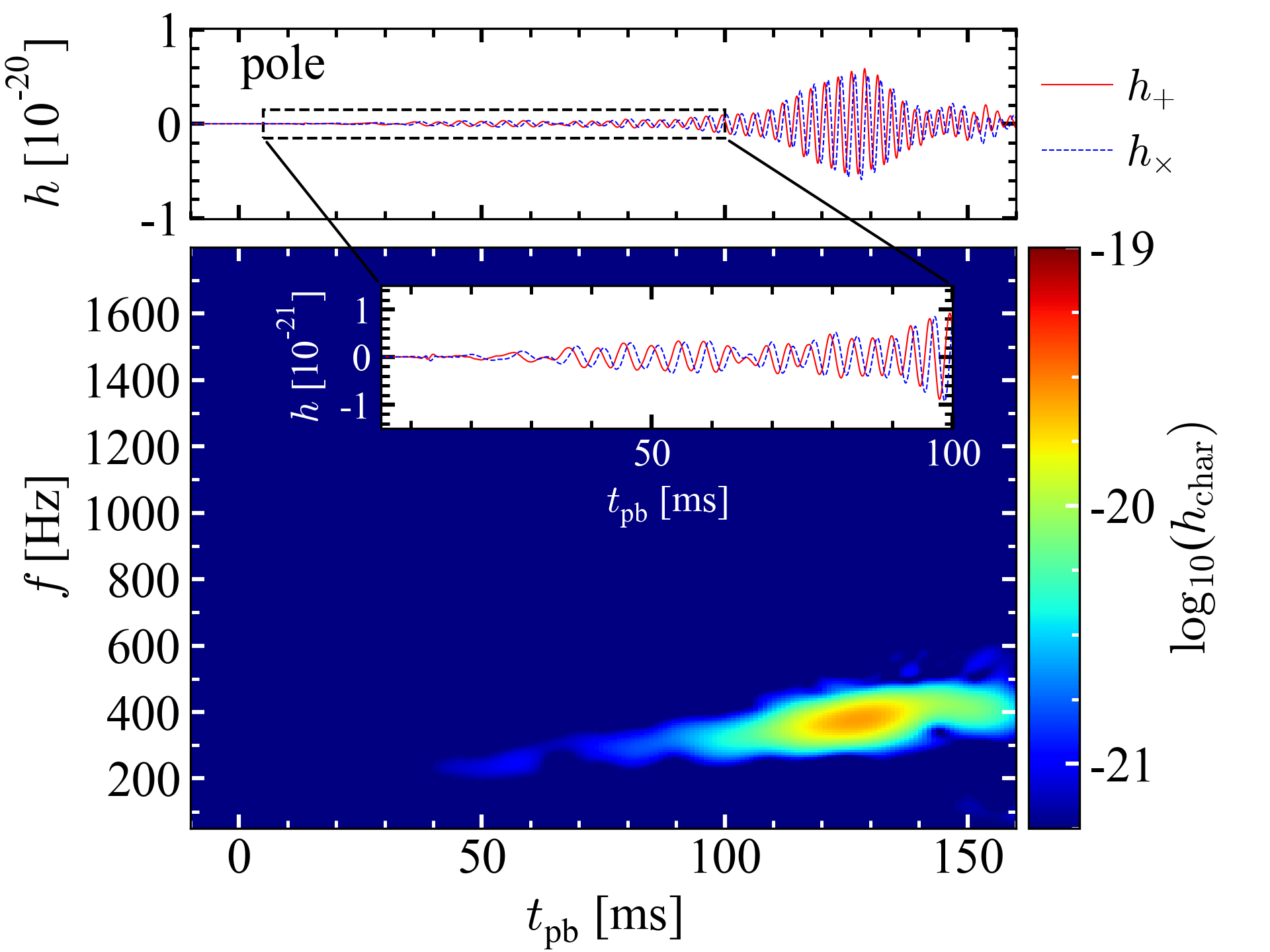}
\begin{center}
\vspace{-0.2cm}
\large{$\Omega_0 = 0$\,rad\,s$^{-1}$}
\end{center}
\includegraphics[ width=0.47\textwidth]{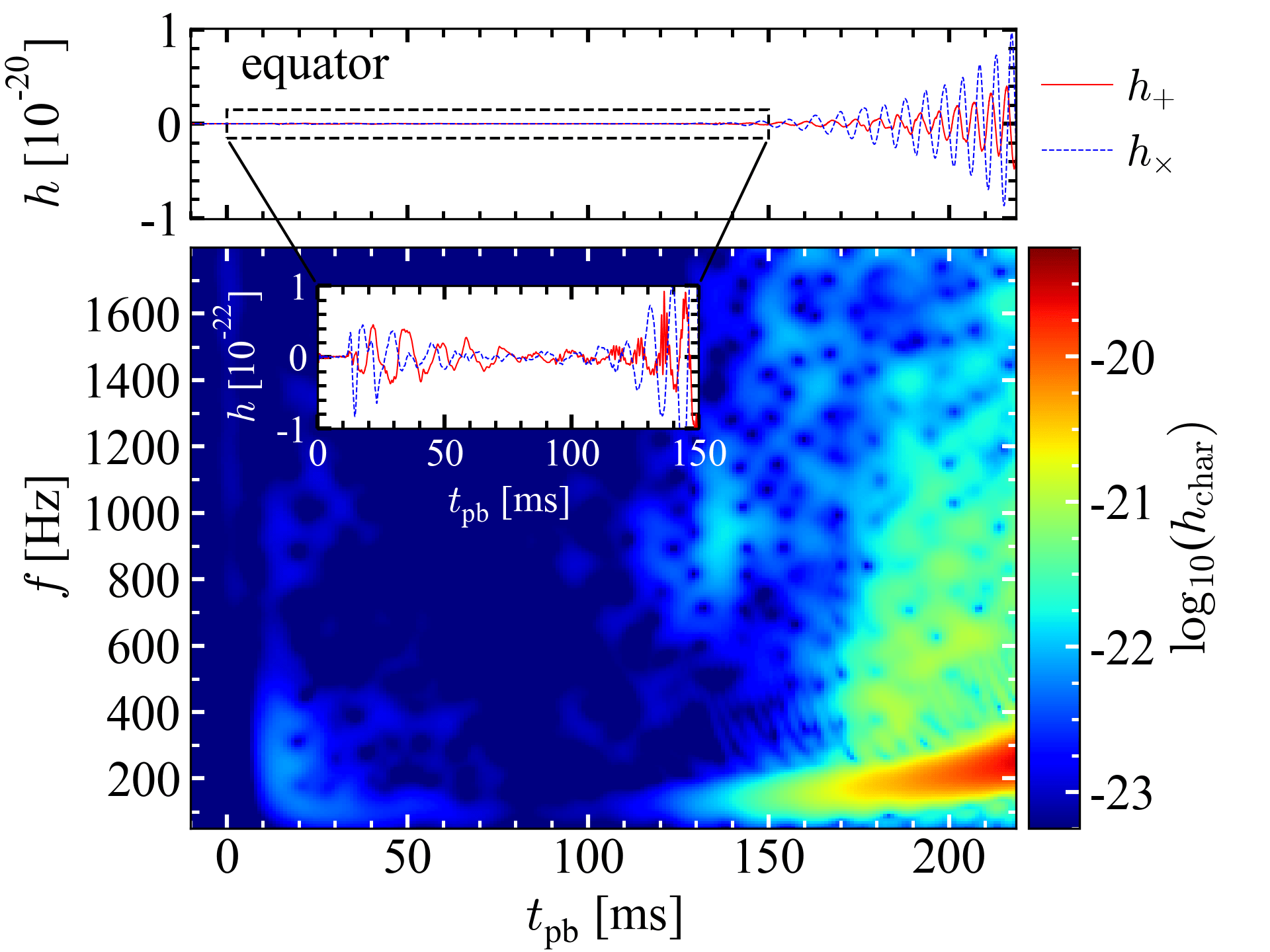}
\includegraphics[ width=0.47\textwidth]{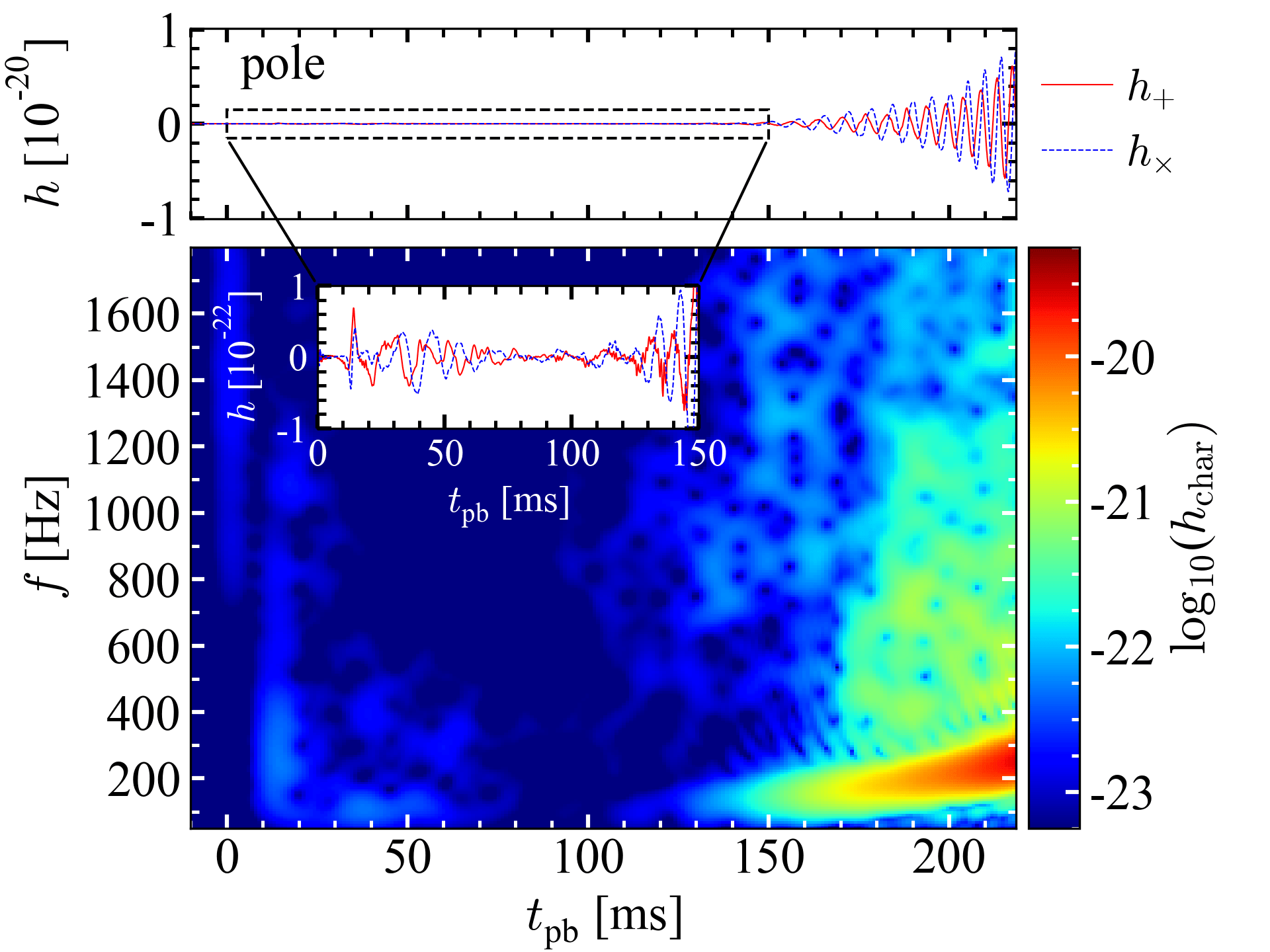}
\caption{GW strains of plus (red solid lines) and cross (blue dashed lines) modes and spectrograms of their characteristic strains for the $\Omega_{0} = 2$\,rad\,s$^{-1}$ (top panels), $\Omega_{0} = 1$\,rad\,s$^{-1}$ (middle panels) and $\Omega_{0} = 0$\,rad\,s$^{-1}$ (bottom panels) models seen along the equator (left panels) and the pole (right panels) at a source distance of 10\,kpc. The inset in each panel zooms into their low-amplitude phase.
\label{fig:gw_spectrogram}}
\end{figure*}

Fig.~\ref{fig:gw_spectrogram} shows the GW waveforms, $h_{+/\times}$ and the spectrograms for the characteristic strain of each model emitted along the equatorial (left column) and the polar directions (right column) for a source distance of $10$\,kpc, respectively.

In the $\Omega_{0} =2$\,rad\,s$^{-1}$ (top panels) and $\Omega_{0} =1$\,rad\,s$^{-1}$ (middle panels) models, the GWs observed along the equatorial direction show the burst signal at core bounce due to the rotational flattening of the core, and also show the oscillation in the ringdown phase shortly after bounce (see insets of the panels).
The rotational flattening deforms the core to be oblate and the aspect ratio
 (of the polar to the equatorial radius of the PNS core) reaches $\sim 1.8$ for the $\Omega_{0} =2$\,rad\,s$^{-1}$ model and $\sim 1.4$ for the $\Omega_{0} =1$\,rad\,s$^{-1}$ model just after core bounce ($t_{\mathrm{pb}}\sim1$\,ms). 
After that, in the $\Omega_{0} =2$\,rad\,s$^{-1}$ model, we see the large-amplitude quasi-sinusoidal oscillation of the GW emitted in both of the equatorial and the polar directions at $t_{\mathrm{pb}}\sim50$--90\,ms, when the low-$T/|W|$ instability develops, leading to the $m=1$ PNS deformation.
After the following quiescent phase until $t_{\mathrm{pb}}\sim110$\,ms, the strong quasi-periodic GW emission due to the $m=2$ PNS deformation becomes active again and lasts until the final simulation time.
In this phase, the peak frequency of the GW increases from 400\,Hz to 800\,Hz with time as shown in the spectrograms.

Similarly, in the $\Omega_{0} =1$\,rad\,s$^{-1}$ model, the GW emission starts to be stronger and the quasi-periodic oscillation at $t_{\rm pb}\sim 100$\,ms, when the $m=2$ PNS deformation develops,  and the oscillation lasts until the final simulation time.
In this phase, the peak frequency of the GW increases from 300\,Hz to 400\,Hz with time.
The peak frequency of the quasi-periodic GWs seen in the above two rotating models corresponds to twice the rotation frequency of the PNS, and the increase of the peak frequency can be interpreted as due to the gradual PNS contraction \citep{Shibagaki20}.

Both of the two rotating models commonly show that the plus mode of the quasi-periodic GW observed in the equatorial plane is roughly two times bigger than the plus and cross modes observed in the polar direction, and the cross mode observed in the equatorial plane is much smaller than the plus mode.
These features are analogous to the GW emission from a rotating ellipsoid \citep[i.e. $h_+ \propto (1+\cos ^2\theta)/2$ and $h_{\times} \propto \cos\theta$; see, for example, Eq. (4.223) in][]{Maggiore08} and consistent with the previous study where the growth of non-axisymmetric instabilities was observed \citep{ott_prl,Scheidegger08,Scheidegger10,KurodaT14,takiwaki18}.

The gravitational waveforms of the $\Omega_{0} =0$\,rad\,s$^{-1}$ model (bottom panel of Fig.~\ref{fig:gw_spectrogram}) show the oscillation shortly after bounce due to prompt convection  (see inset).
At later times, when the spiral SASI is dominated after $t_{\mathrm{pb}} \sim 150$\,ms, the quasi-periodic oscillation of the GWs becomes stronger and its peak frequency is the range of $\sim200 - 300$\,Hz.
In this non-rotating model, no significant differences between the GW signals in the polar and the equatorial directions are observed.

\begin{figure}
\centering
\includegraphics[ width=0.48\textwidth]{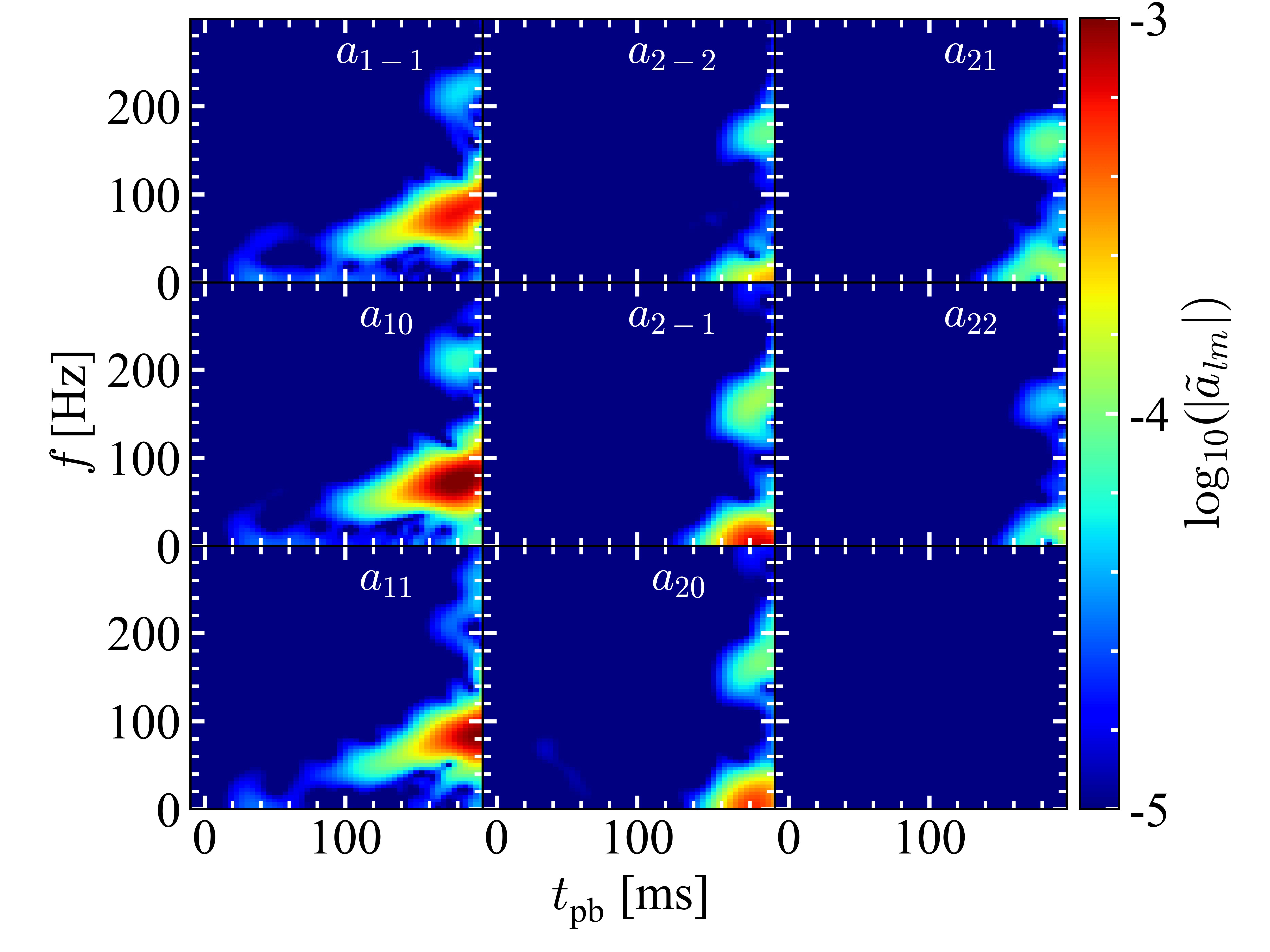}
\caption{Spectrograms of the spherical harmonic coefficients, $a_{l,m}$, of the shock surface with $l=1$ and $l=2$ for the $\Omega_{0} = 0$\,rad\,s$^{-1}$ model. \label{fig:SASI_spectrogram}}
\end{figure}

The quasi-periodic oscillation of the GW amplitudes observed in the $\Omega_{0} =0$\,rad\,s$^{-1}$ model is likely to originate from the SASI since their developments are almost simultaneous.
To confirm this, we perform the time-frequency analysis by computing the spectrograms of the normalized spherical harmonics coefficients of the deformed shock surfaces as shown in the bottom panel of Fig.~\ref{fig:shock_Ylm}.
The spectrograms are shown in Fig.~\ref{fig:SASI_spectrogram}.
After $t_{\mathrm{pb}}\sim 150$\,ms, the peak frequencies are concentrated in the frequency range of  60--110\,Hz in the $l=1$ mode.
This value is roughly half the peak frequency of the GW frequency, and this correspondence is consistent with the observed quasi-periodic GW that originates from the SASI activity \citep{andresen19,Vartanyan19b}. The high-end of the peak GW frequency (extending to 
$\sim 300$ Hz) can also stem from the $m = 2$ shock deformation as shown in the bottom panel of Fig. \ref{fig:shock_Ylm}.

\begin{figure}
\centering
\includegraphics[width=0.45\textwidth]{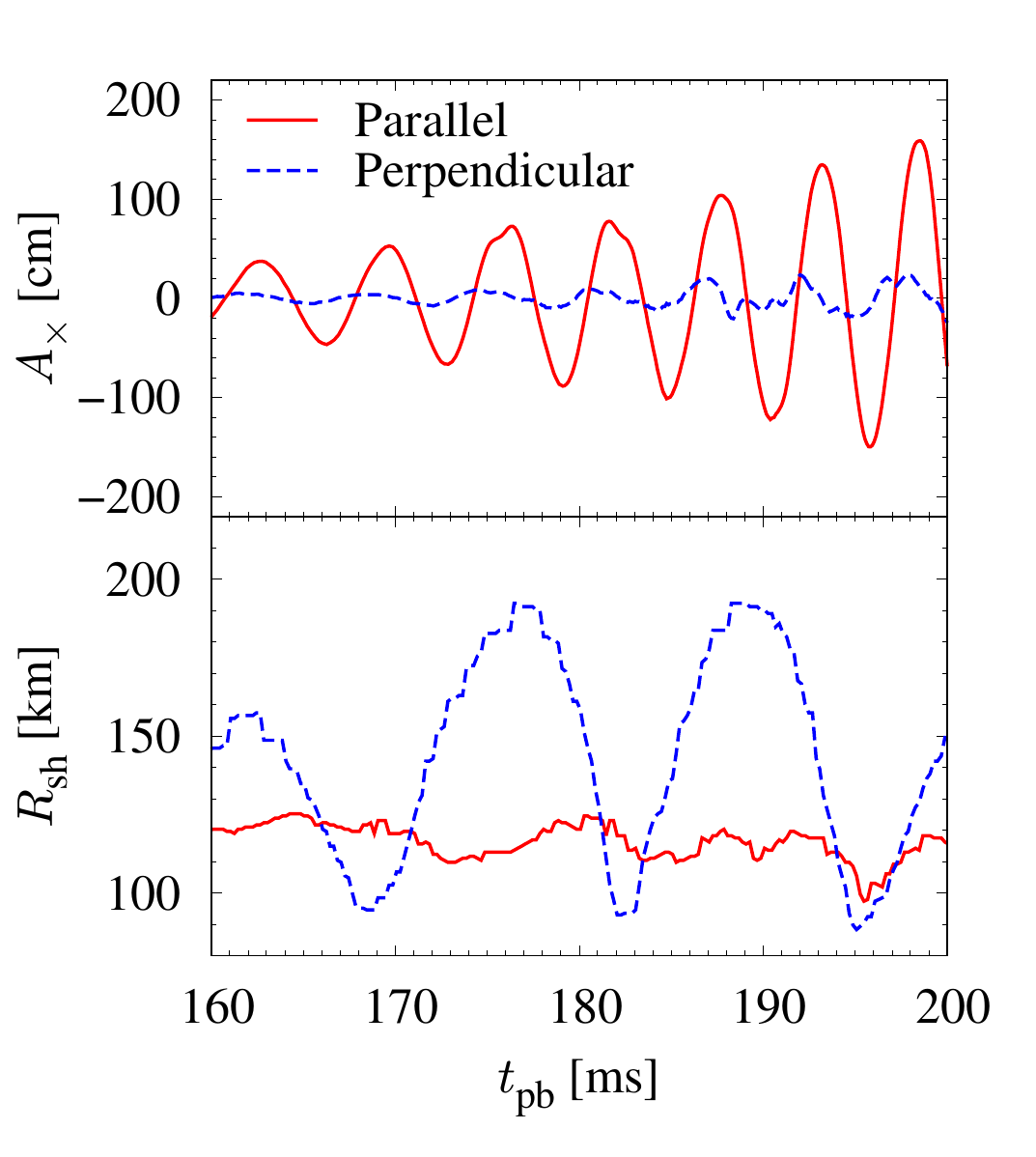}
\caption{GW amplitudes (top panel) and shock radii (bottom panel) for the $\Omega_{0} =0$\,rad\,s$^{-1}$ model as a function of time after bounce in direction parallel to the rotation axis of the spiral SASI ($(\theta,\phi) \sim (60^{\circ},-60^{\circ})$, red solid line) and perpendicular to the rotation axis of the spiral SASI ($(\theta,\phi) \sim (150^{\circ},-60^{\circ})$, blue dashed line). \label{fig:SASI_direction}}
\end{figure}

Furthermore, similarly to
\citet{KurodaT17},
we determine the rotation axis of the spiral SASI for the $\Omega_{0} =0$\,rad\,s$^{-1}$ model, 
and discuss the directional correlation between the GW emission and the shock deformation.
We define the direction vector of the rotation axis of the spiral SASI by focusing on the $l=1$ with $m=-1,0,1$ modes of the shock surface as follows:
\begin{eqnarray}
  \mathbf{L}=\left<\mathbf{r}_{\mathrm{sh}}\right> \times \left<\dot{\mathbf{r}}_{\mathrm{sh}}\right>,
\end{eqnarray}
where $\left<\mathbf{r}_{\mathrm{sh}}\right>=a_{11}\hat{\mathbf{x}}+a_{1-1}\hat{\mathbf{y}}+a_{10}\hat{\mathbf{z}}$ representatively denotes the most expanded direction
of the shock surface in the Cartesian coordinates ($\hat{\mathbf{x}},\hat{\mathbf{y}},\hat{\mathbf{z}}$ are the unit vectors), $\left<\dot{\mathbf{r}}_{\mathrm{sh}}\right>$ does its time derivative, and $\times$ is the cross product. Note  $\mathbf{L}$ is defined as the angular momentum vector whose direction corresponds to the rotational axis, provided that   $\left<\mathbf{r}_{\mathrm{sh}}\right>$ denotes the position of a point particle in the classical dynamics.
Analogously, we convert $\mathbf{L}$ to the azimuthal and polar angles, $\theta$ and $\phi$,
 which points to the rotation axis of the spiral SASI in spherical polar coordinates. We find that the SASI axis direction, $(\theta,\phi)$, moves from $\sim (65^{\circ},-75^{\circ})$ 
to $\sim (50^{\circ},-50^{\circ})$ 
between $t_{\mathrm{pb}}\sim 160$--200\,ms.
Fig.~\ref{fig:SASI_direction}
shows the GW amplitudes for the cross mode
$A_{\times}=Dh_{\times}$ (top panel)
and the shock radii $R_{\rm sh}$ (bottom panel)
in the direction parallel/vertical to the representative rotation axis, $(\theta,\phi) \sim (60^{\circ},-60^{\circ})$. 
The shock radius in the direction vertical to the rotation axis of the spiral SASI, $(\theta,\phi) \sim (150^{\circ},-60^{\circ})$, oscillates at the SASI frequency ($\sim 80-90$\,Hz, bottom panel, blue dashed line), while the GW amplitude emitted parallel to the rotation axis of the spiral SASI, $(\theta,\phi) \sim (60^{\circ},-60^{\circ})$, oscillates roughly at twice
 the SASI frequency ($\sim 160-180$\,Hz, top panel, red line).
By contrast, the shock radius observed parallel to the spiral SASI axis (red line in the bottom panel) and the GW amplitude emitted vertical to the spiral SASI axis (blue dashed line in the top panel) show no clear oscillation.
This directional dependence is also consistent with the quasi-periodic GW that originates from the spiral SASI.

\begin{figure}
\centering
\includegraphics[width=0.47\textwidth]{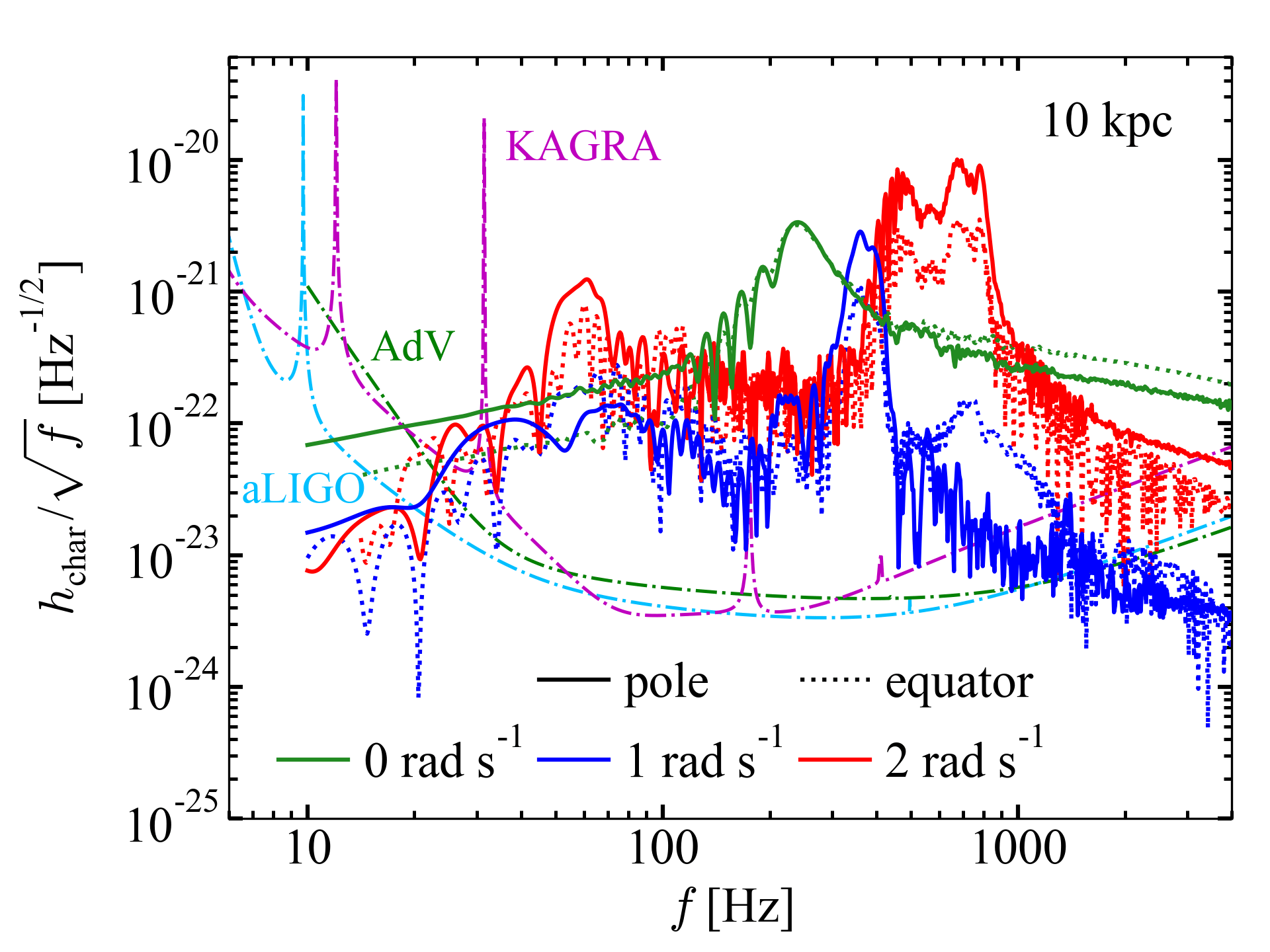}
\includegraphics[ width=0.47\textwidth]{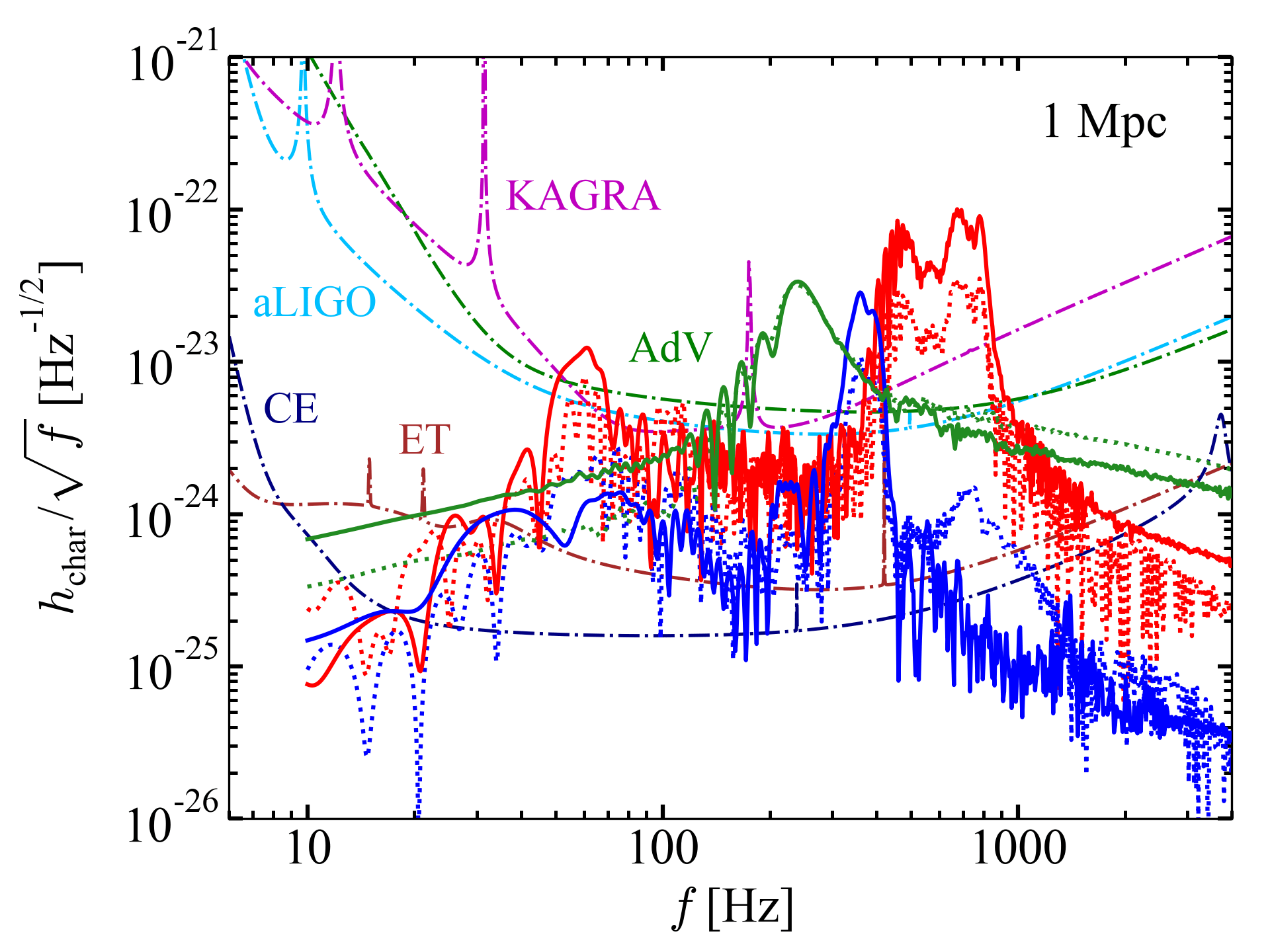}
\caption{Characteristic GW spectral amplitudes of for the $\Omega_{0} =2$\,rad\,s$^{-1}$ (red lines), $\Omega_{0} =1$\,rad\,s$^{-1}$ (blue lines) and $\Omega_{0} = 0$\,rad\,s$^{-1}$ (green lines) models seen along the pole (solid lines) and along the equator (dotted lines) as a source distance of 10\,kpc (top panel) and 1\,Mpc (bottom panel) relative to  the noise amplitudes of advanced LIGO (aLIGO; cyan line), advanced VIRGO (AdV; green line), KAGRA (magenta line) from \citet{abbott18det}, Einstein Telescope \citep[ET; orange;][]{ET}, and Cosmic Explorer \citep[CE; navy;][]{CE}. The detector noise amplitudes are indicated by dash-dotted lines. \label{fig:gw_spectrum}}
\end{figure}

 We proceed to briefly discuss the detectability of the GW signals.
The top panel of Fig.~\ref{fig:gw_spectrum}
shows the GW spectral amplitudes for each model seen from the polar (solid lines) and the equatorial (dotted lines) observer at a distance of 10\,kpc relative to the sensitivity curves of the advanced LIGO, advanced VIRGO, and KAGRA \citep{abbott18det}.
Here the width of the Hann window used in Eq. \eqref{eq:dedomegadf} is taken as the simulation time for each model.

One can see that the GW signals for all of our models emitted in either direction within our Galaxy are within the detection limits of the current-generation GW detectors. 
The bottom panel of Fig.~\ref{fig:gw_spectrum} compares the GW spectral amplitudes observed at a source distance of 1\,Mpc with the sensitivity curves of the current-generation GW detectors as well as the third-generation GW detectors of Einstein Telescope \citep{ET} and Cosmic Explorer \citep{CE}. 
The GW signals could be detectable out to Mpc distance scale by not only the third-generation detectors but also the current-generation detectors.
For example, if they are observed along the polar direction at a distance of 1\,Mpc, 
the signal-to-noise ratio (SNR) for the current-generation detectors is 
$\sim20$ for the $\Omega_{0} =2$\,rad\,s$^{-1}$ model and 
$\sim6$ for the $\Omega_{0} =1$ and 0\,rad\,s$^{-1}$ models at their peak frequency\footnote{Here we simply estimate the SNR by taking the ratio of the signal prediction to the sensitivity curves, although much more detailed analysis is  needed for a more quantitative discussion (e.g. \citet{logue12,Hayama15,Gossan16,Powell16}).}.
 Although the comparison of the GW spectral amplitudes between our models is less meaningful due to the different final simulation time, our results demonstrate, at the bottom line, the importance of third-generation detectors, which extends the detection horizon up to a factor of ten comparing to the current-generation detectors for our models. 

\begin{figure*}
\center
\begin{center}
\large{$\Omega_0 =2$\,rad\,s$^{-1}$}
\vspace{0.2cm}
\end{center}
\includegraphics[ width=0.47\textwidth]{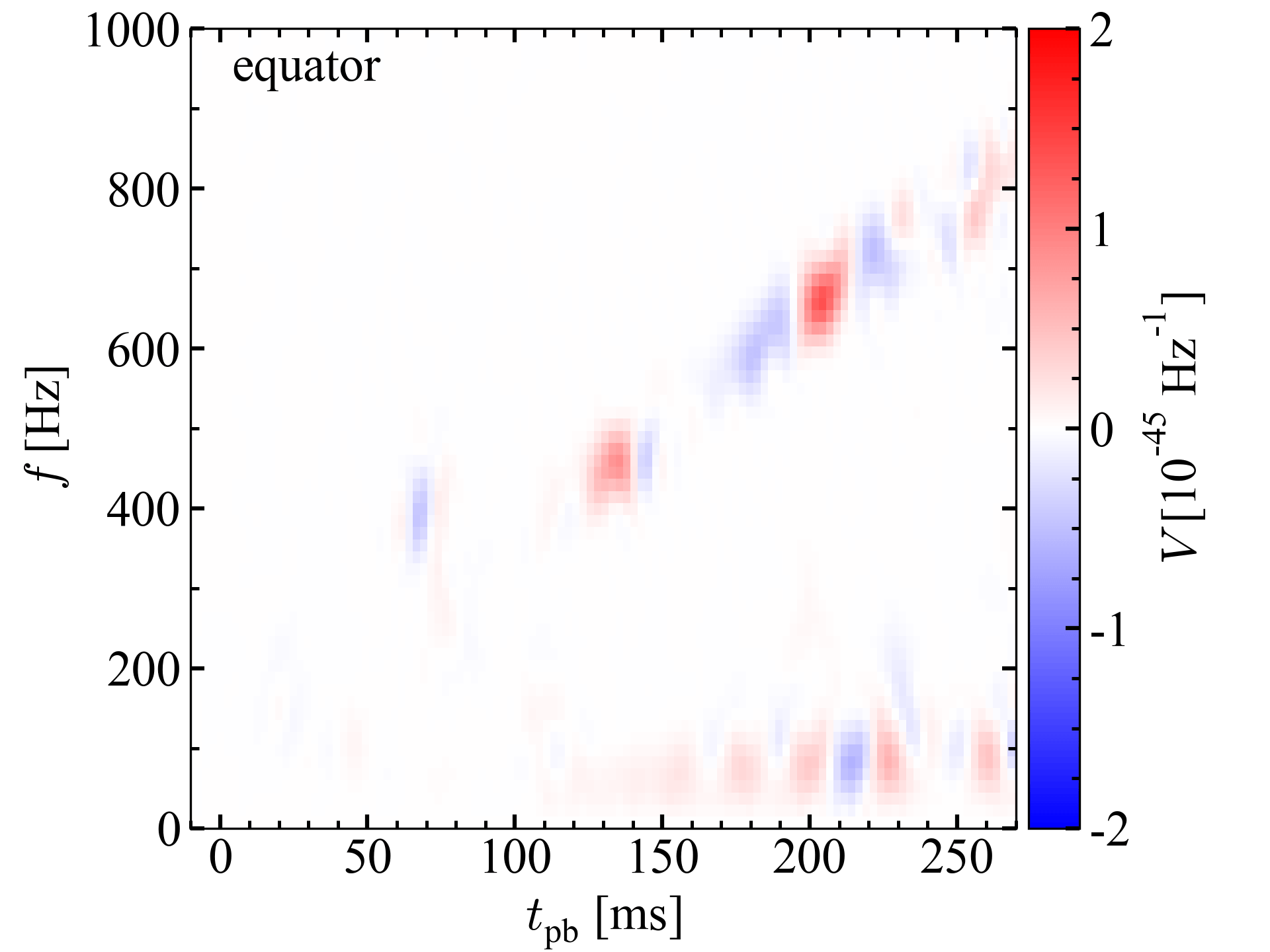}
\includegraphics[ width=0.47\textwidth]{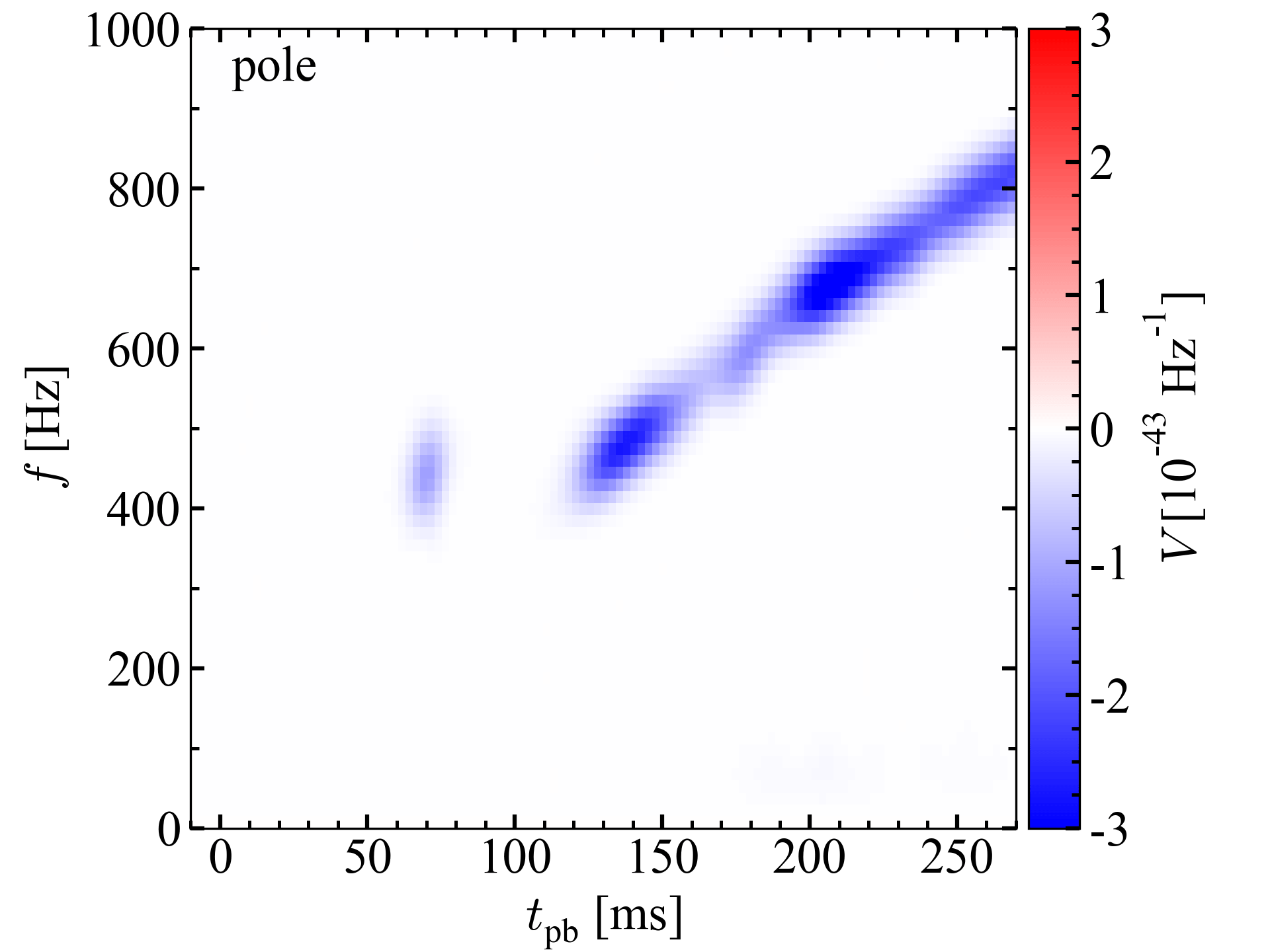}
\begin{center}
\vspace{-0.2cm}
\large{$\Omega_0 = 1$\,rad\,s$^{-1}$}
\end{center}
\includegraphics[ width=0.47\textwidth]{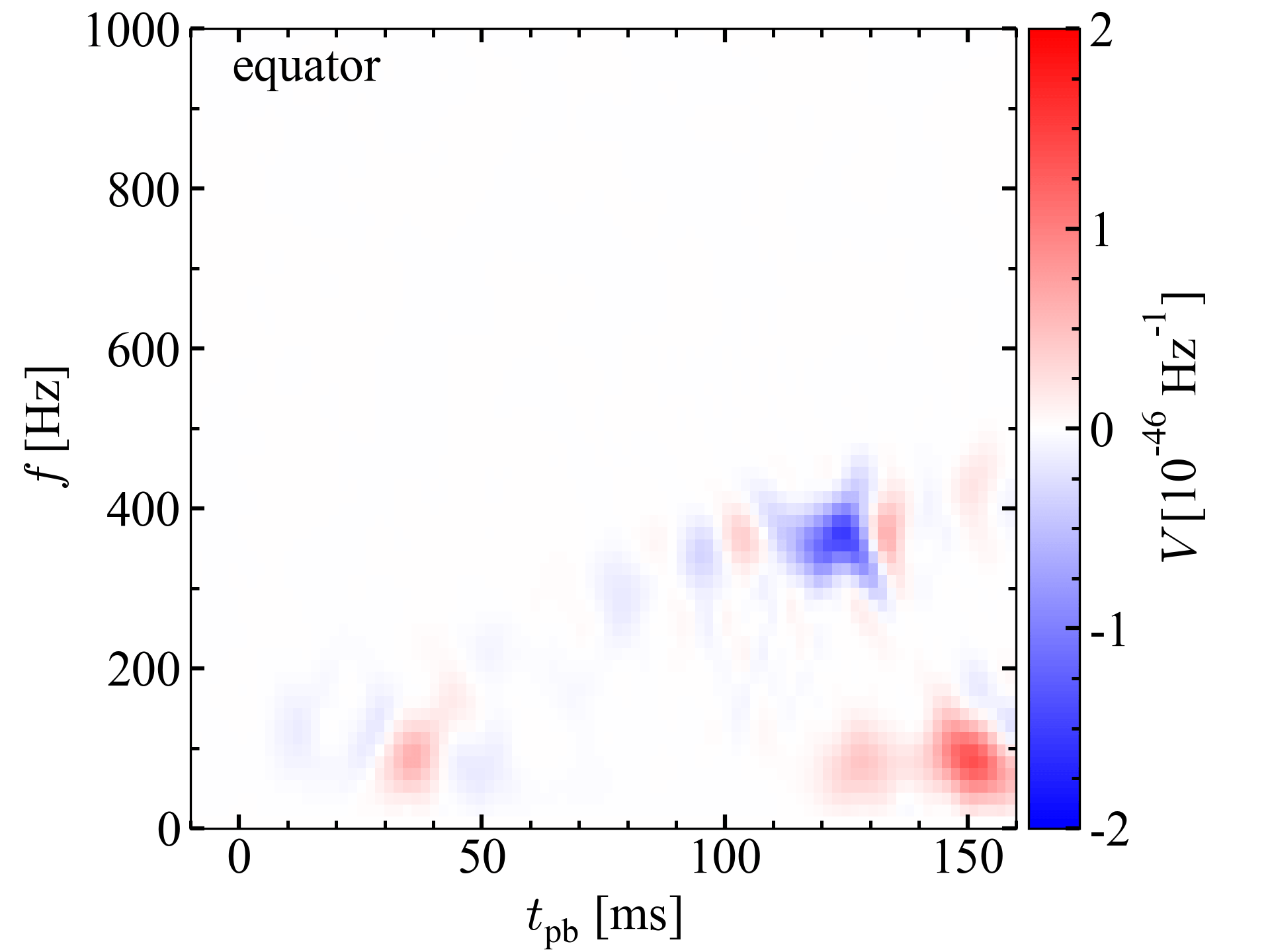}
\includegraphics[ width=0.47\textwidth]{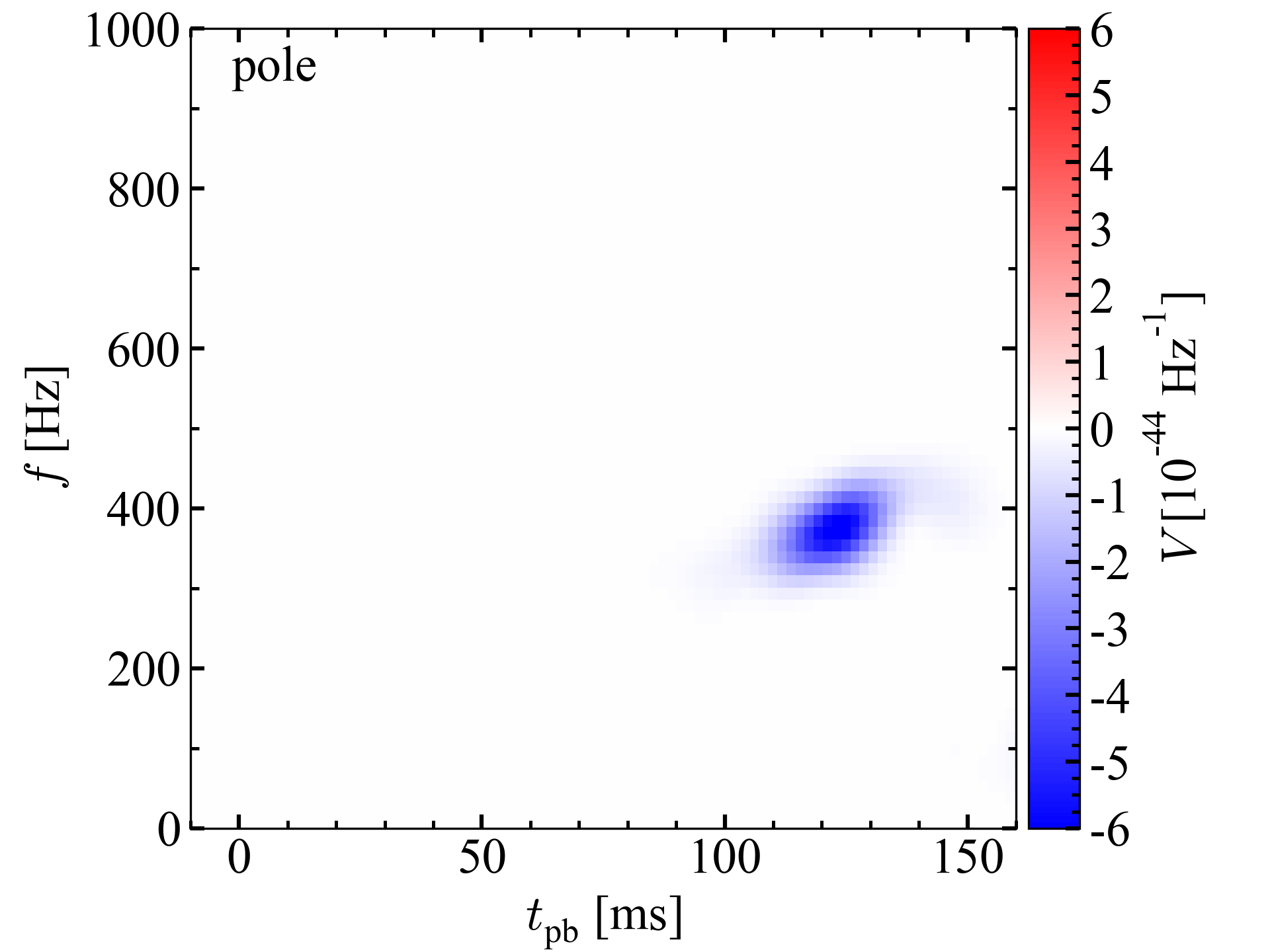}
\begin{center}
\vspace{-0.2cm}
\large{$\Omega_0 =0$\,rad\,s$^{-1}$}
\end{center}
\includegraphics[ width=0.47\textwidth]{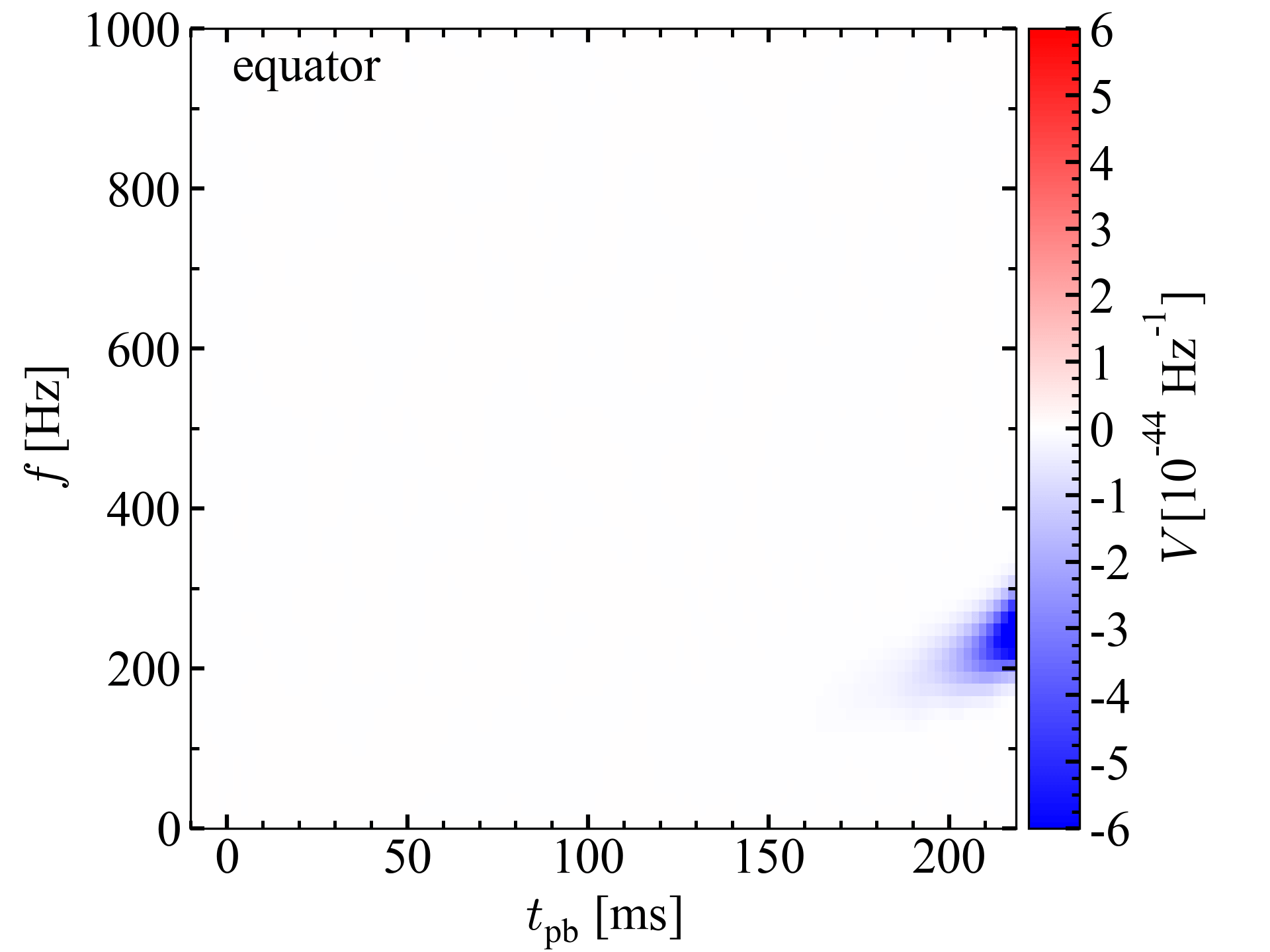}
\includegraphics[ width=0.47\textwidth]{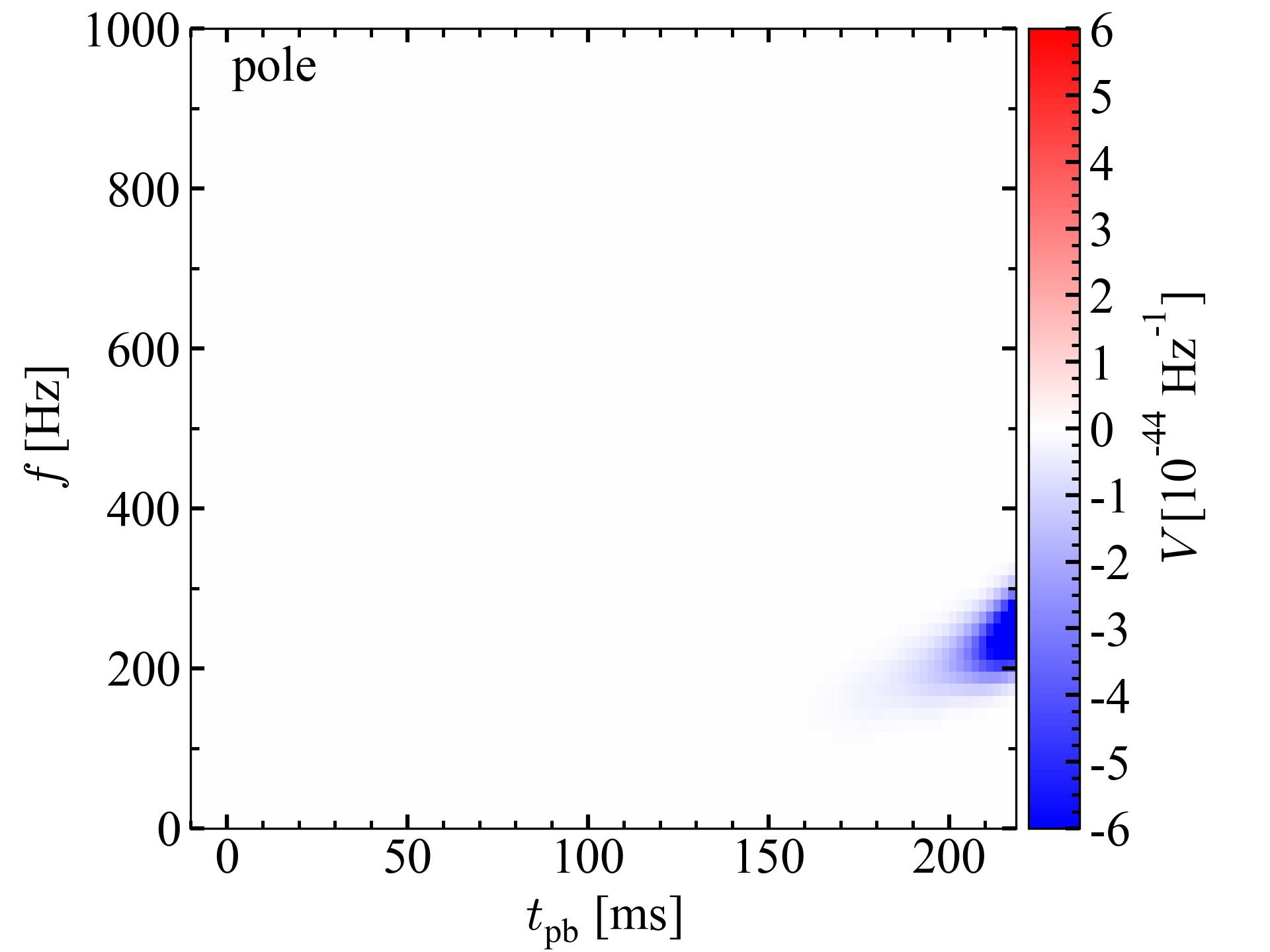}
\caption{Spectrogram of V-mode spectral GW amplitudes for the $\Omega_{0} =2$\,rad\,s$^{-1}$ (top panels), $\Omega_{0} = 1$\,rad\,s$^{-1}$ (middle panels) and $\Omega_{0} =0$\,rad\,s$^{-1}$ (bottom panels) models seen along the equator (left panels) and the pole (right panels) at a source distance of 10\,kpc. \label{fig:gw_V_spectrogram}}
\end{figure*}

Following \citet{hayama16}, we estimate the Stokes parameter $V$ to quantify the circular polarization of the GW signals (see also \citet{kawahara18}).
Here we plot the spectrograms for $V$ of each model observed along the equator (left panels) and the pole (right panels) in Fig.~\ref{fig:gw_V_spectrogram}.  
Observed along the pole (top and middle right panels), our models show the large $|V|$ during the activity of the low-$T/|W|$ instability or the SASI.
Its negative sign means the right-handed polarization.
In contrast, observed along the equator, the two rotating models show two orders of magnitude smaller $|V|$.
 This is because the amplitude of $h_{\times}$ is negligibly small compared to that of $h_{+}$ as shown in the left column of Fig.~\ref{fig:gw_spectrogram}.
If one observes the two rotating models from the equator, the spectrogram of $V$ stochastically changes its sign with time and thus bears essentially no signature of rotation.
The spectrograms of $V$ in the non-rotating model are similar for the equatorial and polar observer directions. More specifically, the equatorial direction is taken to the $x$ axis.
Note that 
this is simply by chance because both of the equatorial and polar observer angles relative to the (spiral) SASI axis happen to  be close ($<90^{\circ}$).

\begin{figure}
\centering
\includegraphics[ width=0.44\textwidth]{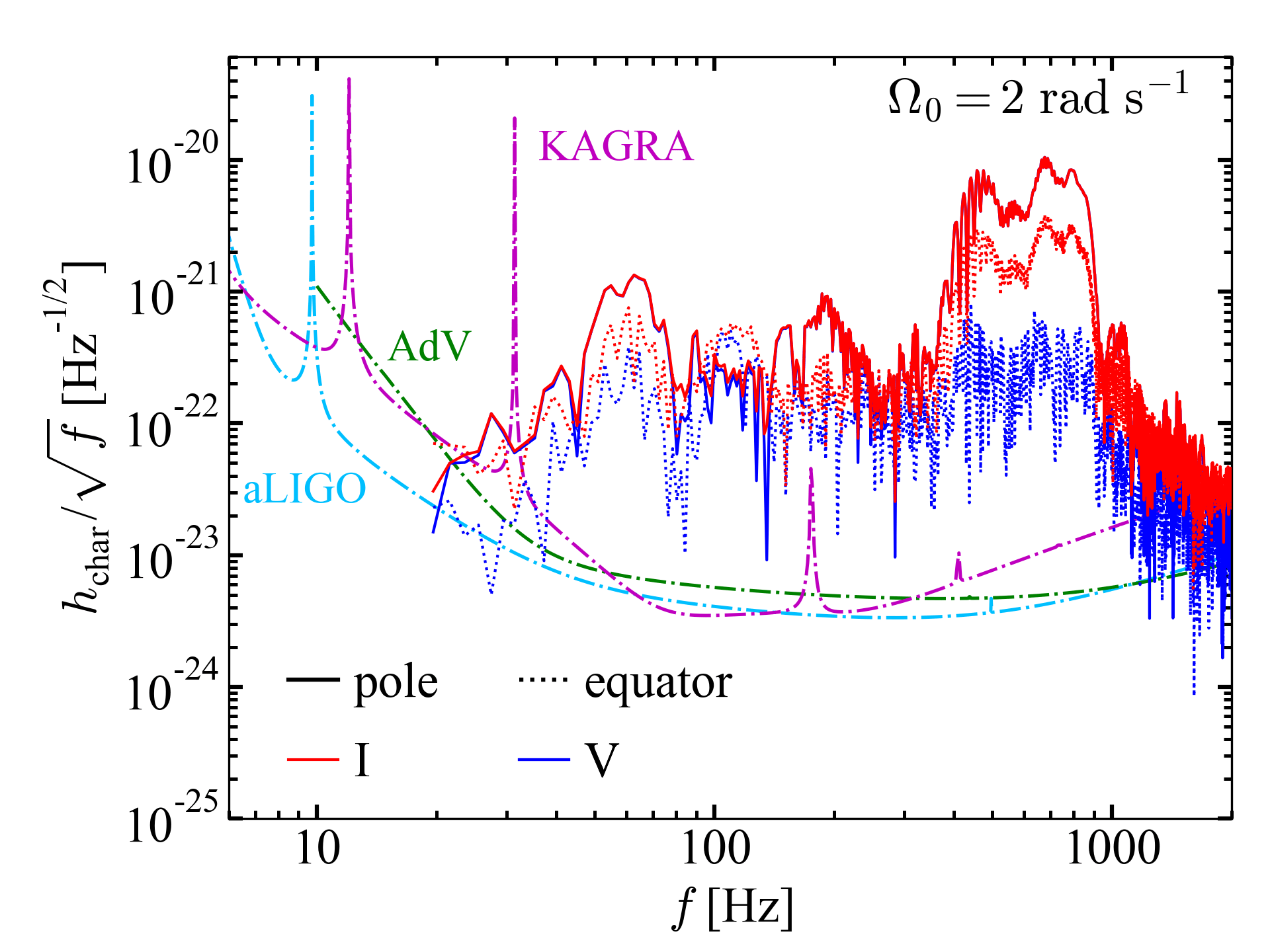}
\includegraphics[ width=0.44\textwidth]{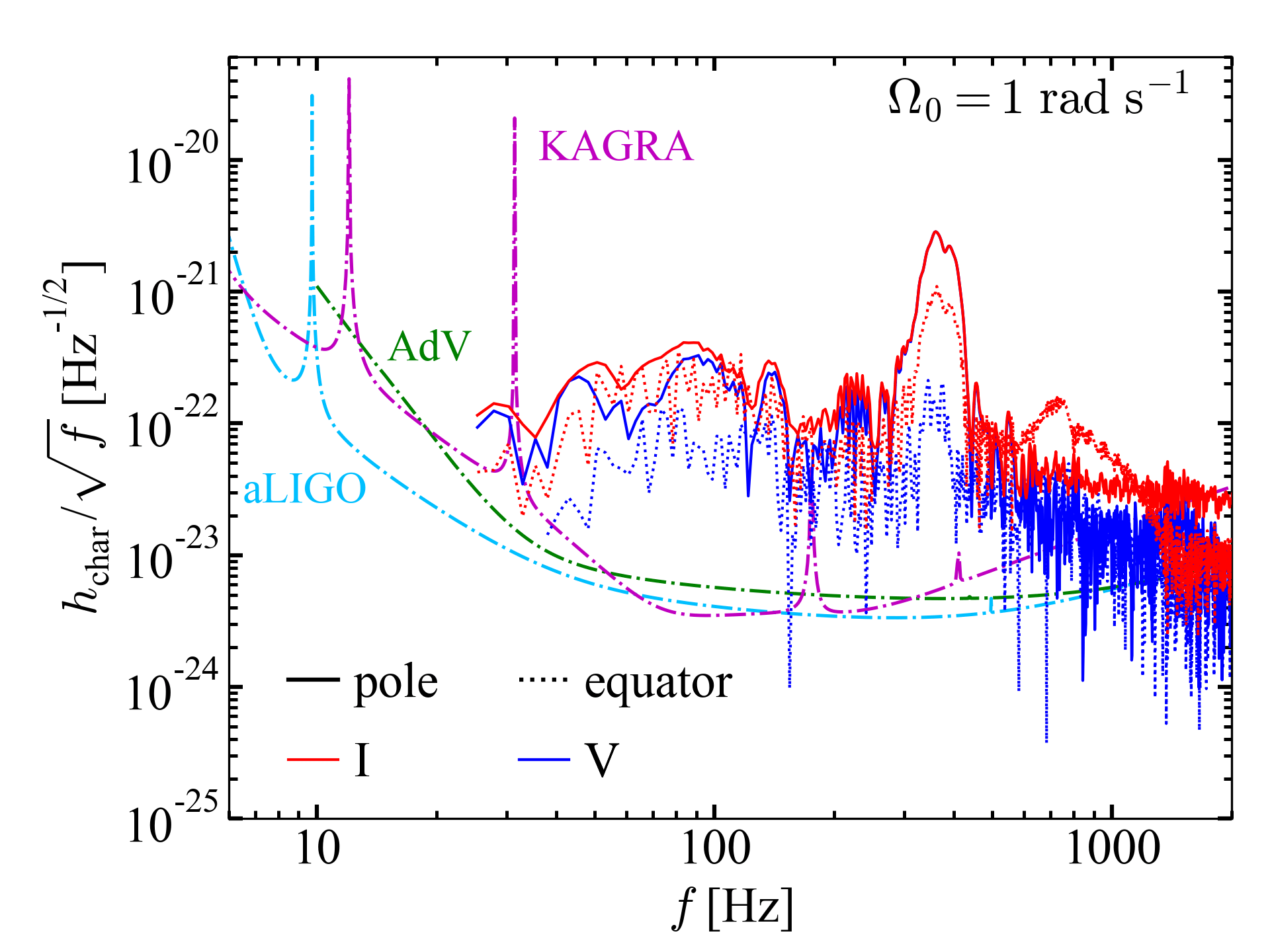}
\includegraphics[ width=0.44\textwidth]{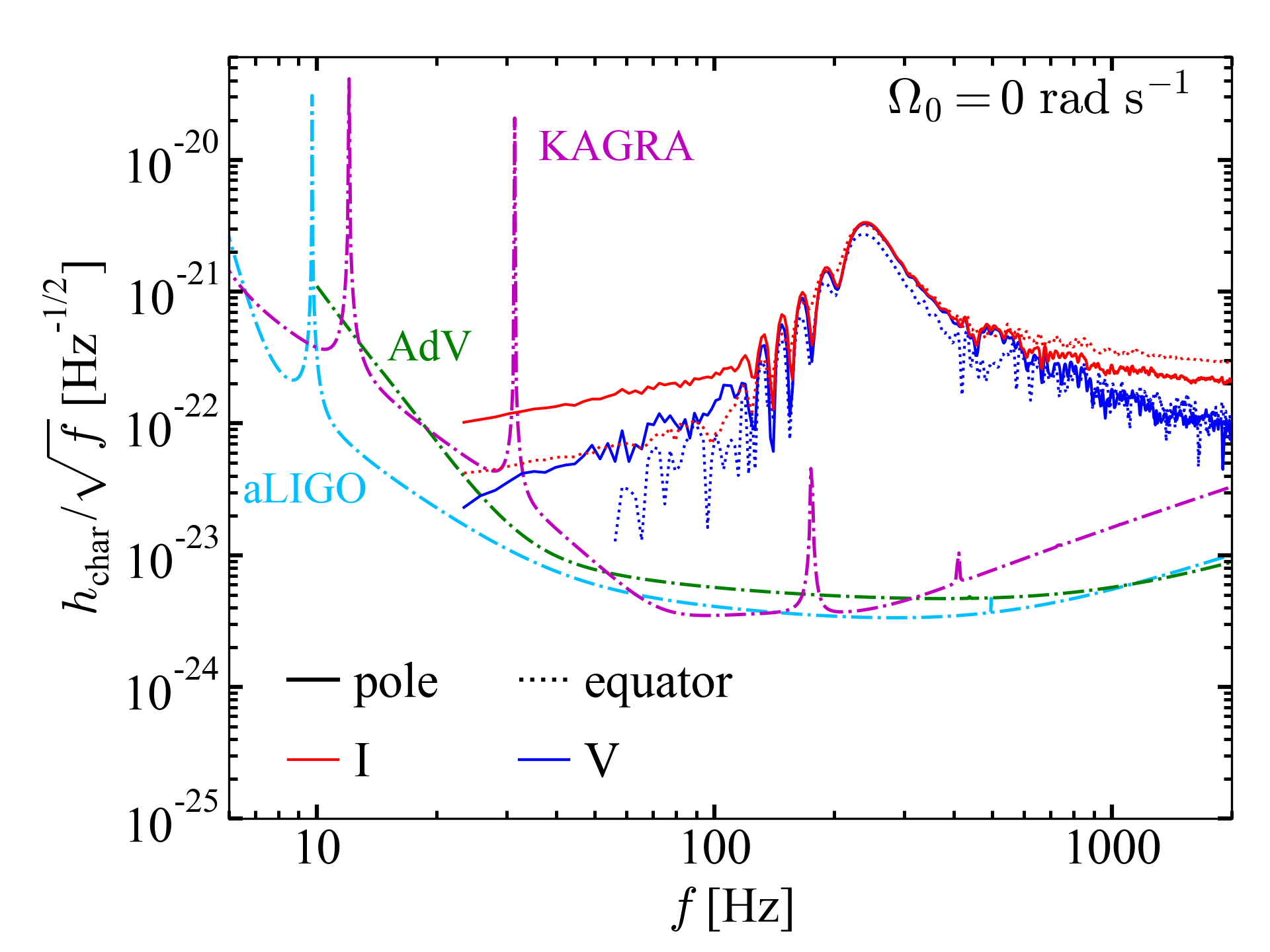}
\caption{Characteristic I-mode (red lines) and V-mode (blue lines) GW spectral amplitudes for the $\Omega_{0} = 2$\,rad\,s$^{-1}$ (top panel), $\Omega_{0} =1$\,rad\,s$^{-1}$ (middle panel) and $\Omega_{0} =0$\,rad\,s$^{-1}$ (bottom panel) models seen along the pole (solid lines) and the equator (dotted lines) as a source distance of 10\,kpc relative to the noise amplitudes of advanced LIGO (aLIGO; cyan line), advanced VIRGO (AdV; green line), KAGRA (magenta line) from \citet{abbott18det}. The detector noise amplitudes are indicated by dash-dotted lines. \label{fig:gw_stokes_spectrum}}
\end{figure}

Next, we explore the detectability of these circularly 
polarized GW signals.
To evaluate a noise spectral density for the Stokes parameters in a simplified manner, we consider an idealized situation, in which two co-located GW detectors have a common feature regarding the (single-sided) noise spectral density 
$S_{n}$
and the two detectors can measure $h_{+}$ and $h_{\times}$ independently.
 Probably, to do this most simply, one detector that measures $h_{\times}$ needs to be located at a position tilted at 45$^\circ$ from another one that measures $h_{+}$. We furthermore assume that the detector noise is Gaussian and uncorrelated. Then the variance $\sigma$ of the noise for the Stokes parameters $I_n$ and $V_n$ can be described as follows:
\begin{eqnarray}
  \sigma_{I_n}^2 &=& \frac{1}{8}S_{n}^2, \label{eq:noiseI}\\
  \sigma_{V_n}^2 &=& \frac{1}{8}S_{n}^2. \label{eq:noiseV}
\end{eqnarray}
See Appendix \ref{appendix:A} for the derivation. Because the noise is Gaussian and detector-independent, the variance for $I_n$ is the same as that for $V_n$. 
We define the characteristic strains for the Stokes parameters of the $I$ and $V$ modes as,
\begin{eqnarray}
  h_{\rm{char},\it{I}}^2 &=& \frac{8f^{2} I}{\Delta f}, \label{eq:hchar_I}\\
  h_{\rm{char},\it{V}}^2 &=& \frac{8f^{2} |V|}{\Delta f}. \label{eq:hchar_V}
\end{eqnarray}
Then, $h_{\mathrm{char},\it{I}}$ is identical to $h_{\mathrm{char}}$ in Eq.~(\ref{eq:hchar}).
If the GW is purely circularly polarized, $h_{\mathrm{char},\it{V}}$ is also identical to $h_{\mathrm{char}}$.
We define the SNR for the $I$ and $V$ modes to match those for $h_{\mathrm{char}}$ as follows.
\begin{eqnarray}
  \frac{\sqrt{8}I}{\sigma_{I_n}}&=&
  \left(\frac{h_{\rm{char},\it{I}}/\sqrt{f}}{\sqrt{S_{n}}}\right)^2\Delta {\rm ln}\,f,     \label{eq:stokes_sn_I}\\
  \frac{\sqrt{8}|V|}{\sigma_{V_n}}&=&
  \left(\frac{h_{\rm{char},\it{V}}/\sqrt{f}}{\sqrt{S_{n}}}\right)^2\Delta {\rm ln}\,f.   \label{eq:stokes_sn_V}
\end{eqnarray}
If we presume that the numerator and denominator in the parentheses of these equations are the signal and noise \citep[see Eq.~(5.2) in ][]{Flanagan:1998a}, respectively, the noise $S_n$ is equivalent for both the $I$ and $V$ modes.
 Note that the $I$ mode is, by definition, larger than or equal to $|V|$ ($I \geq |V|$, e.g., \citet{hayama16}). Then one expects that the $V$ mode basically shows smaller SNR, making the detection of the $V$ mode more difficult compared to the $I$ mode for the detector configuration mentioned above. The coherent network analysis in \citet{Hayama18} showed higher SNR ratio for $V$ than for $I$. This difference is likely to come from the effect of realistic antenna pattern and the position of the multiple detectors
  (LIGOx2, Virgo, and KAGRA), which  is not taken into account in our analysis above and we leave for the future work \citep[e.g.][]{Man20}.

We apply the above expressions of $h_{\mathrm{char},\mathit{I}}$ and $h_{\mathrm{char},\mathit{V}}$ to our models and plot their signals and noises in Fig.~\ref{fig:gw_stokes_spectrum}.
Observed along the pole (solid lines), $h_{\mathrm{char},\mathit{I}}$ (red lines) and $h_{\mathrm{char},\mathit{V}}$ (blue lines) of each model are nearly overlapped at their peak.
This indicates that the GWs are almost purely circularly polarized.
Observed along the equator (dotted lines), the two rotating models show smaller $h_{\mathrm{char},\mathit{V}}$ (dotted blue lines), whose peak is one order of magnitude smaller than that of $h_{\mathrm{char},\mathit{I}}$ (dotted red lines).
In the non-rotating model, the amplitudes of the GW circular polarization observed along the pole and the equator are similar since the tilted angles of the polar and the equatorial (the $x$-axis) directions are comparable relative to the rotation axis of the spiral SASI.
The bottom panel of Fig.~\ref{fig:gw_stokes_spectrum} shows that the SNR of $h_{\mathrm{char},\mathit{V}}$ is comparable or one order-of-magnitude smaller than that of $h_{\mathrm{char},\mathit{I}}$, depending on the GW frequencies.

\subsection{Characteristic time variability of neutrino signals} \label{sec:neu}
\begin{figure}
\centering
\includegraphics[width=0.45\textwidth]{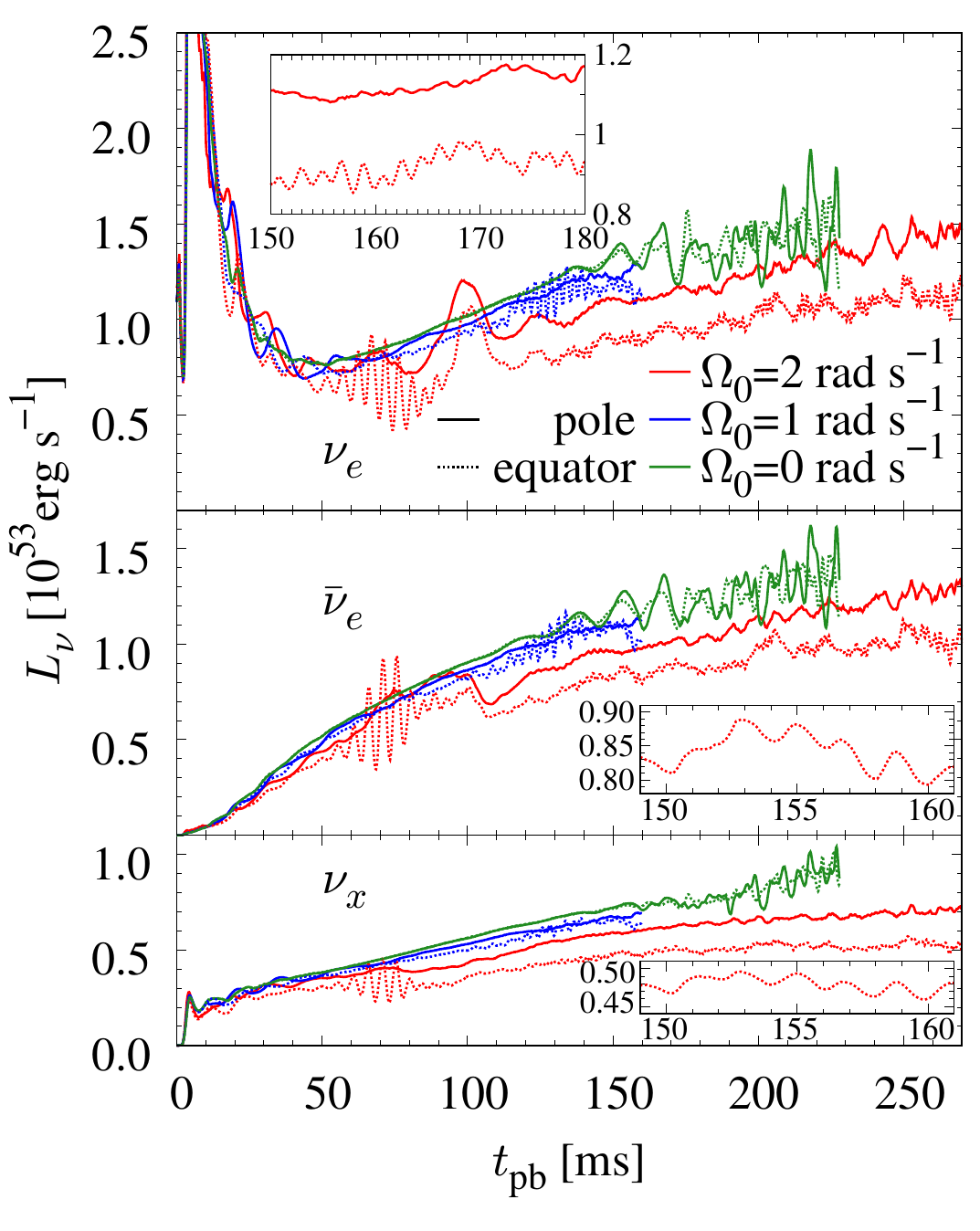}
\caption{Neutrino luminosity of $\nu_e$ (top panel), $\bar\nu_e$ (middle panel) and $\nu_x$ (bottom panel) for the $\Omega_{0} = 2$\,rad\,s$^{-1}$ (red lines), $\Omega_{0} = 1$\,rad\,s$^{-1}$ (blue lines) and $\Omega_{0} = 0$\,rad\,s$^{-1}$ (green lines) models seen along the pole (solid lines) and the equator (dotted lines). The inset of each panel zooms into the small modulation of each neutrino luminosity for the $\Omega_{0} = 2$\,rad\,s$^{-1}$ model seen along the equator. \label{fig:lneu}}
\end{figure}

In this section, we investigate the neutrino emission properties of our models and discuss the detectability.
Following \citet{Tamborra14}, we compute the neutrino luminosities of each flavour in the two selected observer directions for our models.
Fig.~\ref{fig:lneu} shows the neutrino luminosities of $\nu_e$ (top panel), $\bar\nu_e$ (middle panel), and $\nu_x$ (bottom panel) in the polar (solid lines) and equatorial (dotted lines) observer directions for the $\Omega_{0} =2$\,rad\,s$^{-1}$ (red lines), $\Omega_{0} =1$\,rad\,s$^{-1}$ (blue lines), and $\Omega_{0} =0$\,rad\,s$^{-1}$ (green lines) models, respectively.

In the $\Omega_{0} =2$\,rad\,s$^{-1}$ model, observing along the equatorial direction, we can see that the strong quasi-periodic modulation of the neutrino luminosity between $50<t_{\mathrm{pb}}<90$\,ms for all the flavours. During this period,  the $m=1$ and 2 deformations of the PNS are significant 
(see the top panel of Fig.~\ref{fig:Rnu_Ylm}).
As previously identified in \citet{takiwaki18},
 this stems from the neutrino lighthouse effect, where the spinning of strong neutrino emission regions around the spin axis leads to quasi-periodic modulation in the neutrino signal, which is most strongly seen from the equatorial direction.
After $t_{\mathrm{pb}} \sim 120$\,ms, the small modulation of the neutrino luminosity can be also seen (see the dotted line in the inset of each panel).
In contrast, observed along the polar direction, no clear quasi-periodic modulation in the neutrino luminosity is identified (see the solid line in the inset of the top panel for reference).
Similarly to the $\Omega_{0} =2$\,rad\,s$^{-1}$ model, the $\Omega_{0} =1$\,rad\,s$^{-1}$ model shows the quasi-periodic modulation only clearly for the equatorial observer (blue dotted line in the top panel) at the time when the $m=2$ PNS deformation sufficiently develops.
In the $\Omega_{0} =0$\,rad\,s$^{-1}$ model, both of the polar and equatorial observers can see the modulation of the neutrino luminosities after $t_{\mathrm{pb}} \sim150$\,ms when the SASI deveolps (see the bottom panel of Fig.~\ref{fig:shock_Ylm}).

\begin{figure}
\centering
\includegraphics[ width=0.46\textwidth]{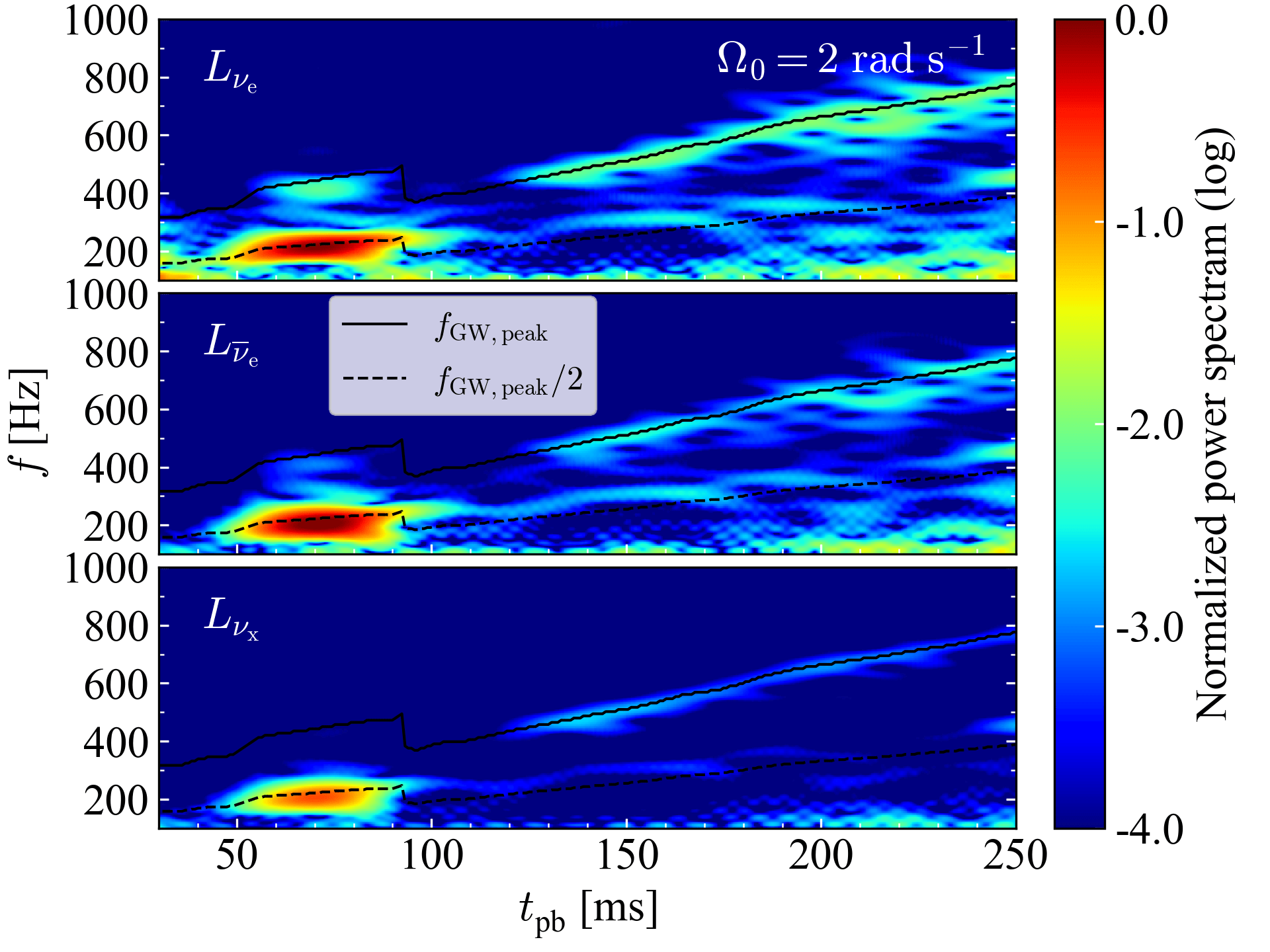}
\includegraphics[ width=0.46\textwidth]{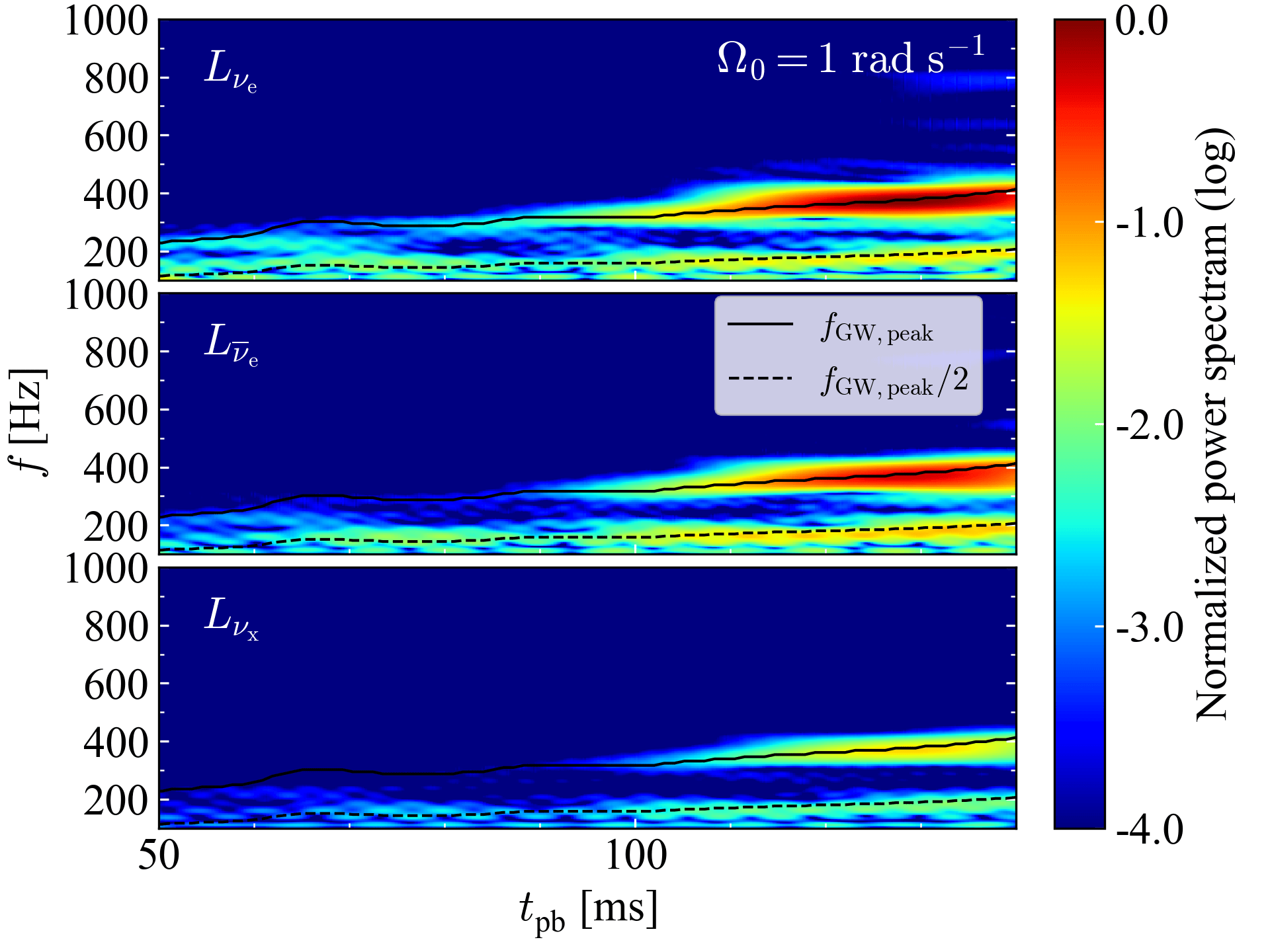}
\includegraphics[ width=0.46\textwidth]{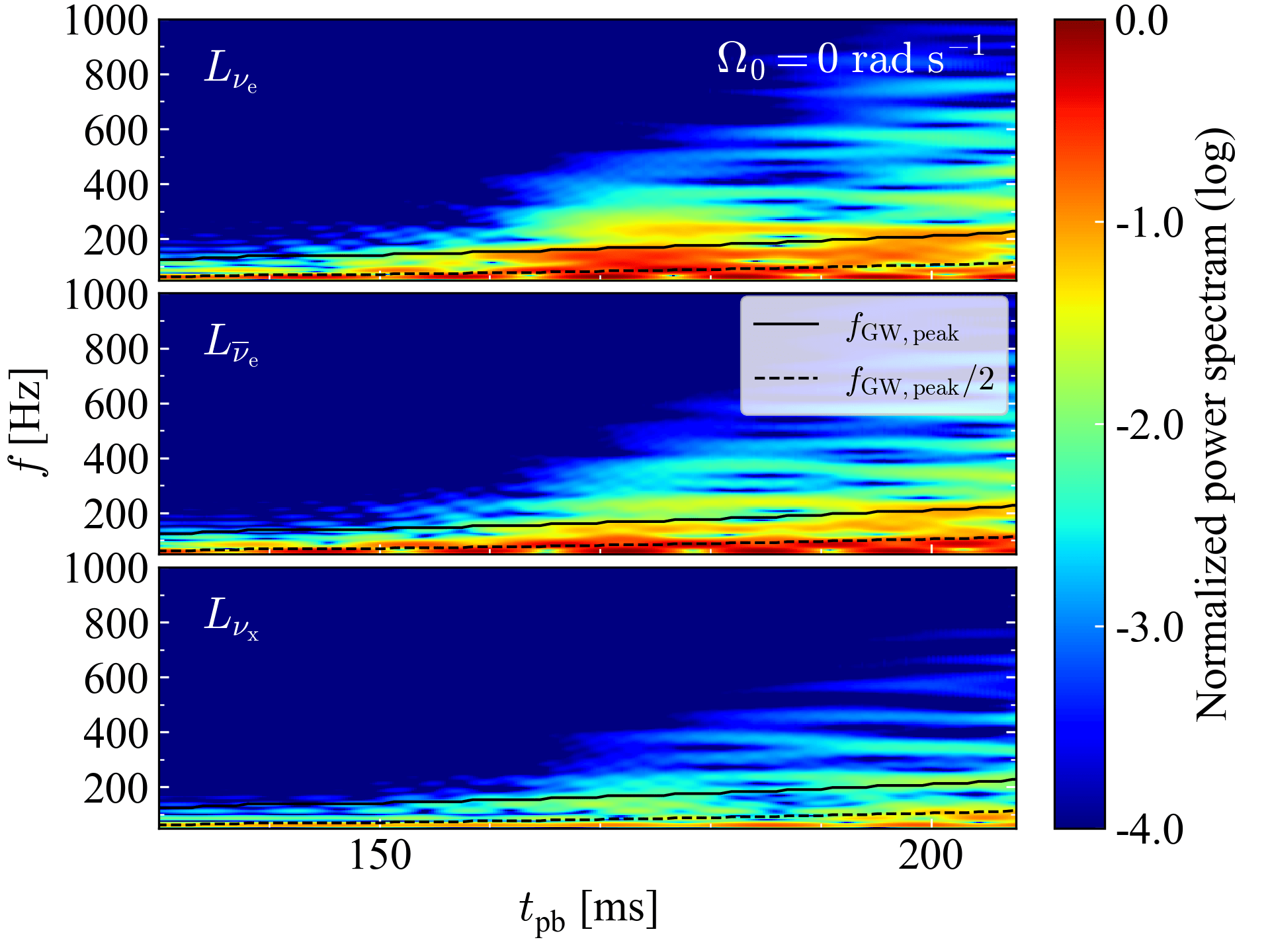}
\caption{Spectrograms of (normalized) neutrino luminosity of $\nu_e$, $\bar\nu_e$ and $\nu_x$ (top, middle and bottom small panels in each panel, respectively) for the $\Omega_{0} = 2$\,rad\,s$^{-1}$ (top panel), $\Omega_{0} = 1$\,rad\,s$^{-1}$ (middle panel) and $\Omega_{0} = 0$\,rad\,s$^{-1}$ (bottom panel) models seen along the equator. The peak frequency of their GWs (black solid lines) and a half of them (black dashed lines) are also plotted. \label{fig:lneu_spectrogram}}
\end{figure}

We perform the spectrogram analysis of the neutrino luminosities to identify the characteristic neutrino modulation frequency.
Here we use a Hann window with the window width of 40\,ms and show the result in Fig.~\ref{fig:lneu_spectrogram}.
The $\Omega_{0} =2$\,rad\,s$^{-1}$ model (top panel) shows the modulation of the neutrino luminosity around $\sim200$ Hz for all flavours between $50<t_{\mathrm{pb}}<100$\,ms.
This frequency corresponds to half of the GW peak frequency $f_{\rm GW,peak}/2$ (see the dotted line in Fig.~\ref{fig:lneu_spectrogram}).
At later times $t_{\rm pb}\gtrsim120$ ms, the modulation frequency increases from 400\,Hz to 800\,Hz with time.
In this period, the characteristic frequency of the neutrino luminosity almost perfectly matches with the GW peak frequency itself $f_{\rm GW,peak}$ (see the solid line).
This is because the rotation frequency of the lighthouse effect depends on the non-axisymmetric deformation mode of the PNS.
This implies that, if a PNS is deformed with the $m$-th mode around its rotation axis, its neutrino luminosity observed by an equatorial observer modulates at $m$ times of the rotation frequency of the PNS ($\sim m/2$ times of the GW peak frequency).
This correspondence can be also seen in the $\Omega_{0} =1$\,rad\,s$^{-1}$ model as well (middle panel).
In this model, the $m=1$ and $m=2$ deformation of the PNS leads to the neutrino time modulation, corresponding to half ($1/2$) and the same ($2/2$) of the GW peak frequency, which is represented by the solid line and dotted line.
The $\Omega_{0} =0$\,rad\,s$^{-1}$ model shows the modulation around $\sim90$\,Hz (bottom panel).
This characteristic frequency also corresponds to about half of the GW peak frequency and closely to the SASI frequency seen from Fig.~\ref{fig:SASI_spectrogram}, since the $l=1$ and $m=-1,0,1$ modes are the dominant modes in the SASI activity.

\begin{figure}
\centering
\includegraphics[ width=0.47\textwidth]{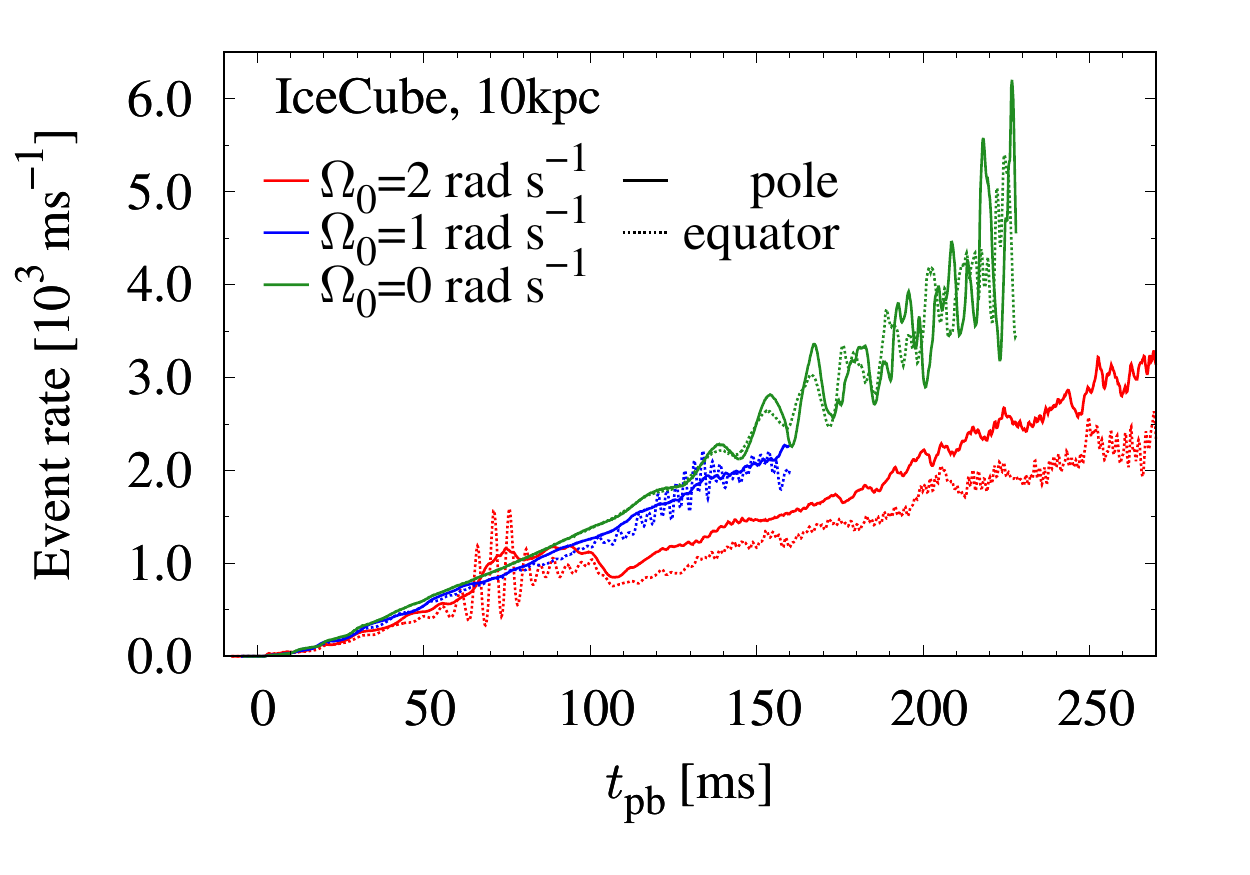}
\includegraphics[ width=0.47\textwidth]{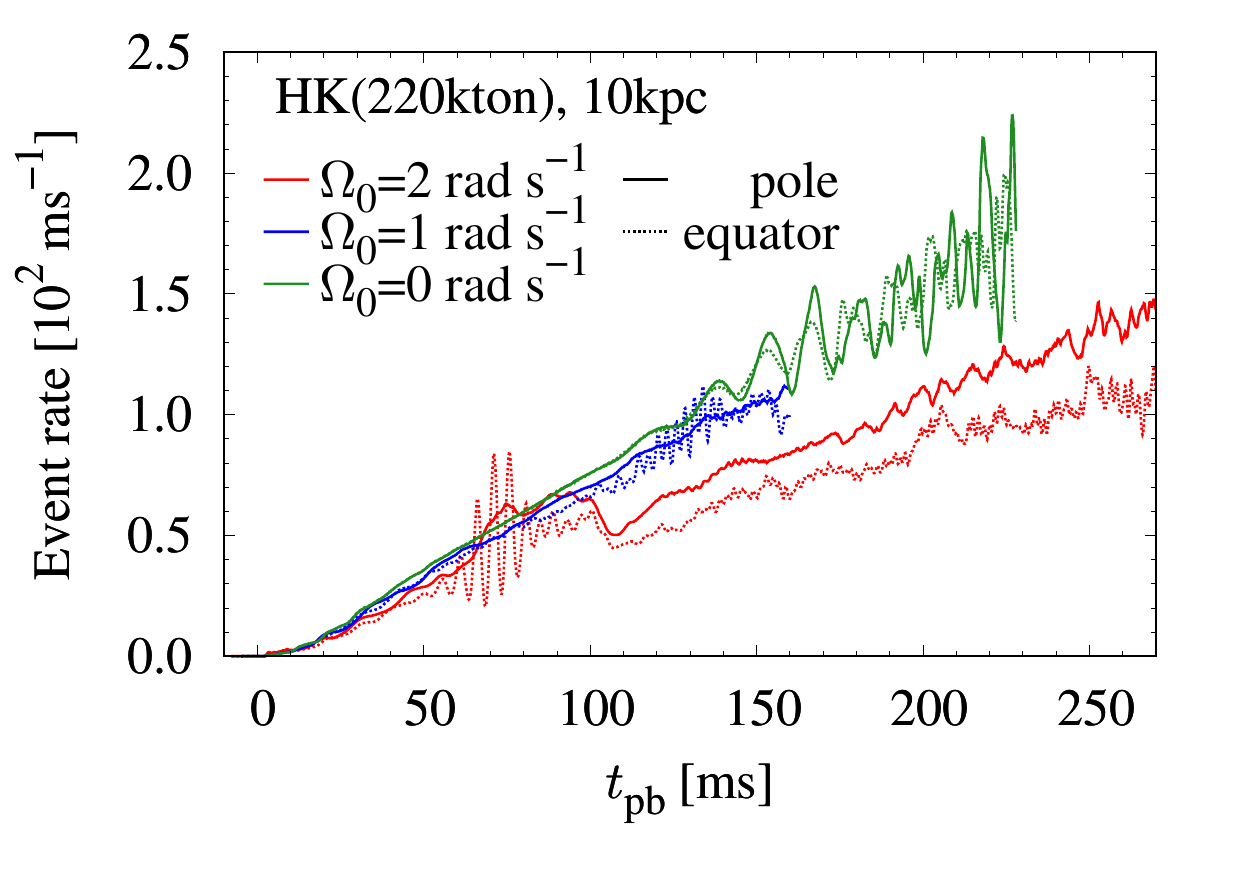}
\caption{Neutrino detection rate of IceCube (top panel) and HK (bottom panel) as a function of time after bounce for the $\Omega_{0} = 2$\,rad\,s$^{-1}$ (red lines), $\Omega_{0} = 1$\,rad\,s$^{-1}$ (blue lines) and $\Omega_{0} = 0$\,rad\,s$^{-1}$ (green lines) models observed at 10\,kpc along the pole (solid lines) and the equator (dotted lines). \label{fig:neu_event}}
\end{figure}

We move on to the discussion about the detectability of these neutrino modulations.
We evaluate the event rates of each model for two neutrino detectors: IceCube \citep{abbasi11,salathe12} and Hyper-Kamiokande (\citet{abe11,HK18}, hereafter HK).
The main channel in these detectors is of anti-electron neutrino ($\bar\nu_e$) with inverse-beta decay. For simplicity, we take into account only this channel to evaluate the event rates\footnote{Consideration of collective neutrino oscillation effects (e.g. \citet{mirizzi16} for a review) especially 
in multi-D simulations is challenging, albeit very important (see \citet{abbar19,shalgar19,johns19,morinaga20,Sasaki20,zaizen20,Cherry20} for collective references therein), which is beyond the scope of this work.}.
We assume that the neutrino energy spectrum is a Fermi-Dirac distribution (see \citet{lund10,takiwaki18} for more detail).
Fig.~\ref{fig:neu_event} shows that the event rates of each model for IceCube (top panel) and HK (bottom panel) at a source distance of 10\,kpc.

The event rate of IceCube for the $\Omega_{0} =2$\,rad\,s$^{-1}$ model observed along the equator (red dotted line) shows the modulation whose amplitude is $\sim1000$\,ms$^{-1}$ between $50<t_{\mathrm{pb}}<100$\,ms and the modulation whose amplitude is $\sim100$\,ms$^{-1}$ after $t_{\mathrm{pb}} \sim 130$\,ms.
In the $\Omega_{0} =1$\,rad\,s$^{-1}$ model, the event rate of IceCube observed along the equator (blue dotted line) is modulated with an amplitude of $\sim500$\,ms$^{-1}$ after $t_{\mathrm{pb}}\sim100$\,ms.
Seen by both polar and equatorial observers, the event rate of IceCube for the $\Omega_{0} =0$\,rad\,s$^{-1}$ model shows the modulation whose amplitude is $\sim1000$\,ms$^{-1}$ after $t_{\mathrm{pb}} \sim 150$\,ms.
One can see that the event rates of HK are about 20 times smaller than those of IceCube.
Although \citet{Tamborra13} pointed out that the SNR of HK is expected to be higher than that of IceCube for a distant supernova,
recent studies \citep{Lin20,Nagakura21} showed that the maximum detectable distance of HK is smaller than that of IceCube.
But it is noteworthy that HK is still beneficial to reconstructing the energy spectra for supernova neutrinos.

\begin{figure}
\centering
\includegraphics[ width=0.44\textwidth]{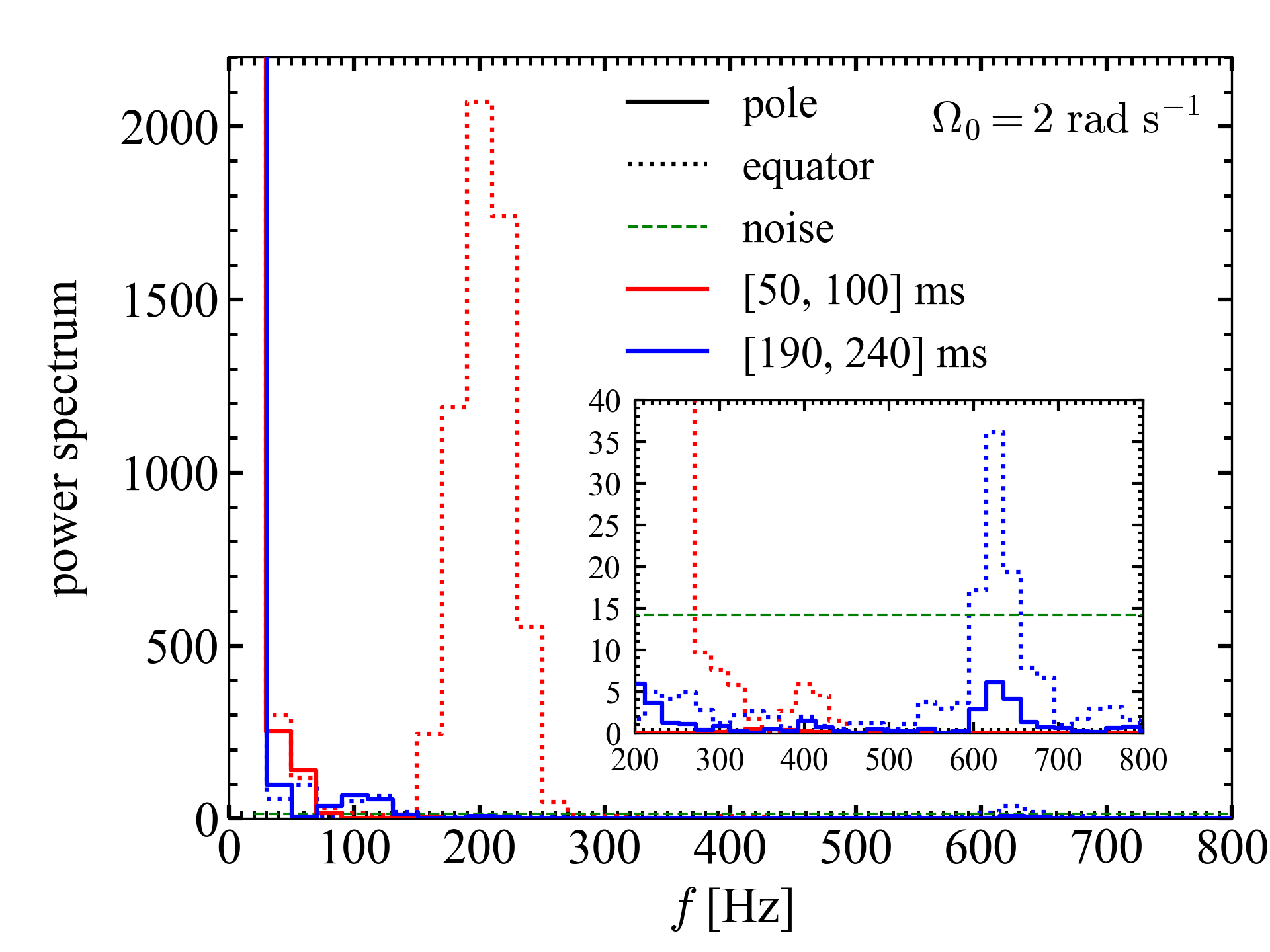}
\includegraphics[ width=0.44\textwidth]{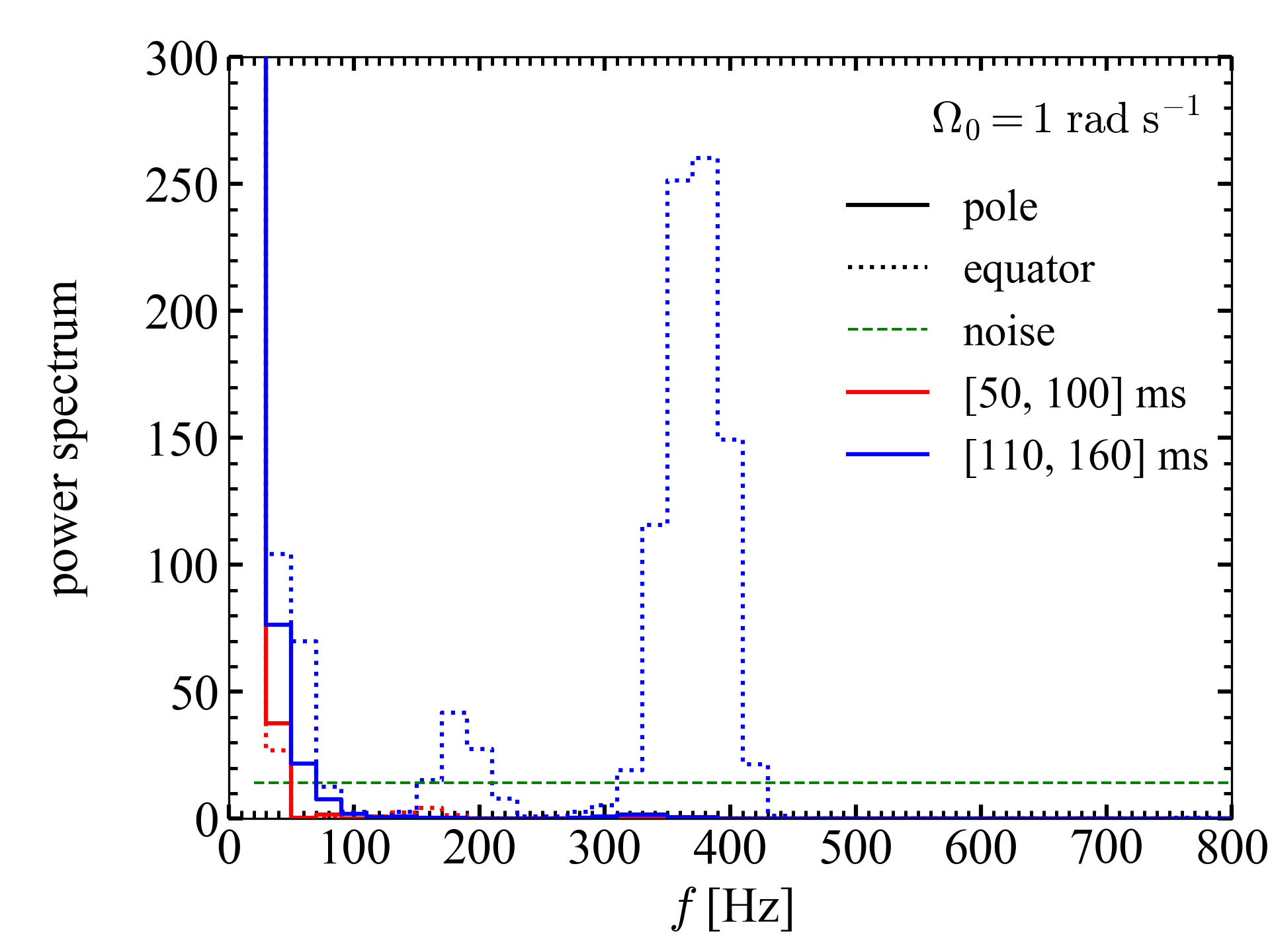}
\includegraphics[ width=0.44\textwidth]{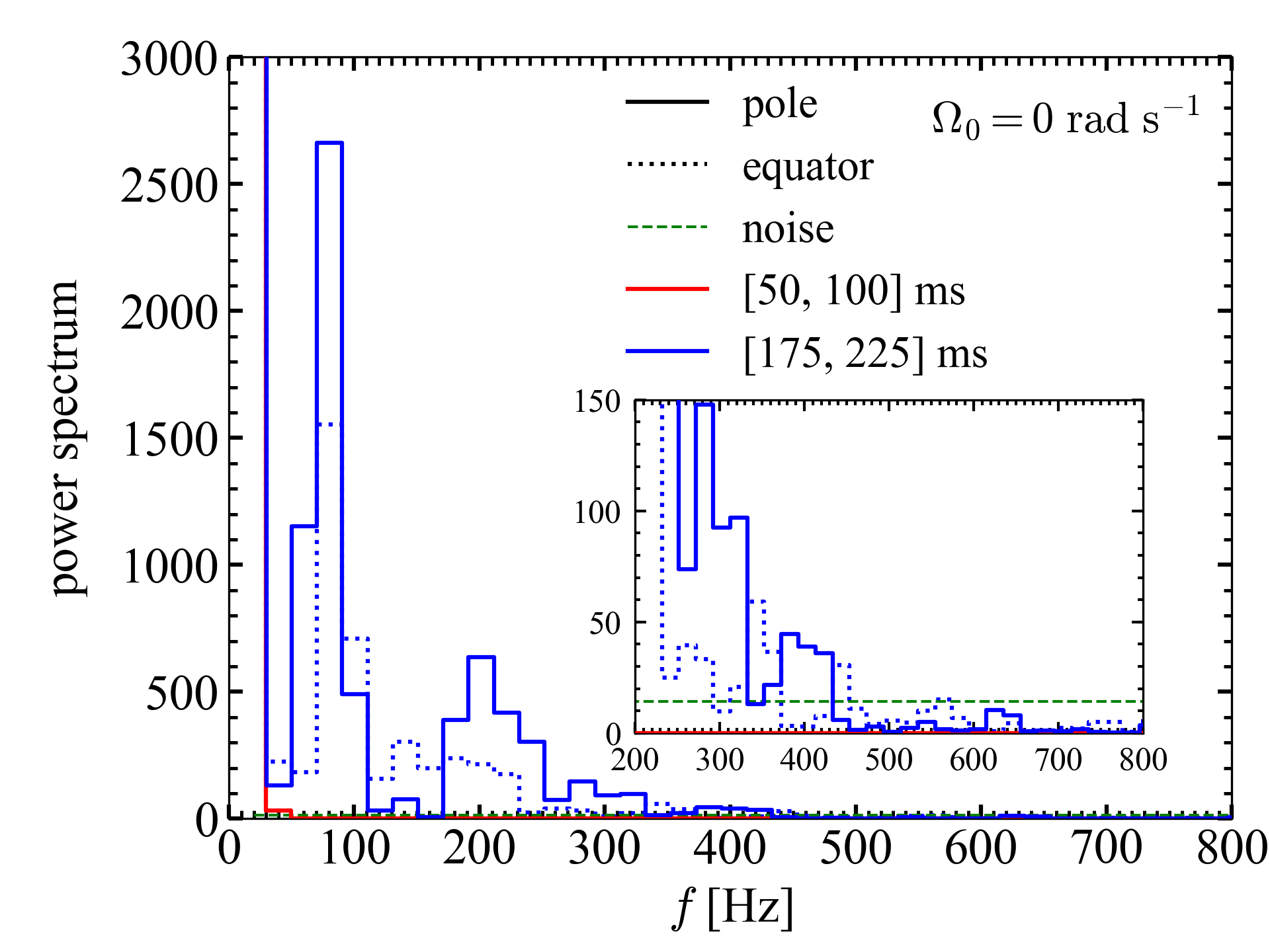}
\caption{Power spectra of the IceCube event rate observed at 10 kpc along the pole (solid lines) and the equator (dotted lines) for the $\Omega_{0} = 2$\,rad\,s$^{-1}$ (top panel), $\Omega_{0} = 1$\,rad\,s$^{-1}$ (middle panel) and $\Omega_{0} = 0$\,rad\,s$^{-1}$ (bottom panel) models relative to the background shot noise (green dashed lines). Fourier transform is applied over the two different time intervals indicated in the legends of each panel. See text for the meaning of the time intervals. The inset in the top and bottom panels zooms into the low peak(s) at high frequencies. \label{fig:neu_event_power}}
\end{figure}

To compare the neutrino modulation amplitudes and the detector noise amplitudes of IceCube, we plot the power spectra and the noises of the event rates for each model observed by the polar and equatorial observers located at a source distance of 10\,kpc in Fig.~\ref{fig:neu_event_power}.
Here we pick up two time intervals, which differs for each model, of 50\,ms to highlight the characteristic neutrino modulation.
In this spectral analysis, we use a Hann window with the window width of 50\,ms to avoid the aliasing effect.

In the top panel of Fig.~\ref{fig:neu_event_power}, we plot the power spectra of the $\Omega_{0} =2$\,rad\,s$^{-1}$ model over the time intervals of the first neutrino modulation ($50<t_{\mathrm{pb}}< 100$\,ms, red lines) and of the the second neutrino modulation ($190<t_{\mathrm{pb}}< 240$\,ms, blue lines). 
Observed along the equatorial direction (dotted line), the characteristic peak frequency in the time interval of the first neutrino modulation can be seen at $\sim200$\,Hz, corresponding to the rotation frequency of the PNS that shows the $m=1$ deformation due to the low-$T/|W|$ instability\footnote{Accordingly, the 
GW spectrogram (top left panel of Fig. \ref{fig:gw_spectrogram}) peaks around $\sim$ 400 Hz for this time  period ($50<t_{\mathrm{pb}}< 100$\,ms).}.
The peak is well above the background noise level ($\sim14$, see the green vertical dashed line in the inset).
In the time interval of a part of the second neutrino modulation ($190<t_{\mathrm{pb}}<240$\,ms), one can see the peak at $\sim620$\,Hz in the spectrum (see the inset), which again nicely matches with the rotation frequency of the PNS that shows the $m=2$ deformation due to the low-$T/|W|$ instability (e.g, top left panel of Fig.~\ref{fig:gw_spectrogram}).
This power spectral amplitude is about two times larger than the noise level.
In contrast, observing along the polar direction (solid line), one cannot find these characteristic spectral peaks, and the spectral amplitude is below the noise level for $f>200$\,Hz. This is simply because of the absence of the neutrino lighthouse effect, seen from the polar direction.

In the middle panel of Fig.~\ref{fig:neu_event_power}, we plot the power spectra of the $\Omega_{0} =1$\,rad\,s$^{-1}$ model before ($50<t_{\mathrm{pb}}< 100$\,ms, red lines) and during the active neutrino modulation epoch ($110<t_{\mathrm{pb}}< 160$\,ms, blue lines), respectively. 
 Seen from the polar direction (solid line), no characteristic peaks above the noise level are identified in both of the time intervals.
 Seen from the equator (blue dotted line), 
 the power spectrum during the active neutrino modulation ($110<t_{\mathrm{pb}}< 160$\,ms) shows the peak well above the noise level at $\sim380$\,Hz. The peak neutrino frequency corresponds to the rotation frequency of the $m=2$ PNS deformation (compare the excess around $\sim 400$\,Hz 
  in the GW spectrogram (middle left panel of Fig. \ref{fig:gw_spectrogram})).

Similar to the middle panel, but the bottom panel of Fig.~\ref{fig:neu_event_power} is for the $\Omega_{0} =0$\,rad\,s$^{-1}$ model, showing before ($50<t_{\mathrm{pb}}< 100$\,ms, red lines) and during the active  neutrino modulation epoch ($175<t_{\mathrm{pb}}< 225$\,ms, blue lines), respectively. 
As already mentioned, the latter epoch is in the midst of vigorous SASI activity. For this time epoch, the power spectra for both of the polar and equatorial observers have the peak well above the noise level at $\sim80$\,Hz, where the peak neutrino frequency matches with the SASI frequency.
Furthermore, one can see the secondary peaks at higher frequencies than the SASI frequency as shown in the previous studies \citep{KurodaT17,Walk18,Walk20}.

\begin{figure*}
    \centering
    \includegraphics[ width=0.8\textwidth]{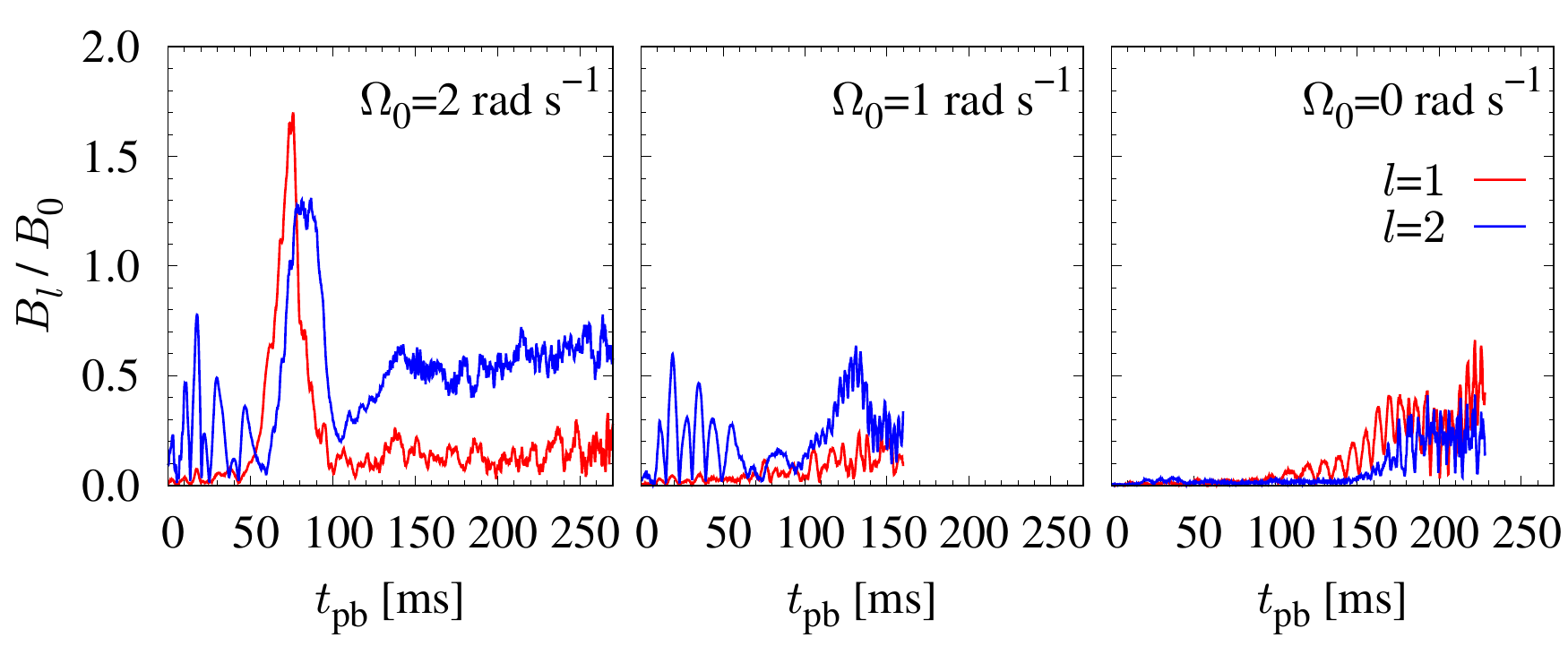}
    \caption{Time evolution of the dipole ($l=1$; red line) and quadrupole ($l=2$; blue line) amplitudes of the ELN emission relative to the monopole amplitude for the $\Omega_{0} = 2$\,rad\,s$^{-1}$ (left panel), $\Omega_{0} = 1$\,rad\,s$^{-1}$ (middle panel) and $\Omega_{0} = 0$\,rad\,s$^{-1}$ (right panel) models.}
    \label{fig:LESA}
\end{figure*}

From the above analysis, the characteristic time modulations in the neutrino signals for each model would be detectable at a distance of 10\,kpc. 
Therefore, our results suggest that a joint observation of the GW and neutrino signals could provide a live information on the rapidly rotating PNS evolution during which the various modes of the non-axisymmetric instabilities develop. When the dipole mode of the spiral SASI is dominant, our non-rotating model shows that the GW frequency is twice as high as the SASI frequency, while the neutrino modulation frequency is almost the same as the SASI frequency (see also \cite{Vartanyan19b}). Very recently, \citet{Walk20} found that the neutrino modulation frequency can be twice as high as the SASI frequency when the SASI quadrupole mode is dominant.
These results lend further support to the speculation that one could decipher the dominant SASI activity via the joint GW-$\nu$ observation from non-rotating stellar core-collapse on the way to the black hole formation.

Finally, we briefly describe asymmetries of the electron lepton number (ELN) emission. \citet{Tamborra14ApJ} first witnessed the LESA, which is characterized by a large-scale asymmetry of the ELN emission. The presence of LESA has been discussed by the subsequent 3D simulations \citep{O'Connor18,Glas19b,Walk19,Vartanyan19b,Walk20,Kuroda20,Stockinger20}.

The previous studies have performed a multipole analysis of the radial ELN flux to measure its asymmetry. As we do in Eqs. (\ref{eq:alm}) and (\ref{eq:clm}), we decompose the ELN flux into spherical harmonics. Fig.~\ref{fig:LESA} displays the square root of the sum of the squares of the spherical harmonic coefficients relative to the monopole one, $B_{l}/B_{0}$, for $l=1$ and 2. Following \citet{Walk19}, we normalize $B_{l}$ by $(2l+1)$.

In the $\Omega_0=2$\,rad\,s$^{-1}$ model (left panel), the dipole ($m=\pm 1$) and quarupole ($m=0$) emissions firstly grow, but weaken soon between $50<t_{\rm pb}<90$\,ms. After $t_{\mathrm{pb}} \sim 110$\,ms, the quadrupole component ($m=0$ mode) becomes larger and dominates over the dipole component. For the $\Omega_0=1$\,rad\,s$^{-1}$ model (middle panel), the quadrupole component ($m=0$ and $\pm2$ modes) starts to develop at $t_{\rm pb}\sim 100$\,ms and become the dominant mode. \citet{janka16} and \citet{Walk19} reported that the dipole ELN emission is weakened by the progenitor rotation and the quadrupole component dominates over the dipole component in their rapidly rotating model, which is consistent with our rotating models. From the coincidence between the development of the low-$T/|W|$ instability (see Fig.~\ref{fig:Rnu_Ylm}) and the growth of asymmetries of the ELN emission in our rotating models, we speculate that the low-$T/|W|$ instability as well as the rotational flattening of the PNS and the aspherical accretion onto the PNS are responsible for the observed asymmetries of the ELN emission in the rotating models.

For the $\Omega_0=0$\,rad\,s$^{-1}$ model (right panel), the dipole component first develops at $t_{\rm pb} \sim 100$\,ms. At later times, the quadrupole component also starts to develop at $t_{\rm pb} \sim 150$\,ms, and reaches comparable amplitudes to the dipole component around $t_{\rm pb} \sim 200$\,ms.
Up to the time of the BH formation, one can see that the large amplitude modulations with the corresponding SASI frequency (see Fig. \ref{fig:SASI_spectrogram}) in the dipole and quadrupole components, while both of the dipole and quadrupole components do not overtake the monopole component.
\citet{Walk20} found that the asymmetric ELN emission and its modulation in their non-rotating BH-forming stellar core-collapse models are induced by the spiral SASI and the effect of the LESA is subdominant. 
Similarly, in the $\Omega_0=0$\,rad\,s$^{-1}$ model, the LESA activity is relatively weak, and the dipole and quadrupole SASI is likely to make the asymmetric ELN emission.
In order to draw a more robust conclusion to what hydrodynamics phenomena drive the asymmetric ELN emission quantitatively, further detailed analysis is needed, which is beyond the scope of this work.

\section{SUMMARY AND DISCUSSION}\label{sec4}
In this paper, we have presented the results of 3D general relativistic radiation hydrodynamic simulations with spectral neutrino transport for a 70\,M$_{\odot}$ star, where we computed three models by imposing the initial central angular velocity of $\Omega_0=0, 1, 2$\,rad\,s$^{-1}$ in a parametric manner. 
In our rotating models, the non-axisymmetric deformation of the PNS due to the development of the low-$T/|W|$ instability is observed. For our most rapidly rotating model ($\Omega_0=2$\,rad\,s$^{-1}$), the dipole PNS deformation firstly grows between $50<t_{\rm pb}<90$\,ms, which is followed by the quadrupole PNS deformation until the final simulation time ($t_{\rm pb} \sim$ 270\,ms). The moderately rotating model ($\Omega_0=1$\,rad\,s$^{-1}$) also shows the growth of the low-$T/|W|$ instability, but only with the two-armed spiral flow.
In the non-rotating model ($\Omega_0=0$\,rad\,s$^{-1}$), the sloshing SASI mode firstly develops between $100<t_{\mathrm{pb}}<150$\,ms, which is followed by the growth of the spiral SASI at $t_{\mathrm{pb}}>150$\,ms.
 For the non-rotating model, the steep rise of the maximum density near at the final simulation time ($t_{\mathrm{pb}}\sim230$\,ms) indicates the black hole formation, whereas such trend is yet to be obtained for the rotating models most likely because of the short simulation time.

We investigated the GW emission from these models in detail.
In the rotating models, the strong quasi-periodic GW emission was observed at the time when the low-$T/|W|$ instability develops.
The peak frequency of the quasi-periodic GW signals increases with time from 400 to 800\,Hz in the $\Omega_0=2$\,rad\,s$^{-1}$ model and from 300 to 400\,Hz in the $\Omega_0=1$\,rad\,s$^{-1}$ model.
In the non-rotating model, the strong quasi-periodic GW signals with the peak frequency of $\sim200-300$\,Hz were obtained when the spiral SASI activity is vigorous.
These GW signals are within the detection limits of the current-generation detectors such as aLIGO, AdV, and KAGRA at their peak frequency at a source distance of 1\,Mpc.

We exploratory investigated the GW circular polarization by computing the Stokes parameters $I$ and $V$.
In the rotating models, the GWs observed along the polar direction show strong circular polarization ($I=V$), whereas the $V$ mode of the GWs observed along the equatorial direction is about two orders of magnitude smaller than $I$.
In the non-rotating model, since the rotation axis of the spiral SASI is tilted ($\theta\sim60^{\circ},\phi\sim-60^{\circ}$), both of the GWs observed along the polar and equatorial directions are circularly polarized.
We formulated the detector noise for the GW Stokes parameters in a simplified situation and compared the characteristic GW strain amplitudes for the Stokes parameters with the detector noises.
We found that both of the Stokes parameters $I$ and $V$ would be detectable at a source distance of 10\,kpc for all of our computed models.

We also evaluated the time variability of the neutrino emission.
In our rotating models, we confirmed the characteristic modulation of the neutrino signals due to the neutrino lighthouse effect. 
Comparing the neutrino modulation frequency with the GW frequency, we found that the neutrino peak frequency is $m/2$ times larger than the GW peak frequency, where $m$ represents the deformation mode of the PNS (around the rotation axis) induced by the low-$T/|W|$ instability.
 In the non-rotating model, the neutrino modulation frequency, as previously identified, is almost the same as the SASI frequency,
  and it is half of the GW frequency.
We discussed the detectability of these neutrino modulations by IceCube and HK at a source distance of 10\,kpc.
If we observe our rotating models along the equatorial direction, the characteristic peak in the power spectra for the event rates at IceCube would be above the noise level.
The characteristic peak due to the SASI in our non-rotating model observed along either polar or equatorial directions is also above the noise level.
This analysis indicates that the joint observation of the GW and neutrino signals would be useful to extract information on growing multiple modes of non-axisymmetric instabilities such as the low-$T/|W|$ instability and the spiral SASI activity, which is otherwise inaccessible by electromagnetic-wave observations.

Finally, we shall discuss several major limitations of our work. Even when the precollapse core is a slow rotator \citep{Heger05} but with a sufficiently weak initial magnetic fields, the magnetic fields may be amplified
 to a dynamically relevant strength by the magnetorotational instability \citep[MRI; e.g.][]{akiy03,Obergaulinger09,masada15,Rembiasz16a}. Although the role of the MRI on the postbounce dynamics remains still unclear, 
  the magnetic fields should play a key role in the angular momentum transport especially in the vicinity of the (differentially rotating) PNS surface. Possible impacts on the growth of the low-$T/|W|$ instability with magnetic fields remain to be clarified. If progenitors have both strong initial magnetic fields ($\sim\mathcal{O}(10^{12})$\,G) and rapid rotation, the magnetic fields are amplified due to the field wrapping to dynamically relevant strength
 shortly after bounce, leading to the formation of MHD jets.
This would also make the postbounce dynamics such as the shock evolution and the post-shock flow very different comparing to those of the non-magnetized models. Such difference should also affect the angular momentum transport, not to mention the onset criteria of the low-$T/|W|$ instability.
Although there are a few recent 3D MHD core-collapse simulations, the development of the low-$T/|W|$ instability has been only reported in \citet {Scheidegger10} and suggested in \citet{Kuroda20} (see, however, \citet{moesta14}).
Also, the convective dynamo may amplify the magnetic fields and modify the dynamics in the late phase of the core collapse \citep[e.g.][]{Raynaud20}.
To answer these questions, one apparently need to perform 3D GR-MHD core-collapse simulations (enough long to follow the growth of non-axisymmetric instabilities) at high numerical resolution, which is nevertheless computationally very expensive at present. 

Another limitation is that we are only able to present a small sample of simulation set, namely for one progenitor model with three different initial angular momentum.
 It remains to be answered how common the growth of the low-$T/|W|$ instability is in the massive stellar core-collapse.
For example, in terms of the progenitor mass, \citet{takiwaki16} and \citet{takiwaki18} performed 3D rapidly core-collapse simulations of 11.2 and 27.0\,M$_{\odot}$ models and found that the 11.2\,M$_{\odot}$ model explodes before the low-$T/|W|$ instability develops, whereas  the rapidly rotating 27.0\,M$_{\odot}$ model explodes assisted by the growth of the low-$T/|W|$ instability.
In terms of the precollapse rotation rates, the $\Omega_0=2$\,rad\,s$^{-1}$ model in this study showed both the one- and two- armed spiral flow, whereas the $\Omega_0=1$\,rad\,s$^{-1}$ model did only the two-armed spiral flow. Probably, a linear analysis of the low-$T/|W|$ instability \citep{Watts05,saijo06,passamonti15,Saijo18}, which has been extensively studied in the context of isolated cold neutron stars, could help understand the onset criteria of the instability in the (much more complicated) stellar core-collapse context.
With regards to the EOSs, the use of different EOSs may affect the development of the low-$T/|W|$ instability \citep{Scheidegger10,Saijo18}. Moreover, although the EOS we used, LS220, is effectively ruled out by constraints from nuclear physics \citep{Tews17}, we applied the LS220 EOS to our simulations since it has been quite widely used in the previous core-collapse simulations. Systematic studies with different EOSs that satisfy such nuclear physics constraints are required for better understanding in the core-collapse of rapidly rotating progenitors and their development of rotational instabilities.
Concerning the GW prediction, the GW  signatures from anisotropic neutrino emission (e.g., \citet{vart20,kotake11}) also remain to be studied.

Provided the expense of 3D-GR simulations with spectral neutrino transport, our simulation covers only $\sim 300$\,ms postbounce. Recently, it 
 is pointed out that galactic supernova
  neutrinos\footnote{Pioneering theoretical work of this topics especially focusing on the black hole formation includes \citet{Keil95,baumgarte96,Sumiyoshi07,Nakazato13,o'connor13} (see \citet{Horiuchi_kneller} for a review),}, depending on the neutron star mass, can be observed over $10$ s after bounce (\citet{suwa19}, see also \citet{nakazato20}). Possible phase transition from hadronic to quark matter in the PNS \citep{Fischer18} can be imprinted in the GW signals \citep{zha20}. Apparently, long-term simulation is needed in the GW-$\nu$ signal prediction by 3D(-GR-MHD), black-hole forming stellar collapse simulations, which is certainly one of the ultimate goals of researchers in this field. 
  
\section*{Acknowledgements}
KK and TT thank Thierry Foglizzo for stimulating discussions about the low-$T/|W|$ instability.
Numerical computations were
carried out on Cray XC50 at the Center for Computational Astrophysics,
National Astronomical Observatory of Japan and on Cray XC40 at YITP
in Kyoto University. This work was supported by Research Institute of
Stellar Explosive Phenomena at Fukuoka University and the associated project (No. 207002), and also by JSPS KAKENHI
Grant Number (JP17H05206, 
JP17K14306, 
JP17H01130, 
JP17H06364, 
 and JP18H01212 
).  
TK was supported by the ERC Starting Grant EUROPIUM-677912.
This research was also supported by MEXT as “Program for Promoting 
researches on the Supercomputer Fugaku” (Toward a unified view of 
he universe: from large scale structures to planets) and JICFuS.

\section*{Data Availability}
The data underlying this article will be shared on reasonable request to the corresponding author.

\bibliographystyle{mnras}
\bibliography{mybib.bib} 

\appendix
\section{Variance of Noise for Stokes Parameters} \label{appendix:A}
In this Appendix, we derive the variance of the noise for the Stokes parameters $I$ and $V$, $\sigma_{I_n}^2$ and $\sigma_{V_n}^2$, shown in Eq.~(\ref{eq:noiseI}) and Eq.~(\ref{eq:noiseV}).
Let us suppose that there are co-located two detectors that have a common feature regarding the (single-sided) noise spectral density, $S_n$, and each of them measures the plus mode and the cross mode of GWs, respectively.
In other words, a detector that measures the cross mode is tilted at 45$^\circ$ from another that measures the plus mode.
We assume that independent white Gaussian noises, $n_+$ and $n_\times$, are generated in each detector.
Then, the detector noises in the frequency space, $\Tilde{n}_+$ and $\Tilde{n}_\times$, follow
\begin{eqnarray}
  \left<|\Tilde{n}_+|^2\right>\Delta f=\left<|\Tilde{n}_\times|^2\right>\Delta f=\frac{1}{2}S_{n},
\end{eqnarray}
and their real and imaginary parts follow a independent normal distribution
\begin{eqnarray}
 P(x)&=&\frac{1}{\sqrt{2\pi a^2}}\exp{\left(-\frac{x^2}{2a^2}\right)},\\
 a^2&=&\frac{S_n}{4\Delta f}. 
\end{eqnarray}
Here $\left<\right>$ means the ensemble average.
Then, the average values of the noises for $I$ and $V$ , $I_n$ and $V_n$, are
\begin{eqnarray}
  \left< I_n \right> &=& \left< \frac{1}{2}\left(|\Tilde{n}_+|^2+|\Tilde{n}_\times|^2\right) \Delta f \right>  \nonumber\\
  &=& \frac{1}{2}S_n,\\
  \left< V_n \right> &=& \left< \frac{i}{2}\left(\Tilde{n}_+\Tilde{n}_\times^* - \Tilde{n}_\times\Tilde{n}_+^* \right)\Delta f \right> \nonumber\\
  &=& 0.
\end{eqnarray}
The average value of the squared noises, $I_n^2$ and $V_n^2$, are
\begin{eqnarray}
  \left< I_n^2 \right> &=& \left< \frac{1}{4}\left(|\Tilde{n}_+|^2+|\Tilde{n}_\times|^2\right)^2\Delta f^2 \right> \nonumber\\
  &=& \frac{\Delta f^2}{4}\left(\left<|\Tilde{n}_+|^4\right>+2\left<|\Tilde{n}_+|^2\right>\left<|\Tilde{n}_\times|^2\right>+\left<|\Tilde{n}_\times|^4\right>\right) \nonumber\\
  &=& \frac{3}{8}S_n^2,\\
  \left< V_n^2 \right> &=& \left< \frac{-1}{4}\left(\Tilde{n}_+\Tilde{n}_\times^* - \Tilde{n}_\times\Tilde{n}_+^* \right)^2\Delta f^2 \right> \nonumber\\
  &=& (\left<\mathrm{Re}\left(\Tilde{n}_+\right)^2\right>\left<\mathrm{Im}\left(\Tilde{n}_\times\right)^2\right> \nonumber\\
  &&+\left<\mathrm{Im}\left(\Tilde{n}_+\right)^2\right>\left<\mathrm{Re}\left(\Tilde{n}_\times\right)^2\right>\nonumber\\
  &&-2\left<\mathrm{Re}\left(\Tilde{n}_+\right)\right>\left<\mathrm{Re}\left(\Tilde{n}_\times\right)\right>\nonumber\\
  &&\ \ \ \ \left<\mathrm{Im}\left(\Tilde{n}_+\right)\right>\left<\mathrm{Im}\left(\Tilde{n}_\times\right)\right> )  \Delta f^2 \nonumber\\
  &=& \frac{1}{8}S_n^2.
\end{eqnarray}
Note that $\int x^2P(x) dx = a^2$, and $\int x^4P(x) dx = 3a^4$.
Thus, we get Eq.~(\ref{eq:noiseI}) and Eq.~(\ref{eq:noiseV}) for $\sigma_{I_n}^2$ and $\sigma_{V_n}^2$:
\begin{eqnarray}
  \sigma_{I_n}^2 &=& \left< I_n^2 \right> -\left< I_n \right>^2 \nonumber\\
  &=& \frac{1}{8}S_{n}^2, \label{eq:noiseIapp}\\
  \sigma_{V_n}^2 &=& \left< V_n^2 \right> -\left< V_n \right>^2 \nonumber\\
  &=& \frac{1}{8}S_{n}^2. \label{eq:noiseVapp}
\end{eqnarray}

\bsp	
\label{lastpage}
\end{document}